\documentclass[11pt,a4paper]{article}
\usepackage{jheppub}
\usepackage{placeins}
\usepackage{braket}
\usepackage{array}
\newcolumntype{P}[1]{>{\centering\arraybackslash}p{#1}}
\usepackage{amsfonts}
\usepackage{mathtools}
\usepackage{mathrsfs}
\usepackage{amssymb}
\usepackage{float}
\usepackage{booktabs}
\usepackage{tikz}
\usetikzlibrary{arrows.meta,shapes.misc}
\usepackage{comment}
\usepackage{yhmath}
\usetikzlibrary{decorations.markings,arrows,calc}
\allowdisplaybreaks

\newcommand{\cA}{\mathcal A}
\newcommand{\cB}{\mathcal B}

\newcommand{\cM}{\mathcal M}

\newcommand{\hL}{\mathsf h_{\rm L}}
\newcommand{\hR}{\mathsf h_{\rm R}}
\newcommand{\cQ}{\mathcal Q}
\newcommand{\cE}{\mathcal E}
\newcommand{\sq}{\mathsf q}
\newcommand{\p}{\partial}

\newcommand{\be}{\begin{equation}}
\newcommand{\ee}{\end{equation}}
\newcommand{\Neff}{N_{\rm eff}}
\newcommand{\TTbar}{T\bar T}

\newcommand{\PV}{\mathrm{PV}}

\title{
A spin chain as a deformed conformal field theory: an ODE/IM approach}

\author[a,b,c]{Sujay K. Ashok,}
\author[c]{Nicol\`o Primi,}
\author[c]{and Roberto Tateo}

\affiliation[a]{The Institute of Mathematical Sciences,
IV Cross Road, C.I.T. Campus, Taramani, Chennai 600113, India}
\affiliation[b]{Homi Bhabha National Institute,
Training School Complex, Anushakti Nagar, Mumbai 400094, India}
\affiliation[c]{Dipartimento di Fisica, Universit\`a di Torino
and INFN, Sezione di Torino, Via Pietro Giuria 1, I-10125 Torino, Italy}

\emailAdd{sashok@imsc.res.in}
\emailAdd{nicolo.primi@unito.it} 
\emailAdd{roberto.tateo@unito.it}

\abstract{
The ODE/IM correspondence relates ordinary differential equations to integrable quantum field theories in the continuum. We ask whether it extends to integrable spin chains of finite length $N$. We analyse the twisted XXZ chain at a special anisotropy and twist, focusing on states in its Lee--Yang scaling sector. Their finite-size corrections are expected to come from the Lee--Yang conformal field theory deformed by irrelevant operators, and with the help of the ODE/IM correspondence we make this deformation explicit for the energy and the local conserved charges. Near each of the two edges of the distribution of the Bethe roots, the chain behaves as one chirality of the Lee--Yang theory in a generalised Gibbs ensemble, whose chemical potentials depend on $N$. The ODE/IM correspondence describes each of these ensembles by a cubic oscillator, whose coefficients we determine as series in $1/N$. The two chiralities are not independent: finite-size effects couple them, so that the charges of each shift the chemical potentials of the other. This turns out to be a generalised $T\bar T$-type interaction. 
Solving this coupling self-consistently, we obtain the energy and the local conserved charges of the chain as series in $1/N$, with no free parameters. We verify these series against the Bethe ansatz, for the ground state and for an excited state.
}

\begin{document}
\maketitle

\section{Introduction and Summary}
\label{sec:intro}

The ODE/IM correspondence relates the spectral data of ordinary differential equations to the functional relations and integral equations of two-dimensional integrable models~\cite{Dorey:1998pt,Bazhanov:1998wj,Dorey:1999uk,Suzuki:1999hu,Bazhanov:2003ni,Dorey:2007zx,Dorey:2025odeim,Fioravanti:2021gbd}. In its original form, it states that the Stokes multipliers of the Schr\"odinger equation with a monomial potential satisfy the functional relations of the conformal limit of a perturbed minimal model. The associated integral equations include the thermodynamic Bethe ansatz
(TBA) and related non-linear integral equations
(NLIEs)~\cite{Zamolodchikov:1989cf,Klassen:1990dx,Klumper:1991jsp,
Klumper:1992pearce,Destri:1992qk,Destri:1994bv,Bazhanov:1996dr}. A key feature of the correspondence is that from the WKB expansion of the ODE's Stokes multipliers, one can read off the  eigenvalues of the local integrals of motion~\cite{Bazhanov:1994ft} of the CFT. The simplest example of the ODE/IM correspondence has the cubic oscillator on the ODE side, and the Lee--Yang minimal model $\cM_{2,5}$ on the CFT side~\cite{Dorey:1998pt}, with the functional relations being the ones studied by Sibuya~\cite{Sibuya:1984fe}.
In~\cite{Ashok:2026gge} the ODE/IM correspondence was extended to generalised Gibbs ensembles (GGEs)~\cite{Vidmar:2016, Essler:2014mya} of the Lee--Yang model. 
In them, the chemical potentials of the higher local conserved charges, which we call times, as is customary for the couplings of the commuting charges of an integrable hierarchy, enter the TBA equation as coefficients of additional terms in the driving function; on the ODE side, these terms are reproduced by a deformation of the cubic oscillator potential, whose coefficients depend on the spectral parameter. These coefficients are fixed, order by order in the times, by requiring that the large-energy WKB period reproduces the driving term of the corresponding TBA equation; the subleading WKB periods then determine the deformed eigenvalues of the local conserved charges.

Exact ODE descriptions of finite spin chains are known in some special cases. In particular, for the anisotropy value $\Delta=\frac12$, the ground state of the XXZ
chain of finite length has a remarkable combinatorial structure~\cite{Fridkin:1999ga,Fridkin:2000fsz,Stroganov:2000ue,Razumov:2000sc},
and its Bethe ansatz equations can be encoded in an ordinary differential equation. This ODE description was
generalised to higher spin in~\cite{Dorey:2003sv}, where the connection with the Heun equation was also discussed.
These finite-size ODE constructions exploit the special structure of the chain at $\Delta=\frac12$. In addition, ODEs also describe the scaling behaviour of the lattice $Q$- and $T$-functions of the XXZ spin chain in the vicinity of the two Fermi points~\cite{Dorey:2007zx,Kotousov:2019nvt,Kotousov:2025qfun}, and of its inhomogeneous generalisations~\cite{Bazhanov:2019xvy,Bazhanov:2020new,Gehrmann:2024kot}. This motivates the search for an ODE description for a finite size spin chain model.

In this paper we adapt the construction of~\cite{Ashok:2026gge} to a spin chain whose scaling limit is the Lee--Yang theory. This model is the twisted XXZ chain of $N$ sites at anisotropy $\gamma=3\pi/5$ and twist $\varphi=2\pi/5$. For these values of the anisotropy and twist, the fusion hierarchy closes, and a suitably normalised transfer-matrix eigenvalue satisfies the Lee--Yang $Y$-system. This $Y$-system, together with the appropriate analyticity properties of the transfer-matrix eigenvalue, can be recast as a lattice version of the Lee--Yang TBA equation. For the ground state, the resulting integral equation has the same kernel as the conformal Lee--Yang TBA equation, while its driving term is replaced by
\be
\nu(\theta)=N\,\ell(\theta)\,,\qquad
\ell(\theta)=\log\frac{\cosh\theta+\frac{\sqrt3}{2}}
{\cosh\theta-\frac{\sqrt3}{2}}\,.
\label{eq:intro-lattice-driving}
\ee

Functional equations have long been used to derive TBA-type integral equations directly on the lattice~\cite{Klumper:1991jsp,Klumper:1992pearce}. The same lattice TBA equation was obtained in the $A_4$ RSOS realisation of the Lee--Yang model in~\cite{Bajnok:2014fca}. We derive it here directly from the analytic properties of the transfer-matrix eigenvalue, in conventions adapted to the twisted XXZ representation, and use it as the starting point for the finite-size ODE reconstruction.

The Bethe roots describing the ground state of the XXZ chain lie in a finite region of the rapidity line, whose two edges define the Fermi points. 
They are also visible in the solution of the lattice TBA equation, where the occupation function changes rapidly across each of them and approaches a step-like profile as $N$ becomes large.
Near either Fermi point, the lattice driving term admits an expansion in exponentials associated with spins $s\equiv1,5\pmod6$. The spin-$1$ term reproduces the driving term $e^{\theta}$ of the conformal Lee--Yang TBA equation, while the higher-spin terms are suppressed by increasing powers of $1/N$. The lattice TBA equation therefore takes, locally, the form of a chiral GGE deformation of the Lee--Yang TBA, with higher-spin chemical potentials that vanish as $N\to\infty$.
This allows us to apply the ODE/IM construction of~\cite{Ashok:2026gge} separately near each Fermi point. The corresponding ODE is a cubic oscillator whose coefficients are determined order by order in $1/N$ and which reduces, as $N\to\infty$, to the conformal cubic oscillator. Its periods determine the chiral conserved charges associated with that Fermi point, which can be organised in terms of the deformed KdV hierarchy~\cite{Karimi:2026kdv,Ashok:2026gge}.

The lattice TBA equation couples the two Fermi points through finite-size effects. In particular, the spin-$s$ time entering the local description near one Fermi point receives a shift proportional to the spin-$s$ conserved charge associated with the other Fermi point. This has the same self-consistent structure as the generalised $T\bar T$ flows~\cite{Smirnov:2016lqw,Cavaglia:2016oda,Conti:2019dxg,Hernandez-Chifflet:2019sua,LeFloch:2019wlf}, with the spin-$1$ sector taking the familiar $T\bar T$-type form. The resulting shifts close into self-consistency equations for the two Fermi-point expansions, which we solve order by order in $1/N$. They determine a cubic oscillator near each Fermi point, whose Stokes data reproduce the local large-$N$ expansion of the $Y$-function order by order in $1/N$. Solving the inverse problem numerically across the band, we find that the reconstructions selected from the two Fermi points are mapped into one another by parity.

Using these self-consistent expansions near the Fermi points, we compute the energy and the higher local charges of the chain as series in $1/N$. The first terms agree with the known effective description of critical spin chains~\cite{Cardy:1987,Reinicke:1987,Alcaraz:1987uk,Lukyanov:1997bp}, in which the finite-size corrections to the conformal spectrum~\cite{Cardy:1984,Cardy:1986,Blote:1986,Affleck:1986} are described by the conformal field theory perturbed by irrelevant operators, while the cubic reconstruction provides a systematic extension to higher orders. We test our results against the exact energy and charges obtained from the Bethe roots, both for the ground state and for the first descendant, in which the left and right regions carry different chiral data.

Finally, we exhibit a simple mathematical structure that underlies the coupling between the left and right regions around the two Fermi points. To each region we associate a generating function, a potential whose derivatives with respect to its times are its charges. We show that the coupling of the two potentials has the form of a Hamilton--Jacobi equation, of the type that governs the energy levels under the $T\bar T$ deformation. Solving this equation order by order in $1/N$ provides a systematic way to generate the finite-size expansion of the energy and higher local charges.

This paper is organised as follows. In Section~\ref{sec:chain}, we describe the spin chain and its lattice TBA equation. In Section~\ref{sec:inverse}, we formulate the reconstruction of the ODE as an inverse monodromy problem. In Section~\ref{sec:chiral}, we review the construction of~\cite{Ashok:2026gge} and apply it near one Fermi point, where the times are small and the cubic and its charges are series in $1/N$. In Section~\ref{sec:gluing}, we analyse the coupling to the second Fermi point. In Section~\ref{sec:energy}, we compute the finite-size energy and the local charges of the chain, and in Section~\ref{sec:gluing-excited} we test them on the first descendant. Finally, in Section~\ref{sec:hj}, we show that the coupling between the two Fermi points can be formulated as a Hamilton--Jacobi equation.
We conclude in Section~\ref{sec:conclusions} with a discussion of our results and some open problems. Appendix~\ref{app:Lambda} collects the analytic properties of the transfer-matrix eigenvalue used to derive the lattice TBA equation, Appendix~\ref{app:classes} the classical and quantum periods of the cubic, and Appendix~\ref{app:sumrule} the derivation of the sum rule.

\section{The twisted XXZ chain and its lattice TBA}
\label{sec:chain}

The Lee--Yang model can be realised as the scaling limit of different
lattice models, including the Ising quantum chain in an imaginary
longitudinal field~\cite{vonGehlen1991} and the twisted XXZ chain
considered here. For particular values of the anisotropy and twist, a suitably normalised ground-state transfer-matrix eigenvalue satisfies the Lee--Yang $Y$-system, which can in turn be expressed as a lattice TBA equation. In this section, we review this derivation, leaving detailed proofs to Appendix~\ref{app:Lambda}.
Our starting points are the TQ relation for the XXZ spin chain \cite{Baxter:1982zz} and the assumption that, in the ground state, the Bethe roots are real, which leads to certain analytic properties of the transfer matrix ground-state eigenvalue. 

\subsection{The Bethe ansatz and conserved charges} 
\label{sec:chain-ba}

We consider a spin-$\tfrac12$ XXZ chain of $N$ sites (with $N$ even) in the critical regime, defined by the Hamiltonian
\be
H=\frac12\sum_{j=1}^{N}\left[\sigma^x_j\sigma^x_{j+1}+\sigma^y_j\sigma^y_{j+1}+\Delta\left(\sigma^z_j\sigma^z_{j+1}-1\right)\right],
\qquad \Delta=\cos\gamma\,,\qquad 0<\gamma<\pi\,,
\label{eq:H}
\ee
with twisted boundary conditions
$\sigma^\pm_{N+1}=e^{\pm i\varphi}\sigma^\pm_1$, $\sigma^z_{N+1}=\sigma^z_1$. The spectrum of $H$ is invariant under the parity transformation $\varphi\to-\varphi$, and we fix the sign of $\varphi$ by using the Bethe equations below. In the sector with $M$ down spins, Bethe eigenstates are parametrised by rapidities $\lambda_1,\dots,\lambda_M$ satisfying the Bethe ansatz equations
\be
\left[\frac{\sinh(\lambda_j+i\gamma/2)}{\sinh(\lambda_j-i\gamma/2)}\right]^N=e^{-i\varphi}\prod_{k\neq j}^M\frac{\sinh(\lambda_j-\lambda_k+i\gamma)}{\sinh(\lambda_j-\lambda_k-i\gamma)}\,.
\label{eq:BAE}
\ee
We can define the momentum and the scattering phase as
\be
p(\lambda)=2\arctan\!\big(\cot\tfrac{\gamma}{2}\tanh\lambda\big)\,,\qquad
\Theta(\lambda)=2\arctan\!\big(\cot\gamma\,\tanh\lambda\big)\,,
\label{eq:pTheta}
\ee
and using these definitions we can rewrite the Bethe equations in logarithmic form:
 \be
Z(\lambda_j)-\varphi=2\pi I_j\,,\quad\text{with}\quad Z(\lambda)=N p(\lambda)-\sum_{k=1}^{M}\Theta(\lambda-\lambda_k)\,,
\label{eq:Zlog}
\ee
where $I_j$ are integers or half-integers, according to the parity of $M$.
The energy of a Bethe state is a sum of one-particle contributions proportional to the derivative of the momentum
\be
\cE=-\sin\gamma\sum_{j=1}^{M}p'(\lambda_j)\,,\qquad
p'(\lambda)=\frac{2\sin\gamma}{\cosh2\lambda-\cos\gamma}\,,
\ee
and the eigenvalues of the higher local charges are given by higher derivatives of this one-particle term:
\be
\cQ_n=-\sin\gamma\sum_{j=1}^M\left.\frac{d^{\,n-1}p(\lambda)}{d\lambda^{n-1}}\right|_{\lambda=\lambda_j}\,,\qquad n\geq 2\,.
\ee
We will focus on the ground state, which lies in the sector with $M=N/2$ and has the quantum numbers $I_j$ distributed symmetrically around zero. The Bethe roots are then real and lie in a finite region of the rapidity line, whose edges define the Fermi points and move logarithmically with the length $N$ of the chain. The twist breaks the symmetry of the roots under $\lambda\to-\lambda$, although the quantum numbers are symmetric.

To see the relation to the Lee--Yang model, we recall that the low-energy limit of the XXZ spin chain is a compact boson, whose radius is set by $\gamma$ and whose effective central charge is lowered below $1$ by the twist $\varphi$~\cite{Alcaraz:1987uk,Alcaraz:1987zr,Klumper:1991}:
\be
c_{\rm eff}=1-\frac{3\,\varphi^2}{2\pi(\pi-\gamma)}\,.
\label{eq:ceff}
\ee
Choosing the values
\be
\gamma=\frac{3\pi}{5}\,,\qquad \varphi=\frac{2\pi}{5}\,,
\ee
 gives the effective central charge of the Lee--Yang minimal model $\cM_{2,5}$:
\be
c_{\rm eff}=c-24\,h_\phi=\frac25\,,\qquad c=-\frac{22}{5}\,,\qquad h_\phi=-\frac15\,.
\ee
Here $\phi$ is the primary of lowest weight. At these values of the anisotropy and twist, the low-energy sector of the twisted six-vertex model is described by the Lee--Yang minimal model. Below we will see that this identification is also reflected in the functional relation satisfied by the ground-state transfer-matrix eigenvalue, which leads to the Lee--Yang $Y$-system.

\subsection{The transfer matrix and the lattice TBA}
\label{sec:chain-tba}

The twisted transfer matrix of the XXZ chain can be built from the six-vertex trigonometric R-matrix and forms a commuting family depending on the spectral parameter $v$. The Hamiltonian~\eqref{eq:H} and the higher local conserved charges are obtained from suitable logarithmic derivatives of this family. The transfer matrix commutes with the Hamiltonian and the higher local conserved charges, and the Bethe states provide simultaneous eigenstates of these commuting operators. We will denote the eigenvalue of the transfer matrix on the ground state of the spin chain as $\Lambda(v)$; this quantity can be expressed in terms of the Bethe roots $\{\lambda_j\}_{j=1}^M$ via the TQ relation~\cite{Baxter:1982zz,Klumper:1991}\footnote{The TQ equation holds for every eigenvalue; in this section we mainly consider the ground state.}
\be
\Lambda(v)=e^{\frac{i\varphi}{2}}\sinh^N\!\big(v+\tfrac{i\gamma}{2}\big)\prod_{j=1}^{M}\frac{\sinh(v-\lambda_j-i\gamma)}{\sinh(v-\lambda_j)}
+e^{-\frac{i\varphi}{2}}\sinh^N\!\big(v-\tfrac{i\gamma}{2}\big)\prod_{j=1}^{M}\frac{\sinh(v-\lambda_j+i\gamma)}{\sinh(v-\lambda_j)}\,.
\label{eq:TQequation}
\ee
In this expression, the potential poles at $v=\lambda_j$ cancel thanks to the Bethe equations~\eqref{eq:BAE}. We will use three properties of $\Lambda$. The first follows from the cancellation of the poles just described; the other two are proved in Appendix~\ref{app:Lambda}:
\begin{itemize}
\item $\Lambda$ is an entire function of the spectral parameter.
\item For the ground state at $\gamma=3\pi/5$ and $\varphi=2\pi/5$, whose Bethe roots are real and for which $M=N/2$, all the zeros of $\Lambda$ lie on the line ${\rm Im}\,v=\pi/2\pmod\pi$.
\item $\Lambda$ satisfies the functional relation
\be
\Lambda(v)\,\Lambda(v+i\gamma)=\sinh^N\!\big(v-\tfrac{i\gamma}2\big)\sinh^N\!\big(v+\tfrac{3i\gamma}2\big)+(-1)^M\sinh^N\!\big(v+\tfrac{i\gamma}2\big)\,\Lambda(v+3i\gamma)\,.
\label{eq:Lambda-relation}
\ee
This is derived in Appendix~\ref{app:closure} starting from the TQ relation, and expresses the root-of-unity closure of the fusion hierarchy~\cite{Bazhanov:1988bb,Belavin:2001,Klumper:1991jsp,Klumper:1992pearce,Takahashi:1999}.
\end{itemize}

\subsubsection{The $Y$-system}

We can now define the lattice\footnote{We call lattice quantities those built from the transfer matrix of the chain at finite $N$, such as $Y$, its TBA equation and its driving term, to distinguish them from their conformal counterparts, in which the driving term is $e^{\theta}$.} $Y$-function from $\Lambda$, using the rapidity variable $\theta$ defined in terms of the spectral parameter $\theta\equiv\pi v/\gamma$, as
\be
Y(\theta)=(-1)^M\,\big[g(v)\big]^N\,\Lambda(v)\,,\qquad
g(v)=\frac{\sinh\!\big(v-\frac{5i\gamma}{6}\big)}{\sinh\!\big(v-\frac{i\gamma}{6}\big)\,\sinh\!\big(v+\frac{i\gamma}{6}\big)}\,.
\label{eq:Y-Lambda}
\ee
At $\gamma=3\pi/5$ and $\varphi=2\pi/5$, the relation~\eqref{eq:Lambda-relation} when written in terms of the function $Y$ defined in~\eqref{eq:Y-Lambda}, takes the form of the celebrated Lee--Yang $Y$-system~\cite{Zamolodchikov:1989cf}:
\be
Y\!\left(\theta+\tfrac{i\pi}{3}\right)Y\!\left(\theta-\tfrac{i\pi}{3}\right)=1+Y(\theta)\,.
\label{eq:Ysystem}
\ee
This is shown in detail in Appendix \ref{app:Ysystem}. 

\subsubsection{The walls}
\label{sec:chain-walls}

From its definition, the function $g^N$ has poles of order $N$ at $\theta=\pm i\pi/6$ and zeros of order $N$ at $\theta=\pm5i\pi/6$. Since $\Lambda$ is entire, the poles of $g^N$ at $\theta=\pm i\pi/6$ can only be cancelled by zeros of $\Lambda$. For the ground state, the zeros of $\Lambda$ lie on ${\rm Im}\,\theta=5\pi/6\pmod{5\pi/3}$ and therefore do not cancel these poles (see the second property of $\Lambda$ in Section~\ref{sec:chain-tba}). Hence $Y$ has poles of order $N$ at $\theta=\pm i\pi/6$. Using the $Y$-system, we can then show that these poles are accompanied by zeros of order $N$ at $\theta=\pm5i\pi/6$. Indeed, evaluating~\eqref{eq:Ysystem} at $\theta=\pm i\pi/2$ gives
\be
Y(\pm 5 i \pi/6)=\frac{1+Y(\pm i \pi/2)}{Y(\pm i \pi/6)}\,.
\ee
For the ground state we prove in Appendix~\ref{app:momentum} that $Y(\pm i\pi/2)=1$. Thus the pole of order $N$ in the denominator of the RHS gives a zero of order exactly $N$ in the LHS. 

\begin{figure}[!tbh]
\centering
\begin{tikzpicture}[x=1.0cm,y=0.6cm,>=stealth,
  pole/.style={draw=red!80!black,line width=1.1pt,cross out,inner sep=0pt,minimum size=7pt},
  zero/.style={draw=blue!70!black,line width=1.1pt,circle,inner sep=0pt,minimum size=6pt}]
  \fill[gray!18] (-3,-2) rectangle (3,2);
  \draw[gray,dashed] (-3,2)--(3,2) (-3,-2)--(3,-2);
  \draw[->] (-3.3,0)--(3.5,0) node[right]{$\mathrm{Re}\,\theta$};
  \draw[->] (0,-5.8)--(0,6.0) node[above]{$\mathrm{Im}\,\theta$};
  \foreach \y/\l in {1/{\pi/6},2/{\pi/3},5/{5\pi/6}}{
     \draw (-3,\y)--(-2.9,\y); \draw (-3,-\y)--(-2.9,-\y);
     \node[left,font=\small] at (-3.02,\y) {$\l$};
     \node[left,font=\small] at (-3.02,-\y) {$-\l$};}
  \node[pole] at (0,1) {}; \node[pole] at (0,-1) {};
  \node[zero] at (0,5) {}; \node[zero] at (0,-5) {};
  \node[right,font=\footnotesize,align=left] at (3.6,5) {zero of order $N$};
  \node[right,font=\footnotesize,align=left] at (3.6,1) {pole of order $N$};
  \node[right,font=\footnotesize,align=left] at (3.6,-1) {pole of order $N$};
  \node[right,font=\footnotesize,align=left] at (3.6,-5) {zero of order $N$};
  \node[font=\footnotesize,align=center] at (1.7,-1.0) {strip\\$|\mathrm{Im}\,\theta|<\pi/3$};
\end{tikzpicture}
\caption{The walls of the lattice $Y$-function in the complex rapidity plane: poles of order $N$ at $\pm i\pi/6$ (crosses) and zeros of order $N$ at $\pm5i\pi/6$ (circles). The integration contour of the lattice TBA lies on the real axis; the shaded strip $|{\rm Im}\,\theta|<\pi/3$ contains no other zeros or poles of $Y$ for the ground state.}
\label{fig:walls}
\end{figure}
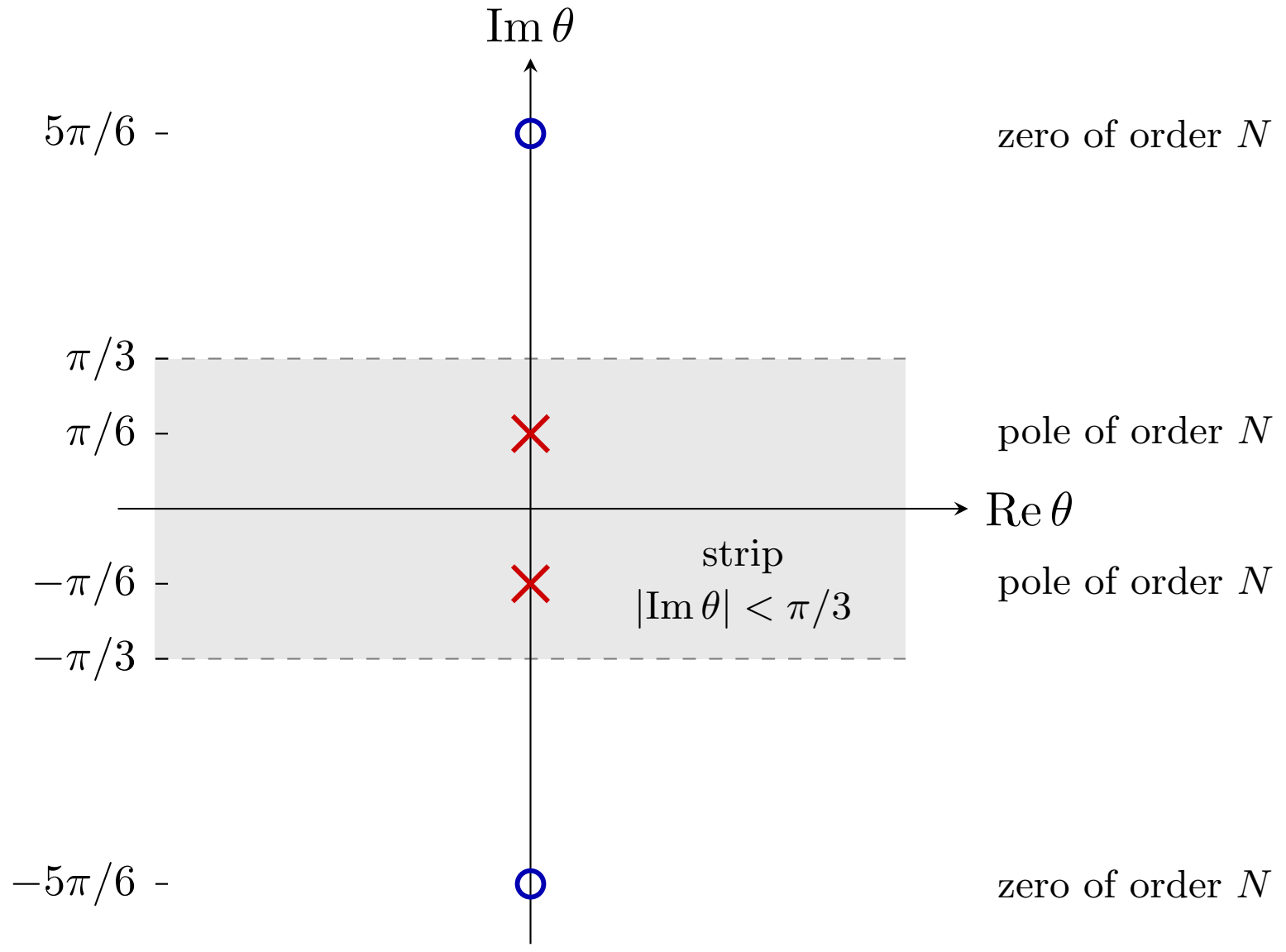

We call the four points $\theta=\pm i\pi/6,\,\theta=\pm 5i\pi/6$ the walls (see Figure~\ref{fig:walls}),
\be
\begin{aligned}
&\text{poles of order }N\text{ at }\theta=\pm\frac{i\pi}{6}\,,\\
&\text{zeros of order }N\text{ at }\theta=\pm\frac{5i\pi}{6}\,.
\end{aligned}
\label{eq:walls}
\ee
The remaining zeros of $Y$ are inherited from $\Lambda$
and lie on the line \({\rm Im}\,\theta=5\pi/6\).

Since the function $Y$ is periodic with period $5i\pi/3$, 
it is natural to define the strip $|{\rm Im}\,\theta|<5\pi/6$. The zeros of $Y$ lie precisely at the boundaries of this strip, and inside it the only singularities of $Y$ are the poles located at $\pm i\pi/6$. This strip contains the physical strip $|{\rm Im}\,\theta|<\pi/3$, inside which the TBA equation is valid.

\subsubsection{The lattice TBA equation}

We derive the lattice TBA equation from the \(Y\)-system~\eqref{eq:Ysystem}, by using the fact that the walls~\eqref{eq:walls} are the only zeros and poles of \(Y\) in the strip \(|{\rm Im}\,\theta|<5\pi/6\), together with its prescribed behaviour at large \(|{\rm Re}\,\theta|\). We first define
\be
f(\theta)=\frac{\cosh\theta+\frac{\sqrt3}{2}}{\cosh\theta-\frac{\sqrt3}{2}}\,,\qquad
f\big(\theta+\tfrac{i\pi}{3}\big)\,f\big(\theta-\tfrac{i\pi}{3}\big)=f(\theta)\,.
\label{eq:driving-f}
\ee
The function $f^N$ is real on the real axis and even; it tends to $1$ at large $|{\rm Re}\,\theta|$, and in the  strip \(|{\rm Im}\,\theta|<5\pi/6\) its poles and zeros are exactly the walls~\eqref{eq:walls}. The rescaled function
\be
\widetilde Y(\theta)=\frac{Y(\theta)}{\big(f(\theta)\big)^N}
\ee
is therefore analytic and non-zero in the period strip $|{\rm Im}\,\theta|<5\pi/6$, which contains the physical strip $|{\rm Im}\,\theta|\le\pi/3$ used below, and it tends to $\frac{1+\sqrt5}{2}$ at large $|{\rm Re}\,\theta|$ (see Appendix~\ref{app:Yasymptotics}). Dividing~\eqref{eq:Ysystem} by $Y$ and using~\eqref{eq:driving-f} gives
\be
\log\widetilde Y\big(\theta+\tfrac{i\pi}3\big)+\log\widetilde Y\big(\theta-\tfrac{i\pi}3\big)-\log\widetilde Y(\theta)=\log\Big(1+\frac1{Y(\theta)}\Big)\,.
\ee
We can solve this equation using the same methods as~\cite{Zamolodchikov:1991et}, obtaining an integral equation 
\be
\log\widetilde Y(\theta)=-\int_{-\infty}^{\infty}d\theta'\,\Phi(\theta-\theta')\,\log\big(1+Y^{-1}(\theta')\big)\,,
\ee
where $\Phi$ is the Lee--Yang kernel~\cite{Zamolodchikov:1989cf}, which is the logarithmic derivative of the Lee--Yang two-body scattering amplitude~\cite{Cardy:1989fw,Smirnov:1989fa}:
\be
\Phi(\theta)=-\frac{2\sqrt3}{\pi}\,\frac{\cosh\theta}{1+2\cosh2\theta}\,.
\label{eq:kernel}
\ee
Adding back $N\log f$ leads to what we call the lattice TBA equation:
\be
\epsilon(\theta)=\nu(\theta)-\int_{-\infty}^{\infty}d\theta'\,\Phi(\theta-\theta')\,L(\theta')\,,\quad\text{with}\quad \epsilon\equiv\log Y\,,\quad L\equiv\log\big(1+e^{-\epsilon}\big)\,,
\label{eq:latTBA}
\ee
where the driving term now takes the form:
\be
\nu(\theta)=N\ell(\theta)\,,\qquad \ell(\theta)=\log f(\theta)\,.
\label{eq:driving_term}
\ee
The same equation was obtained for the $A_4$ RSOS model in~\cite{Bajnok:2014fca}, by the method of~\cite{Klumper:1991jsp,Klumper:1992pearce}.

\subsection{The Fermi points and the conformal limit}
\label{sec:chain-fermi}

The driving term of the lattice TBA equation is $\nu(\theta)=N\ell(\theta)$, with
\be
\ell(\theta)=\log\frac{\cosh\theta+\frac{\sqrt3}{2}}{\cosh\theta-\frac{\sqrt3}{2}}\,.
\label{eq:ell-def}
\ee
It can be rewritten as an infinite sum over odd spins:
\begin{align}
\nu(\theta)&=4N\sum_{s\ {\rm odd}}\cos\frac{\pi s}{6}\,\frac{e^{-s|\theta|}}{s}\notag\\
&=N\left(2\sqrt3\,e^{-|\theta|}-\frac{2\sqrt3}{5}e^{-5|\theta|}-\frac{2\sqrt3}{7}e^{-7|\theta|}+\frac{2\sqrt3}{11}e^{-11|\theta|}+\dots\right)\,.
\label{eq:f-series}
\end{align}
It is even in $\theta$ and decays exponentially at large $|\theta|$. Only the odd integers with $\cos(\pi s/6)\neq0$ appear, that is $s\equiv1,5\pmod6$. These are the spins of the non-trivial local conserved charges of the Lee--Yang model, which can be switched on in its GGE~\cite{Ashok:2026gge}. The kernel~\eqref{eq:kernel} of the lattice TBA equation has an expansion of the same form at large $|\theta|$, with the same spins:
\be
\Phi(\theta)\simeq-\frac{\sqrt3}{\pi}\sum_{s}\chi(s)\,e^{-s|\theta|}\,,\qquad
\chi(s)=\begin{cases}+1 & s\equiv1 \pmod 6\\ -1 & s\equiv 5\pmod 6\end{cases}\,.
\label{eq:kernel-tail}
\ee
The leading term of the driving term is $2\sqrt3\,Ne^{-|\theta|}$, so at large $N$ it becomes of order one at $|\theta|=a$, with
\be
a=\log(2\sqrt3N)\,.
\label{eq:a-def}
\ee
The points $\theta=\pm a$ give the leading large-$N$ positions of the two Fermi points, one for each chirality of the low-energy theory. As $N\to\infty$ the Bethe roots of the ground state fill the interval between them, up to a range of order one in $\theta$ at each end, as in Section~\ref{sec:chain-ba}. We call the interval $|\theta|<a$ the band.
\begin{figure}
\centering
\includegraphics[width=.80\textwidth]{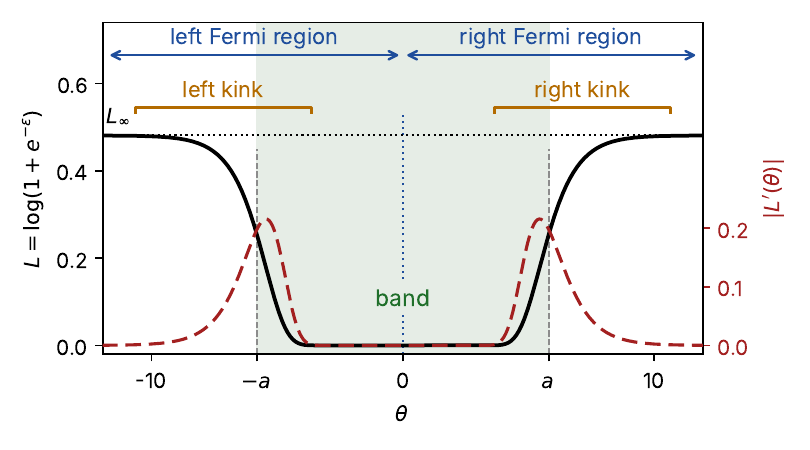}
\caption{The function $L=\log(1+e^{-\epsilon})$ of the lattice TBA equation for $N=96$ (solid, left axis) and $|L'(\theta)|$ (dashed, right axis). In the band $|\theta|<a$, with $a=\log(2\sqrt3\,N)$, $L$ is exponentially small, while outside it tends to $L_\infty=\log\frac{1+\sqrt5}{2}$ (dotted horizontal line). The kinks, bracketed in the figure, are the two regions where $L$ changes between these values and where $L'$ is not zero. The dashed vertical lines mark the leading-order positions $\theta=\mp a$ of the Fermi points, and the dotted vertical line at $\theta=0$ splits the real line into the left and right Fermi regions.}
\label{fig:fermi-points}
\end{figure}
Deep inside the band, away from the two Fermi-point regions, the driving term is large, $\epsilon(\theta)\simeq N\ell(\theta)$ grows linearly with $N$ at fixed $\theta$, and the occupation function $L(\theta)=\log(1+e^{-\epsilon})$ defined in~\eqref{eq:latTBA} is exponentially small. Outside the band $\nu$ is small, and $Y$ tends to the value $\frac{1+\sqrt5}{2}$ found above, so that $\epsilon=\log Y$ and $L$ both tend to $L_\infty=\log\frac{1+\sqrt5}{2}$. The function $L$ therefore develops a rapid crossover near each Fermi point, as shown in Figure~\ref{fig:fermi-points}. 

We split the real line at the centre of the band, where
$L$ is smallest. We refer to the regions $\theta<0$ and
$\theta>0$ as the left and right Fermi regions,
respectively. The kink itself is the crossover region around the corresponding
Fermi point. Near the left Fermi point the driving term is $\nu\simeq e^{\theta+a}$, and into the band, for $\theta+a\gg1$, the occupation function falls off as
\be
L(\theta)\simeq\exp\!\big(-e^{\theta+a}\big)\,.
\ee
The transition between the two regimes takes place over a range of order one in $\theta$ around the Fermi point, and it is a kink of the kind found in the TBA of~\cite{Zamolodchikov:1989cf}. The occupation function is smallest at the centre of the band, the point farthest from both Fermi points, where
\be
L(0)\sim e^{-N\ell(0)}=e^{-N\log(7+4\sqrt3)}\simeq e^{-2.63N}\,.
\label{eq:L0_inband}
\ee
This is smaller than any power of $1/N$. Our analysis in the following sections is an expansion in $1/N$ around each Fermi point, in which the middle of the band does not enter, so that corrections of this size, which appear in the exact sum rules we derive later, do not affect any coefficient of the series.

\subsubsection{The chiral CFT limit}

We define the left kink coordinate $\theta'=\theta+a$. In the series~\eqref{eq:f-series} this gives
\be
\nu(\theta)=\sum_{s=1,5\,(\text{mod } 6)} t_s\,e^{s\theta'}\,,\qquad
t_s=\frac{4N}{s}\cos\frac{\pi s}{6}\,\big(2\sqrt3N\big)^{-s}\,.
\label{eq:lattice-times}
\ee
Hence, near the left Fermi point $\theta=-a$, the lattice driving term has the form of a GGE driving term, in which the chemical potential of the spin-$s$ charge is $t_s$, which we call the lattice GGE times. They decay at large $N$ as $N^{1-s}$, and $t_1=1$. As $N\to\infty$ all the higher spins are turned off, and the lattice TBA expanded near the left Fermi point reduces to the chiral Lee--Yang TBA equation~\cite{Zamolodchikov:1989cf,Klassen:1990dx}:
\be
\epsilon_0(\theta')=e^{\theta'}-\int_{-\infty}^{\infty}d\theta''\,\Phi(\theta'-\theta'')\,L_0(\theta'')\,,\qquad L_0=\log\left(1+e^{-\epsilon_0}\right).
\label{eq:cftTBA}
\ee
Its solution is the conformal kink of one chirality, where $L_0$ vanishes at large positive $\theta'$, and at large negative $\theta'$ the function $Y_0=e^{\epsilon_0}$ approaches the golden ratio. Its contribution to the Casimir energy is
\be
\int_{-\infty}^{\infty}\frac{d\theta'}{2\pi}\, e^{\theta'}L_0(\theta')=\frac{\pi c_{\rm eff}}{12}=\frac{\pi}{30}\,.
\label{eq:cftCasimir}
\ee
In this limit the right Fermi region decouples from the left chiral equation. At finite $N$, the coupling to the right region is restored through the tail~\eqref{eq:kernel-tail} of the kernel, evaluated over the separation $2a=2\log(2\sqrt3N)$ between the leading Fermi-point positions, and is therefore suppressed by powers of $1/N$. Thus, as $N\to\infty$, each Fermi region is described independently by its own chiral TBA. The two Fermi regions are related by the parity transformation $\theta\to-\theta$, under which $\nu$ and the kernel are even, so we analyse the left one in detail and obtain the right one by parity.

\section{The cubic oscillator and the lattice inverse problem}
\label{sec:inverse}

The conformal Lee--Yang ODE/IM correspondence provides the reference point for the construction below. So, we first recall the direct Stokes problem for the cubic oscillator. For the finite chain, we pose the inverse problem of reconstructing the cubic from the lattice $Y$-function, solve it numerically, and set up the large-$N$ analysis near the Fermi points.

\subsection{The cubic oscillator and Stokes data}
\label{sec:inv-cubic}

\begin{figure}[!tbp]
\centering
\begin{tikzpicture}[scale=0.85,>=stealth]
\foreach \a in {36,108,180,252,324}{\draw[dashed,gray!70,line width=0.5pt] (0,0)--(\a:2.9);}
\foreach \a/\k in {0/0,72/1,144/2,216/-2,288/-1}{
  \node at (\a:2.45) {\small $S_{\k}$};
  \node[blue!70!black] at (\a:1.6) {$\psi_{\k}$};
}
\draw[->,red!70!black,line width=1.0pt] (12:0.85) arc (12:60:0.85);
\node[red!70!black] at (36:0.45) {\small $\omega$};
\filldraw (0,0) circle (1.2pt);
\end{tikzpicture}
\caption{The five Stokes sectors $S_k$ of the cubic oscillator, separated by the dashed rays, with the solution $\psi_k$ that decays in sector $S_k$. The rotation $x\to\omega x$ maps each sector to the next.}
\label{fig:sectors}
\end{figure}
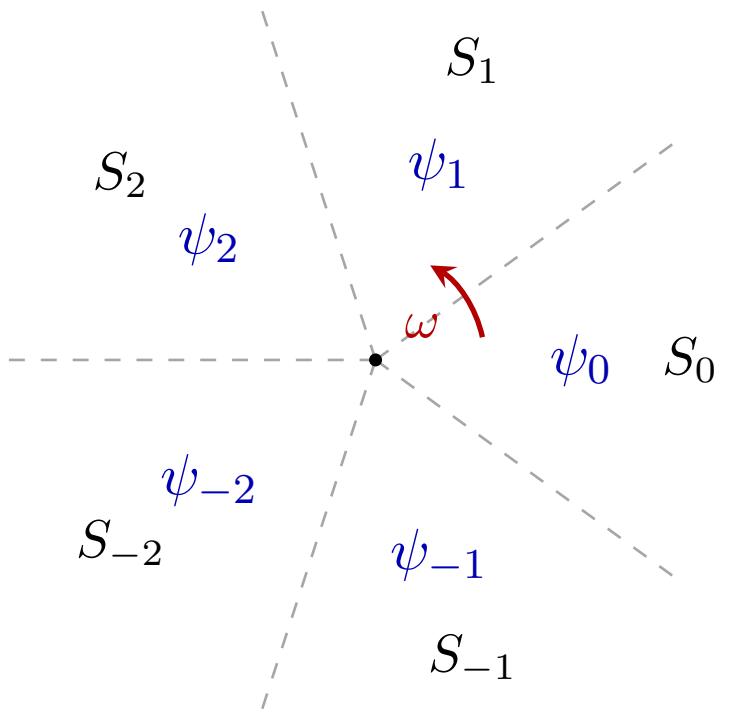

We begin with the direct Stokes problem for the general cubic oscillator~\cite{Dorey:1998pt, Sibuya:1984fe, Ashok:2026gge}:
\be
\left[-\frac{d^2}{dx^2}+x^3+Q\,x-P\right]\psi(x)=0\,,
\label{eq:cubic}
\ee
where $Q$ and $P$ may depend on the spectral parameter $E$. 
At large $|x|$ the potential divides the complex plane into five Stokes sectors, labelled by $k\in\mathbb Z_5$ (Figure~\ref{fig:sectors}).
In each sector there is a canonical solution $\psi_k$ that decays at infinity. The rotation $x\to\omega x$, with $\omega=e^{2\pi i/5}$, maps one sector to the next. Since $\omega^5=1$, if $\psi(x)$ solves~\eqref{eq:cubic} with coefficients $(Q,P)$, then $\psi(\omega^{-1}x)$ solves it with coefficients $(\omega^{-3}Q,\omega^{-2}P)$. Writing the inverse of this map as
\be
R:\ (Q,P)\longmapsto(\omega^3Q,\,\omega^2P)\,,
\label{eq:cubic-rotation}
\ee
the solution of~\eqref{eq:cubic} with coefficients $(Q,P)$ that decays in sector $k$ is
\be
\psi_k(x)=\omega^{k/2}\,
\psi_0\bigl(\omega^{-k}x\,;R^k(Q,P)\bigr)\,,
\ee
where $\psi_0$ is the solution that decays along the positive real axis. We normalise it by its WKB behaviour,
\be
\psi_0\simeq V^{-1/4}\exp\Big(-\int^x\sqrt{V(y)}\,dy\Big)\,.
\ee
Here $V(x)=x^3+Qx-P$, and the branch of $\sqrt{V}$ is chosen such that
$\sqrt{V}\sim x^{3/2}$ along the positive real axis. The integral
$\int^x\sqrt{V(y)}\,dy$ is defined by expanding
$\sqrt{V}=x^{3/2}+\frac{Q}{2}x^{-1/2}-\frac{P}{2}x^{-3/2}+\cdots$
at large $x$ and integrating term by term, with no additive constant.
This prescription fixes the normalisation of $\psi_0$ uniquely.
With this normalisation, the Wronskians of adjacent canonical solutions are
$W[\psi_k,\psi_{k+1}]=2i$. The Stokes multiplier relevant below is
\be
{\cal T}(E)=\frac{W[\psi_{-1},\psi_{1}]}{2i}\,,
\label{eq:Tdef}
\ee
in the conventions of~\cite{Dorey:1998pt}. When useful, we write
${\cal T}(E\,|\,Q,P)$ to display explicitly the dependence on the coefficients of the cubic.

At the conformal point, $Q=0$ and $P=E$, the rotation~\eqref{eq:cubic-rotation}
reduces to the spectral rotation $E\to\omega^2E$. Since $\omega^5=1$,
three successive rotations give $E\to\omega^6E=\omega E$. With the
parametrisation $e^\theta\propto E^{5/6}$, this corresponds to the shift
$\theta\to\theta+i\pi/3$ appearing in the
$Y$-system~\eqref{eq:Ysystem}. The ODE/IM correspondence identifies the
Stokes multiplier ${\cal T}(E)$ with the Lee--Yang
$Y$-function~\cite{Dorey:1998pt,Dorey:2007zx},
\be
Y(\theta)={\cal T}(E)\,,\qquad
e^{\theta}=\cA\,E^{5/6}\,,\qquad
\cA=\frac{2\sqrt3}{3}\,
B\!\left(\tfrac13,\tfrac32\right)
=\frac{2}{\sqrt3}\,
\frac{\Gamma(\frac13)\Gamma(\frac32)}
{\Gamma(\frac{11}{6})}\,,
\label{eq:YT-dict}
\ee
where $\cA$ is the classical WKB period of $x^3-E$ around the pair of
complex turning points, after factoring out the overall $E^{5/6}$
dependence. The fractional power $E^{5/6}$ is defined by analytic
continuation from positive real $E$, where it is taken to be positive.
On this branch,
$(\omega^{\pm1}E)^{5/6}=e^{\pm i\pi/3}E^{5/6}$.
The Stokes multiplier satisfies Sibuya's functional
relation~\cite{Sibuya:1984fe,Dorey:1998pt},
\be
{\cal T}(\omega^{-1}E)\,{\cal T}(\omega E)
=
1+{\cal T}(E)\,.
\label{eq:Sibuya-functional}
\ee
Through~\eqref{eq:YT-dict} and the branch prescription above,
$E\to\omega^{\pm1}E$ corresponds to
$\theta\to\theta\pm i\pi/3$, so this relation is precisely the
Lee--Yang $Y$-system~\eqref{eq:Ysystem}.

\subsection{The inverse problem for the chain}
\label{sec:inv-problem}

For the chain at finite $N$, the $Y$-function is known exactly from the transfer-matrix eigenvalue, $Y(\theta)=(-1)^M g^N\Lambda$, as in~\eqref{eq:Y-Lambda}. On the real
axis the same function is obtained from the lattice TBA equation, which we use for the analytic analysis. For the ground state the Bethe roots
are real, and therefore
\be
Y(\bar\theta)=\overline{Y(\theta)}\,.
\label{eq:Y-reality}
\ee
In particular, $Y(\theta)$ is real for real $\theta$. Under the dictionary~\eqref{eq:YT-dict}, the rotations $E\to\omega^{\pm1}E$ shift $\theta$ by $\pm i\pi/3$, so the neighbouring Stokes multipliers ${\cal T}(\omega^{\pm1}E)$ correspond to the shifted values, which we denote by
\be
Y^\pm(\theta):=
Y\!\left(\theta\pm\frac{i\pi}{3}\right)\,.
\label{eq:Ypm}
\ee
For real $\theta$, reality and the $Y$-system imply
\be
Y^-(\theta)=\overline{Y^+(\theta)}\,,
\qquad
Y^+(\theta)Y^-(\theta)=1+Y(\theta)\,.
\label{eq:Ypm-relations}
\ee
The values $Y^\pm(\theta)$ lie on the boundary of the physical strip.
They are unambiguously defined by the transfer-matrix expression. When
they are evaluated from the lattice TBA, they are understood as analytic
continuations to the boundary; in this continuation a pole of the kernel
reaches the integration contour and the corresponding residue has to be
included. The conformal correspondence provides the boundary condition for the
inverse problem. As shown in Section~\ref{sec:chain-fermi}, near the left Fermi point,
with $\theta'=\theta+a$ kept fixed as $N\to\infty$, the lattice TBA
reduces to the chiral Lee--Yang TBA. The lattice function therefore tends
to the conformal kink $Y_0(\theta')$, which Section~\ref{sec:inv-cubic} identifies with
the Stokes multiplier of the cubic with $Q=0$ and $P=E$. The right
Fermi point gives the parity-related conformal limit. At finite $N$ we do not assume that the lattice $Y$-function is already
the Stokes multiplier of a cubic. Instead, we pose this as an inverse
monodromy problem: we seek a spectral family of cubic oscillators
\be
\left[-\frac{d^2}{dx^2}+x^3+Q(E)x-P(E)\right]\psi(x)=0
\ee
whose Stokes data reproduce the lattice data. Using the same covariant
labelling of neighbouring Stokes multipliers as in the conformal
problem, we require
\be
{\cal T}(E\,|\,Q,P)=Y(\theta)\,,
\qquad
{\cal T}(\omega E\,|\,Q,P)=Y^+(\theta)\,.
\label{eq:lattice-Stokes-data}
\ee
Once the two
relations in~\eqref{eq:lattice-Stokes-data} hold, Sibuya's functional
relation~\eqref{eq:Sibuya-functional}, together with
\eqref{eq:Ypm-relations}, gives automatically
\be
{\cal T}(\omega^{-1}E\,|\,Q,P)=Y^-(\theta)\,.
\label{eq:lattice-Stokes-minus}
\ee
At fixed $\theta$, the inverse problem for the spin chain is therefore the reconstruction
\be
\bigl(Y(\theta),Y^+(\theta)\bigr)
\longmapsto
\bigl(Q(E),P(E)\bigr)\,.
\label{eq:lattice-inverse-map}
\ee
For real $\theta$, the modulus of $Y^+(\theta)$ is fixed by~\eqref{eq:Ypm-relations}, so the pair $(Y,Y^+)$ contains two independent real quantities, matching the number of real coefficients $Q$ and $P$ of the cubic. This counting alone does not guarantee either local invertibility or the existence of a real solution.
Equations~\eqref{eq:lattice-Stokes-data} therefore define the inverse problem for the chain. They do not assume that a solution exists globally for every $\theta$; we examine this numerically in Section~\ref{sec:inv-numerics}.
For the analytic solution near the Fermi points we need the logarithms of the Stokes data. They are related to the periods of the cubic, which we introduce in Section~\ref{sec:inv-periods}.

\subsection{Numerical solution across the band}
\label{sec:inv-numerics}

Before looking for an analytic solution, we solve the inverse problem~\eqref{eq:lattice-Stokes-data} numerically at finite $N$, point by point along the real $\theta$ axis. At each $\theta$ the Stokes multipliers are computed at high precision from the subdominant solutions, fixed by their WKB behaviour at large $|x|$, by forming the Wronskians~\eqref{eq:Tdef}.  Then, $(Q,P)$ are adjusted until they reproduce the two lattice data.

The inverse map is multivalued: the logarithm of the second Stokes multiplier is fixed only up to multiples of $2\pi i$, and each choice gives a different branch. We select the branch that reduces to the conformal solution, $Q\to0$ and $P\to E$, near the left Fermi point as $N\to\infty$, and follow it continuously in $\theta$, each solution serving as the starting point for the next. The resulting coefficients, $Q_L(\theta)$ and $P_L(\theta)$, are real and smooth along the whole sampled interval, and an independent integration of the reconstructed cubic reproduces the lattice data.

\begin{figure}[!tbp]
\centering
\includegraphics[width=0.98\textwidth]{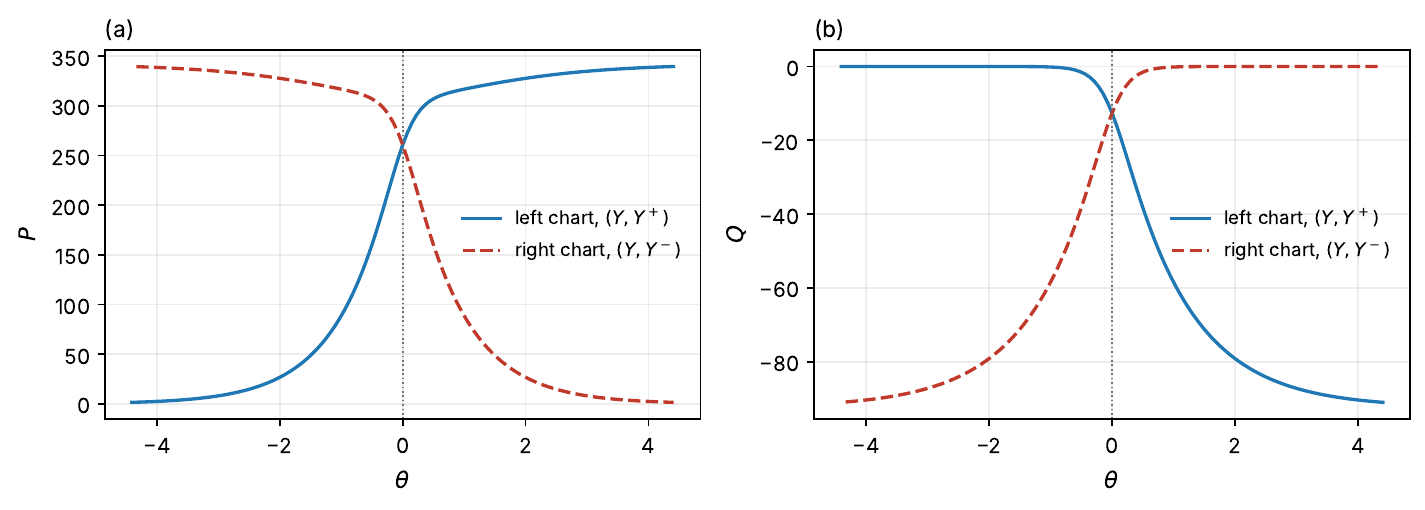}
\caption{The coefficients of the cubic from the numerical solution of the inverse problem for $N=96$: (a) $P$, (b) $Q$. Solid: the branch selected at the left Fermi point from $(Y,Y^+)$; dashed: its parity image, selected at the right Fermi point from $(Y,Y^-)$. The range shown, $|\theta|\le4.4$, lies inside the band $|\theta|<a$.}
\label{fig:physicalparity}
\end{figure}

For the ground state the lattice solution is parity invariant,
\be
Y(-\theta)=Y(\theta)\,,
\label{eq:lattice-parity}
\ee
but parity exchanges its two shifted values,
\be
Y^+(-\theta)=Y^-(\theta)\,,\qquad Y^-(-\theta)=Y^+(\theta)\,.
\label{eq:Ypm-parity}
\ee
Parity therefore maps the reconstruction from $(Y,Y^+)$ into the one from $(Y,Y^-)$, with the two adjacent Stokes multipliers exchanged. The branch selected at the right Fermi point, which reduces to the conformal cubic there, is the parity image of the one selected at the left,
\be
Q_R(\theta)=Q_L(-\theta)\,,\qquad P_R(\theta)=P_L(-\theta)\,.
\label{eq:QP-parity}
\ee
The coefficients of either reconstruction are thus not separately even in $\theta$; this does not break parity, however, since parity also reverses the orientation of the neighbouring Stokes sector used in the inverse problem. What matters is that the Stokes multipliers computed from the left and the right cubic agree with each other and with the lattice $Y$-function. Continued to the right, the left branch ends far from the conformal cubic (Figure~\ref{fig:physicalparity}).

The numerical solution is real and smooth across the band, and the descriptions at the two Fermi points are related by parity. It does not show, however, how the coefficients depend on $N$, nor give the charges. For that we want an analytic solution as a series in $1/N$, and the natural tool is the expansion of the Stokes multipliers in terms of the periods of the cubic.

\subsection{The WKB periods of the cubic}
\label{sec:inv-periods}

To analyse the inverse problem~\eqref{eq:lattice-Stokes-data}
perturbatively, it is convenient to lift the Stokes multipliers to
logarithmic coordinates. Along a chosen reconstruction we take the
logarithms continuously and define
\be
\Pi_{0}:=\log {\cal T}(E)\,,\qquad
\Pi_{\pm}:=\log {\cal T}(\omega^{\pm1}E)\,.
\label{eq:T-periods}
\ee
These logarithms are lifts of the Stokes data: changing a branch shifts
the corresponding $\Pi$ by an integer multiple of $2\pi i$,
without changing the Stokes multiplier itself. For the lattice
reconstruction the lifts are fixed by continuity from the corresponding
Fermi-point limit. Near the left Fermi point we use
$\Pi_{0}=\log Y(\theta)$ and
$\Pi_{+}=\log Y^+(\theta)$, with the branches continuously
connected to the conformal kink.

The shifted logarithms can be obtained directly from the real-axis
lattice TBA. Continuing~\eqref{eq:latTBA} to the upper boundary of the
physical strip, one of the poles of the kernel reaches the integration
contour. For real $\theta$, with the logarithmic branch fixed as above,
the continuation gives
\be
\log Y^+(\theta)
=
\nu\!\left(\theta+\frac{i\pi}{3}\right)
-\PV\!\int_{-\infty}^{\infty}d\theta'\,
\Phi\!\left(\theta-\theta'+\frac{i\pi}{3}\right)L(\theta')
+\frac12 L(\theta)\,.
\label{eq:Yplus-boundary}
\ee
The integral is understood as a principal value, and the last term is
the half-residue contribution from the pole on the contour. For real $\theta$, the lower boundary value is obtained by complex conjugation,
\be
\log Y^-(\theta)=\overline{\log Y^+(\theta)}\,,
\label{eq:Yminus-boundary}
\ee
so that its branch is the conjugate of the one fixed above for $\log Y^+$.

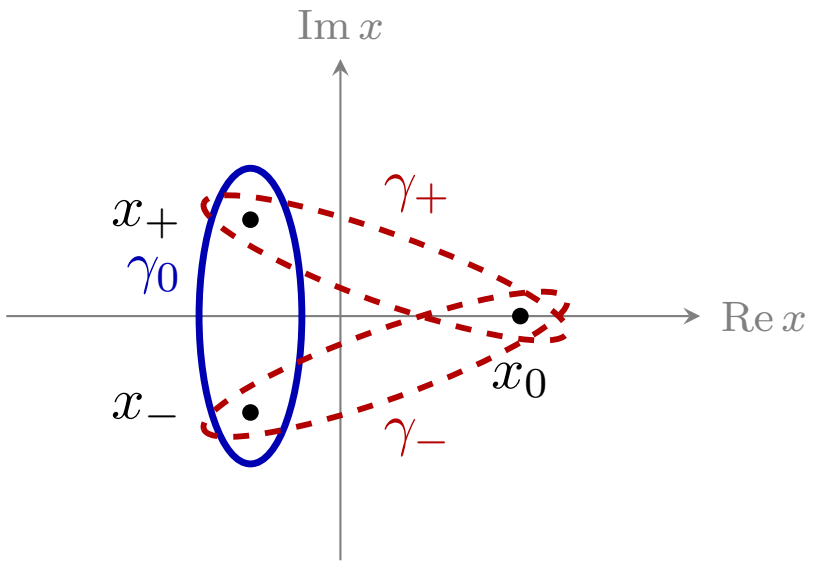
\begin{figure}[!tbp]
\centering
\begin{tikzpicture}[scale=0.85,>=stealth]
\draw[->,gray] (-2.6,0) -- (2.8,0) node[right] {\scriptsize ${\rm Re}\,x$};
\draw[->,gray] (0,-1.9) -- (0,2.0) node[above] {\scriptsize ${\rm Im}\,x$};
\coordinate (x0) at (1.4,0); \coordinate (xp) at (-0.7,0.75); \coordinate (xm) at (-0.7,-0.75);
\draw[blue!70!black,line width=1.3pt] (-0.7,0) ellipse (0.40 and 1.15);
\draw[red!70!black,dashed,line width=1.1pt,rotate around={160.35:(0.35,0.375)}] (0.35,0.375) ellipse (1.50 and 0.27);
\draw[red!70!black,dashed,line width=1.1pt,rotate around={-160.35:(0.35,-0.375)}] (0.35,-0.375) ellipse (1.50 and 0.27);
\filldraw (x0) circle (1.6pt); \node[below=5pt] at (x0) {$x_0$};
\filldraw (xp) circle (1.6pt); \node[left=9pt] at (xp) {$x_+$};
\filldraw (xm) circle (1.6pt); \node[left=9pt] at (xm) {$x_-$};
\node[blue!70!black] at (-1.45,0.32) {$\gamma_0$};
\node[red!70!black] at (0.6,0.95) {$\gamma_+$};
\node[red!70!black] at (0.6,-0.95) {$\gamma_-$};
\end{tikzpicture}
\caption{The turning points of the real cubic and the cycles $\gamma_0$ (solid) and $\gamma_\pm$ (dashed). The real turning point $x_0$ and the complex-conjugate pair $x_\pm$ satisfy ${\rm Re}\,x_\pm=-x_0/2$.}
\label{fig:cycles}
\end{figure}

Consider the real cubic
$V=x^3+Qx-P$ in the chamber $P>0$ and
$4Q^3+27P^2>0$, away from the discriminant locus. It has one real
turning point $x_0$ and a complex-conjugate pair $x_\pm$. We denote by
$\gamma_0$ the cycle around $x_\pm$, and by $\gamma_\pm$ the cycles
around $x_0$ and $x_\pm$, respectively, with the orientations defined
in Appendix~\ref{app:classes}. They satisfy
$\gamma_++\gamma_-=\gamma_0$ (Figure~\ref{fig:cycles}).
For each cycle one can define the formal WKB series
\be
\Pi^{\rm WKB}_\gamma
=
\Pi^{(0)}_\gamma+\Pi^{(2)}_\gamma+\Pi^{(4)}_\gamma+\cdots\,,
\label{def:WKBseries}
\ee
whose leading term is the classical period
\be
\Pi^{(0)}_\gamma=\oint_\gamma\sqrt{V}\,dx\,.
\label{classicalperiod}
\ee
The higher terms $\Pi^{(2k)}_\gamma$ are the periods of the even terms
of the Riccati expansion of the WKB solution, as reviewed in
Appendix~\ref{app:classes}. The odd terms are total derivatives, apart
from the first one which gives the WKB prefactor already included in
the normalisation of the canonical solutions. In the large-period WKB
regime, the exact logarithmic coordinates~\eqref{eq:T-periods} have the
asymptotic expansions
\be
\Pi_0\simeq\Pi^{\rm WKB}_{\gamma_0}\,,\qquad
\Pi_{\pm}
\simeq\Pi^{\rm WKB}_{\gamma_\pm}\,.
\label{eq:Stokes-WKB}
\ee
It is useful to distinguish these exact coordinates from their formal
WKB expansions. In exact WKB theory, suitable Borel resummations of
the formal periods exponentiate to Voros symbols
\cite{Voros:1983,DelabaerePham:1999,KawaiTakei:2005,ItoMarinoShu:2018},
which are related in each chamber to the Stokes multipliers by the
corresponding connection formulae. We will not need these resummations
explicitly: throughout this work, $\Pi^{\rm WKB}_{\gamma_0,\gamma_\pm}$ is used only
as the large-period asymptotic expansion of the exact coordinates
$\Pi_{0,\pm}$. With logarithmic branches chosen consistently, Sibuya's functional
relation~\eqref{eq:Sibuya-functional} becomes
\be
\Pi_{+}+\Pi_{-}
=
\log\bigl(1+e^{\Pi_{0}}\bigr)
=
\Pi_{0}
+\log\bigl(1+e^{-\Pi_{0}}\bigr)\,.
\label{eq:qWronskian}
\ee
On the other hand, the homology relation
$\gamma_++\gamma_-=\gamma_0$ implies
\be
\Pi^{\rm WKB}_{\gamma_+}
+
\Pi^{\rm WKB}_{\gamma_-}
=
\Pi^{\rm WKB}_{\gamma_0}
\label{eq:WKB-homology}
\ee
term by term, since every coefficient of the WKB series is a period
over the corresponding cycle. The difference between
\eqref{eq:qWronskian} and~\eqref{eq:WKB-homology} is the term
$\log(1+e^{-\Pi_{0}})$, which is exponentially small when
$\Re\Pi_{0}$ is large and positive and is therefore invisible
to the formal WKB expansion.

\subsubsection{Closed form of the periods}

We now give the periods of the cubic in closed form, as functions of $(Q,P)$; these are the expressions used in the large-$N$ analysis near the Fermi points. Much of this analysis was carried out in~\cite{Ashok:2026gge}, and some details are collected in Appendix~\ref{app:classes}. To compute the classical period~\eqref{classicalperiod}, we expand $\sqrt{x^3+Qx-P}$ in powers of $Q$. The cycle integrals of the resulting terms $x^n(x^3-P)^{1/2-n}$ are Euler Beta functions. Introducing $\kappa=Q/P^{2/3}$, the series separates into the three residue classes $n=3k+r$, with $r=0,1,2$. The class $r=2$ vanishes, while the other two resum to hypergeometric functions of $z=-4\kappa^3/27=-4Q^3/(27P^2)$. This gives
\be
\begin{aligned}
\Pi^{(0)}_{\gamma_0}(Q,P)
&=
\cA\,P^{5/6}\,
{}_2F_1\!\left(
-\tfrac{5}{12},\tfrac{1}{12};\tfrac23;z
\right)
+\frac{\cB}{2}\,Q\,P^{1/6}\,
{}_2F_1\!\left(
-\tfrac{1}{12},\tfrac{5}{12};\tfrac43;z
\right),
\\
\Pi^{(0)}_{\gamma_\pm}(Q,P)
&=
e^{\pm i\pi/3}\,\cA\,P^{5/6}\,
{}_2F_1\!\left(
-\tfrac{5}{12},\tfrac{1}{12};\tfrac23;z
\right)
+e^{\mp i\pi/3}\,\frac{\cB}{2}\,Q\,P^{1/6}\,
{}_2F_1\!\left(
-\tfrac{1}{12},\tfrac{5}{12};\tfrac43;z
\right),
\\
&\hspace{7cm}
z=-\frac{4Q^3}{27P^2}\,.
\end{aligned}
\label{eq:Pi-series}
\ee
Here
\be
\cA=\frac{2}{\sqrt3}\,B\!\left(\tfrac13,\tfrac32\right),\qquad
\cB=\frac{2}{\sqrt3}\,B\!\left(\tfrac23,\tfrac12\right),
\label{eq:B-period}
\ee
where $\cA$ is the constant already introduced in~\eqref{eq:YT-dict}. The two terms of $\Pi^{(0)}_{\gamma_0}$ are the two Frobenius solutions, with exponents $0$ and $1/3$ in $z$, of the Picard--Fuchs equation of this family of curves.

The higher WKB terms can be written as differential operators in
$(Q,P)$ acting on the classical period. The first ones are
\be
\begin{gathered}
\Pi^{(2)}_\gamma
=
\frac12\,\p_Q\p_P\,\Pi^{(0)}_\gamma\,,
\qquad
\Pi^{(4)}_\gamma
=
\frac{7}{40}\,(\p_Q\p_P)^2\,\Pi^{(0)}_\gamma\,,
\\
\Pi^{(6)}_\gamma
=
\frac{31}{560}\,(\p_Q\p_P)^3\,\Pi^{(0)}_\gamma
+
\frac{1}{1260}\,\p_P^5\,\Pi^{(0)}_\gamma\,.
\end{gathered}
\label{eq:PF}
\ee
More generally, $\Pi^{(2k)}_\gamma$ is a linear combination of
operators $\p_Q^a\p_P^b$ with $2a+3b=5k$.
\subsection{Strategy for the reconstruction}
\label{sec:strategy}

We would like to solve the inverse problem analytically at large $N$. As Figure~\ref{fig:fermi-points} shows, the occupation function $L$ changes from its plateau outside the band to exponentially small values across each kink. The exponentially weighted moments of $L$ that control the finite-size corrections are therefore dominated by the neighbourhoods of the two Fermi points, and we solve the problem in their  neighbourhood.

As seen in Section~\ref{sec:chain-fermi}, near a Fermi point the driving term $N\ell(\theta)$ expands in exponentials with exactly the spins of the Lee--Yang local charges, with coefficients~\eqref{eq:lattice-times}. The lattice TBA equation~\eqref{eq:latTBA} then takes the form of the TBA equation of a chiral generalised Gibbs ensemble, whose times, which we call the bare times, are functions of $N$. For such an ensemble the corresponding cubic can be constructed order by order in the times, following~\cite{Ashok:2026gge}.

This description is local. Near the left Fermi point the higher terms of the driving series are suppressed by powers of $1/N$, but towards the centre of the band they are all of order $N$, so the expansion cannot be carried there. This limitation concerns the expansion, not the inverse problem itself, which we solved numerically across the band in Section~\ref{sec:inv-numerics}.

The two Fermi points are not independent, however: the kernel has a tail with the same exponentials~\eqref{eq:kernel-tail}, through which each Fermi region contributes to the driving term seen by the other. The construction therefore proceeds in two steps. In Section~\ref{sec:chiral} we construct the cubic near the left Fermi point as a function of the times; in Section~\ref{sec:gluing} we include the contribution of the opposite Fermi region and determine the times self-consistently.

\section{Reconstruction near a Fermi point: the GGE cubic and the large-\texorpdfstring{$N$}{N} expansion}
\label{sec:chiral}

We now carry out the first step of this strategy and construct the cubic near the left Fermi point as a function of the times.

\subsection{The chiral TBA and the construction of the cubic}
\label{sec:chiral-gge}

We first consider the chiral GGE equation with the bare times of~\eqref{eq:lattice-times}. In the left kink coordinate $\theta'=\theta+a$ it reads:
\be
\epsilon(\theta')=\sum_{s\equiv1,5\,(6)}t_s\,e^{s\theta'}-\int_{-\infty}^{\infty} d\theta''\,\Phi(\theta'-\theta'')\,L(\theta'')\,,\qquad t_s=\frac{4N}{s}\cos\frac{\pi s}{6}\,\big(2\sqrt3N\big)^{-s}\,.
\label{eq:chiralTBA}
\ee
The construction of~\cite{Ashok:2026gge} associates to this chiral TBA a cubic oscillator~\eqref{eq:cubic} whose Stokes multiplier is the $Y$-function, ${\cal T}(E)=Y=e^{\epsilon}$. Its coefficients $Q(E)$ and $P(E)$ are polynomials in $E$ at each order in the higher times, and reduce to $Q=0$, $P=E$ at the conformal point, where all times but $t_1$ vanish (see~\eqref{eq:YT-dict}). We determine them order by order, using $t_s\sim N^{1-s}$ to organise the large-$N$ expansion; through relative order $N^{-8}$ we retain the terms linear in $t_5$ and $t_7$ and quadratic in $t_5$. At each order, we match the growing part of the formal WKB period around the complex pair of turning points to the driving term, coefficient by coefficient:
\be
\oint_{\gamma_0} S(x;E)\,dx=\sum_{k\ge0}\Pi^{(2k)}\big(Q(E),P(E)\big)=\sum_{s}t_s\,e^{s\theta'}+O(e^{-\theta'})\,.
\label{eq:defining}
\ee
Here $S$ is the formal WKB solution of the Riccati equation and $\Pi^{(2k)}\equiv\Pi^{(2k)}_{\gamma_0}$ is the WKB period of order $2k$ around the complex pair of turning points, defined in Section~\ref{sec:inv-periods}. Once the growing part has fixed the coefficients of the cubic, the decaying terms, proportional to $e^{-s\theta'}$, give the chiral moments and hence the eigenvalues of the local charges of the deformed KdV hierarchy~\cite{Ashok:2026gge,Karimi:2026kdv}.

\subsubsection{Spin counting}
\label{sec:spincounting}

If $\psi(x)$ solves the cubic~\eqref{eq:cubic} with coefficients $(Q,P)$, then $\psi(\omega x)$, with $\omega=e^{2\pi i/5}$, solves the cubic with coefficients $(\omega^3Q,\omega^2P)$. In the undeformed case this is the same equation with energy $\omega^2E$. For the same to hold in the GGE case, we need to have 
\be Q(\omega^2E)=\omega^3Q(E) \quad \text{and} \quad  P(\omega^2E)=\omega^2P(E).\ee Therefore $Q$ can only contain powers $E^n$ with $n\equiv4\pmod5$, and $P$ only powers with $n\equiv1\pmod5$. Following~\cite{Ashok:2026gge}, we construct $P(E)$ and
$Q(E)$ order by order in the higher times, with polynomial
dependence on $E$ at each order. The corresponding
solutions of the cubic are therefore entire in $E$
order by order. Hence
\be
Q(E)=\sum_{t\ge0}q_t\,E^{4+5t}\,,\qquad P(E)=\sum_{t\ge0}p_t\,E^{1+5t}\,.
\label{eq:allowed}
\ee
Each monomial carries a definite spin, which follows from the leading term of the period,
\be
\Pi^{(0)}\simeq\cA\,P^{5/6}+\frac{\cB}{2}\,Q\,P^{1/6}\,.
\ee
Near the conformal point, $P=E+\delta P$ and $Q$ is small, so to first order
we get \be
\Pi^{(0)}\simeq\cA\,E^{5/6}+\frac56\,\cA\,E^{-1/6}\,\delta P+\frac{\cB}{2}\,E^{1/6}\,Q\,.
\ee
So the monomial $E^n$ contained in $\delta P$ contributes to $\Pi^{(0)}$ a term proportional to $E^{n-1/6}$, and the same monomial in $Q$ a term proportional to $E^{n+1/6}$. Since $e^{s\theta'}\propto E^{5s/6}$, these are terms of spin
\be
s=\frac{6n-1}{5}\quad\text{if contained in }P\,,\qquad
s=\frac{6n+1}{5}\quad\text{if contained in }Q\,.
\ee
Looking at \eqref{eq:allowed}, we see that $Q$ carries only monomials of form $E^{4+5t}$, which have spin $5+6t$; while $P$ carries only monomials of form $E^{1+5t}$, having spin $1+6t$. Overall we see that
\be
Q \text{ carries spins } s\equiv 5\pmod6\,,\quad P \text{ carries spins } s\equiv 1\pmod6\,.
\ee
The same counting applies to every term of the WKB series. The leading term $\cA P^{5/6}$ has spin $1$, each monomial of spin $s$ that it contains raises the spin\footnote{Under $x\to E^{1/3}x$ the three terms of the potential scale in the same way if $Q\sim E^{2/3}$ and $P\sim E$. A monomial $E^n$ of $P$ is therefore of relative size $E^{n-1}$ compared with its conformal value $P=E$, and one of $Q$ of relative size $E^{n-2/3}$ compared with its natural scale; for a monomial of spin $s$ both are $E^{5(s-1)/6}$.} by $s-1$, and each order $k$ of the WKB series lowers it by $2k$. A term of $\Pi^{(2k)}$ containing monomials of spins $s_1,\dots,s_m$ therefore carries the spin
\be
r=1+\sum_{i=1}^m(s_i-1)-2k\,,
\ee
and it contributes to the coefficient of $e^{r\theta'}$ in~\eqref{eq:defining}. For each positive $r$ the total coefficient must equal $t_r$ if $e^{r\theta'}$ appears in the driving term, and vanish otherwise. These conditions fix the coefficients of $Q$ and $P$. The terms of negative spin in the period are not constrained by this condition, and they are the predictions of the construction.

The matching conditions determine the coefficients of
$P(E)$ and $Q(E)$ as functions of the higher times.
Comparing the coefficients of $e^{s\theta'}$ on the
two sides of~\eqref{eq:defining} gives one equation
for each positive spin.
At each order in the higher times, the matching equations
form a linear triangular system for the new corrections
to the coefficients. We solve the system from the highest spin downwards~\cite{Ashok:2026gge}, which we do explicitly in Section~\ref{sec:chiral-dict}.

\subsection{Expansion of the ODE potential in the times}
\label{sec:chiral-dict}

The construction of~\cite{Ashok:2026gge} was worked out in detail for a GGE with a single time, $t_5$. For the spin-chain, all the times of~\eqref{eq:chiralTBA} are present, so here we solve the defining condition for several times at once, through order $N^{-8}$. The times that contribute are $t_5$, $t_7$ and the product $t_5^2$, together with $t_1=1$. By the spin counting argument, the unknowns are then the coefficients of the leading monomials possessing the same positive spins\footnote{In the convention of~\cite{Ashok:2026gge}, the coefficient of $E$ in $P$ is not corrected and the leading term of the period is $D(\kappa)E^{5/6}$, so that $e^{\theta'}=D(\kappa)E^{5/6}$. With several times it is more convenient to keep $e^{\theta'}=\cA E^{5/6}$ to all orders and to let the spin-one condition fix the correction $p_1$ of the coefficient of $E$ in $P$. The two descriptions are related by a rescaling of $E$, and the charges computed from the WKB periods do not depend on the choice.}:
\be
Q=Q_4E^4+\dots\,,\qquad P=E\,(1+p_1)+p_6E^6+\dots\,,
\label{def:selectedspins}
\ee
since the next monomials of~\eqref{eq:allowed}, $E^9$ in $Q$ and $E^{11}$ in $P$, have spins $11$ and $13$.

Through relative order $N^{-8}$, terms of degree at most
two in $Q$ can contribute to the growing part of the
WKB period in~\eqref{eq:defining}. The highest WKB
period required is $\Pi^{(8)}$, through its term quadratic
in $Q$, which contributes at spin $1$. The terms needed
for this matching, including those independent of $Q$,
are:
\be
\begin{aligned}
\Pi^{(0)}&=\cA\,P^{5/6}+\frac{\cB}{2}\,Q\,P^{1/6}+\dots\,,\qquad
\Pi^{(2)}=\frac{\cB}{24}\,P^{-5/6}-\frac{35}{2592}\,\cA\,Q^2P^{-13/6}+\dots\,,\\
\Pi^{(4)}&=\gamma_4\,\cA\,Q\,P^{-19/6}+\frac{1309}{62208}\,\cB\,Q^2P^{-23/6}+\dots\,,\\
\Pi^{(6)}&=\mathfrak c_3\,\cA\,P^{-25/6}+\dots\,,\qquad
\Pi^{(8)}=\gamma_8\,\cA\,Q^2P^{-43/6}+\dots\,,
\end{aligned}
\label{eq:Pi4-term}
\ee
with
\be
\gamma_4=\frac{637}{31104}\,,\qquad
\mathfrak c_3=-\frac{21983}{1119744}\,,\qquad
\gamma_8=-\frac{1556782955}{2902376448}\,.
\ee
These expressions follow from the Picard--Fuchs form~\eqref{eq:PF} of the quantum periods applied to the expansion of the classical period in $Q$ (see Section~\ref{sec:inv-periods} and Appendix~\ref{app:classes}). Using \eqref{def:selectedspins}, we equate for each spin $s$ the coefficient of $E^{5s/6}$ on the two sides, where the right-hand side is $t_s\cA^s$, since $t_se^{s\theta'}=t_s\cA^sE^{5s/6}$. This gives
\be
\begin{aligned}
&\text{spin }7:&&\frac56\,p_6-\frac{35}{2592}\,Q_4^2=\cA^6\,t_7\,,\\
&\text{spin }5:&&Q_4\Big(1+\frac{p_1}{6}\Big)-\frac{5}{72}\,p_6+\frac{1309}{31104}\,Q_4^2=\frac{2\cA^5}{\cB}\,t_5\,,\\
&\text{spin }1:&&1+\frac56\,p_1-\frac{5}{72}\,p_1^2-\frac{25}{6}\,\mathfrak c_3\,p_6+\gamma_4\,Q_4\Big(1-\frac{19}{6}\,p_1\Big)+\gamma_8\,Q_4^2=t_1\,.
\end{aligned}
\label{eq:spin-conditions}
\ee
Here the conditions of spins $7$ and $1$ have been divided by $\cA$, and that of spin $5$ by $\cB/2$. Since $t_1=1$, the constant terms in the spin-$1$
condition cancel, leaving an equation that determines
$p_1$ order by order in terms of $Q_4$ and $p_6$.
At each perturbative order, the spin-$7$ condition
determines the new correction to $p_6$, followed by
$Q_4$ from the spin-$5$ condition and $p_1$ from the
spin-$1$ condition. Note that the nonlinear terms involve only
corrections already determined at lower orders.

Solving this triangular system of equations order by order, we obtain the complete dictionary through relative order $N^{-8}$:
\be
\begin{aligned}
Q(E)&=E^4\Big[\frac{2\cA^5}{\cB}\,t_5-\frac{637}{4320}\,\frac{\cA^{10}}{\cB^2}\,t_5^2+\frac{\cA^6}{12}\,t_7\Big]+O(N^{-10})\,,\\
P(E)&=E\Big[1+C_1t_5+\varrho_2\,\frac{\cA^{10}}{\cB^2}\,t_5^2+p_7\,t_7\Big]+E^6\Big[\frac65\,\cA^6\,t_7+\frac7{108}\,\frac{\cA^{10}}{\cB^2}\,t_5^2\Big]+O(N^{-10})\,,\\
\varrho_2&=\frac{5168743489}{2015539200}\,,\qquad C_1=-\frac{637}{12960}\,\frac{\cA^5}{\cB}\,,\qquad p_7=-\frac{55913}{466560}\,\cA^6\,.
\end{aligned}
\label{eq:lattice-dict}
\ee
The constants $C_1$, $\varrho_2$ and $p_7$ are the coefficients of the spin-one term, $p_1=C_1t_5+\varrho_2\,\frac{\cA^{10}}{\cB^2}\,t_5^2+p_7\,t_7$, which keeps the coefficient of $e^{\theta'}$ equal to one. 
At linear order the spin-5 part of~\eqref{eq:lattice-dict} is the first-order series of~\cite{Ashok:2026gge}, and the terms of second order in $t_5$ follow from their normalisation $D(\kappa)$ and the map $\kappa_{\rm TBA}(\kappa)$ of that reference.

\subsection{The charges of the chiral GGE: deformed KdV}
\label{sec:chiral-charges}

Equation~\eqref{eq:defining} fixes the growing part at $\theta'\rightarrow\infty$ of the period. For the chiral TBA~\eqref{eq:chiralTBA} the tail~\eqref{eq:kernel-tail} of the kernel gives, at large $\theta'$,
\be
\epsilon(\theta')=\sum_{s\equiv 1,5\, \text{mod} \, 6}t_s\,e^{s\theta'}+\frac{\sqrt3}{\pi}\sum_{s}\chi(s)\,M_s[t]\,e^{-s\theta'}\,,\qquad M_s[t]\equiv\int_{-\infty}^{\infty}d\theta'\,e^{s\theta'}L(\theta')\,.
\label{eq:chiral-tail}
\ee
On the ODE side the same coefficients are the negative-spin terms of the WKB period, so the cubic predicts the moments $M_s[t]$ as functions of the times; these quantities are the eigenvalues of the local charges of the GGE~\cite{Ashok:2026gge}. 
Here the spin label is restricted to $s\equiv1,5\pmod6$,
as in the kernel expansion~\eqref{eq:kernel-tail}.
When the higher times vanish, these moments reduce
to their conformal vacuum values $M_s^{(0)}$.
The charges that we will use are
\begin{align}
M^{(0)}_1&=\frac{\pi}{\sqrt3}\,\frac{\cA\cB}{24}=\frac{\pi^2}{15}\,,&
M^{(0)}_5&=\frac{\pi}{\sqrt3}\,\frac{21983}{1119744}\,\cA^6\,, \notag\\[4pt]
M^{(0)}_7&=\frac{\pi}{\sqrt3}\,\frac{26317819}{376233984}\,\cA^7\cB\,,&
M^{(0)}_{11}&=\frac{\pi}{\sqrt3}\,\frac{70877384469605}{13792092880896}\,\cA^{12}\,,
\label{vacuummoments}
\end{align}
where in the first we have used the relation $\cA\cB=8\pi\sqrt3/5$.

To make the discussion simpler, we now specialise to a GGE with $t_1=1$ and $t_5$ as
the only non-zero higher time.
The moments of a GGE with a spin-5 time are the eigenvalues of a set of local charges, constructed in~\cite{Karimi:2026kdv}.  In the perturbative regime
of small chemical potential,  the modular transformation that exchanges the two cycles of the torus~\cite{Downing:2024nfb,Ashok:2026cyi,Downing:2025huv} turns the GGE trace $\mathrm{Tr}\,\hat q^{L_0-c/24}e^{\alpha I_5}$ into a trace $\mathrm{Tr}\,q^{\mathcal I^{(5)}_1}$ over the same CFT Hilbert space, with the dual nome $q$ and a deformed Hamiltonian $\mathcal I^{(5)}_1$ (the exact transform would also contain non-perturbative contributions, that we shall ignore in this asymptotic analysis). The deformed Hamiltonian belongs to a commuting hierarchy of deformed charges. The first two such charges are~\cite{Karimi:2026kdv}:
\be
\begin{aligned}
\mathcal I^{(5)}_1&=I_1+\varsigma\,J_{1,5}+\frac{\varsigma^2}{2}\,J_{1,9}+\frac{\varsigma^3}{8}\,J_{1,13}+\dots\,,\\
\mathcal I^{(5)}_5&=I_5+\varsigma\,J_{5,9}+\frac{\varsigma^2}{2}\,J_{5,13}+\dots\,,
\end{aligned}
\qquad
\varsigma=\frac{125\cdot1729}{648}\,\frac{\cA^5}{\cB}\,t_5\,,
\label{eq:dijkgraaf-H}
\ee
where $I_s$ are the KdV charges and $J_{s,s+4m}$ is the zero mode of a quasi-primary field of weight $s+4m+1$, fixed by requiring that the deformed charges commute.
The moments of the spin-$5$ GGE studied
in~\cite{Ashok:2026gge} are related to the vacuum
eigenvalues of the deformed charges by
\be
M_s[t]=\mathcal N_s\,\mathcal I^{(5)}_s\{\phi\}\,,
\qquad \mathcal N_1=-4\pi^2\,,
\ee
where the normalisation constants $\mathcal N_s$ depend on neither the state nor the times; $\mathcal N_1$ is fixed by $I_1\{\phi\}=-\frac1{60}$ and $M^{(0)}_1=\frac{\pi^2}{15}$. 

The key point is that the deformed charges~\eqref{eq:dijkgraaf-H} are not built only from the Lee--Yang KdV charges: $J_{5,9}$, of spin $9\equiv3\pmod6$, is not a KdV charge, yet it contributes to the vacuum eigenvalue of the deformed charge $\mathcal I^{(5)}_5$~\cite{Karimi:2026kdv}, and the WKB periods of the deformed cubic reproduce this contribution~\cite{Ashok:2026gge}. 
The first non-KdV contribution to $M_1[t]$ occurs at
order $t_5^2$, through
$\frac{\varsigma^2}{2}J_{1,9}=\varsigma^2J_{5,9}$.
At first order in $t_5$, this contribution gives the linear response: 
\be
\frac{1}{M^{(0)}_5}\,\frac{\partial M_5}{\partial t_5}\Big|_{t=0}=\frac{\varsigma}{t_5}\,\frac{J_{5,9}\{\phi\}}{I_5\{\phi\}}=-\frac{35\cdot1729^2}{2^6\,3^4\,89}\,\frac{\cA^5}{\cB}\,.
\label{eq:M5-response}
\ee
This analysis concerns a single chiral GGE with only one time switched on, but it carries a general lesson: a GGE with KdV times has moments that do not contain only KdV charges, and thus a description built only from KdV charges and their bilinears misses some terms. 

For the spin-chain, these responses generate power-law
finite-size corrections through the scaling
$t_s\sim N^{1-s}$ of the higher times. In particular,
$J_{5,9}$ contributes through the $t_5$-dependence of
$M_1[t]$ and $M_5[t]$, without requiring an independent
spin-$9$ time in the driving term.

\subsubsection{Parametric responses of the charges}

Equation~\eqref{eq:M5-response} gives the normalised
linear response of $M_5[t]$ to $t_5$ at the conformal
point, expressed in terms of the spin-$9$ charge
$J_{5,9}$.
Derivatives of this
kind are the data of a single kink that the coupling of the two kinks
requires in the next section, and we collect them here. They are the
coefficients of the expansion of the charges in the times:
\be
M_s[t]=M^{(0)}_s+\sum_{p}t_p\,\frac{\partial M_s}{\partial t_p}\Big|_{t=0}+\frac12\sum_{p,p'}t_p\,t_{p'}\,\frac{\partial^2M_s}{\partial t_p\,\partial t_{p'}}\Big|_{t=0}+\dots\,.
\ee
Here $t=0$ denotes the conformal point, with all higher
times set to zero and $t_1=1$. The sums run over
$p,p'>1$ with $p,p'\equiv1,5\pmod6$. The linear term contains derivatives such as the spin-9
response~\eqref{eq:M5-response}, which is the case $s=p=5$, and mixed
derivatives such as $\partial M_5/\partial t_7=\partial M_7/\partial t_5$,
of spin 11, which are symmetric in the two times. The quadratic term
contains second derivatives such as $\partial^2M_5/\partial t_5^2$, of
spin 13. 

To obtain the responses displayed below, we insert
the dictionary~\eqref{eq:lattice-dict} into the WKB
periods and expand in the higher times. We extract
$M_s[t]$ from the coefficient of $e^{-s\theta'}$
using the normalisation in~\eqref{eq:chiral-tail}.
This gives
\be
\begin{aligned}
    \frac{\partial M_5}{\partial t_7}\Big|_{t=0}=\frac{\partial M_7}{\partial t_5}\Big|_{t=0}&=-\frac{43891461005}{54144288}\,\frac{\cA^5}{\cB}\,M^{(0)}_7\,,\\
\frac{\partial^2M_5}{\partial t_5^2}\Big|_{t=0}&=\frac{2819745755315125}{3587659776}\,\Big(\frac{\cA^5}{\cB}\Big)^{2}M^{(0)}_5\,.
\end{aligned}
\ee
In summary, the spin-chain enters this section only through the values of the times, $t_s\propto N^{1-s}$. The data of a single kink needed in the next section are (i) the cubic for these times, given by the dictionary~\eqref{eq:lattice-dict}, (ii) the vacuum moments $M^{(0)}_s$ in \eqref{vacuummoments}, and (iii) the responses of the moments to the times discussed above. The coefficients of the cubic and the moments are series in $1/N$.

\section{Coupling the left and right kinks}
\label{sec:gluing}

The previous section determined the cubic coefficients
and moments as functions of the times of an isolated
chiral GGE in the left Fermi region. The two kinks are not independent, however; the TBA kernel connects them, and we shall see that the times of one depend on the state of the other. This section makes the dependence precise.

\subsection{The two kinks and their coupling}
\label{sec:gluing-fold}

The occupation function $L$ of the ground state of the lattice TBA has the form shown in Figure~\ref{fig:fermi-points}. 
At fixed real $\theta$ as $N\to\infty$, we have that
$\epsilon(\theta)\simeq N\ell(\theta)$ and $L(\theta)$
is exponentially small, as illustrated at $\theta=0$
by~\eqref{eq:L0_inband}. Outside the band $\nu$ is small and so $L$ tends to the constant solution $\log\frac{1+\sqrt5}{2}$. As discussed in Section~\ref{sec:chain-fermi}, the
left kink solution tends to the conformal kink
of~\eqref{eq:cftTBA} as $N\to\infty$ with the kink
coordinate $\theta'=\theta+a$ held fixed. Recall also that we split the rapidity regions into left and right Fermi regions, depending on whether $\theta <0$ or $\theta >0$. 

With this splitting in mind, we define the moments of the two kinks as:
\be
I_s^{\rm L}=\int_{-\infty}^{0}d\theta\,e^{s\theta}\,L(\theta)\,,\qquad
I_s^{\rm R}=\int_{0}^{\infty}d\theta\,e^{-s\theta}\,L(\theta)\,.
\label{eq:moments-LR}
\ee
Consider now the convolution term in the lattice TBA equation \eqref{eq:latTBA}: 
\be 
\int_{-\infty}^{\infty}d\theta'\,\Phi(\theta-\theta')\,L(\theta')\,.
\ee 
In the left Fermi region, i.e. for $\theta<0$, we split the convolution at $\theta'=0$. In the part with $\theta'>0$ the argument of the kernel, $\theta-\theta'$, is negative, so we can use the convergent tail expansion~\eqref{eq:kernel-tail},
\be 
\Phi(\theta-\theta')=-\frac{\sqrt3}{\pi}\sum_s\chi(s)\,e^{s(\theta-\theta')}\,.
\ee 
Expanding also the driving term as in~\eqref{eq:f-series}, the full TBA equation for $\theta<0$ can be rewritten, exactly, as
\be
\epsilon(\theta)=\sum_{s}\tau^{\rm L}_s\,e^{s\theta}-\int_{-\infty}^{0}d\theta'\,\Phi(\theta-\theta')L(\theta')\,,
\label{eq:folded}
\ee
with
\be
\tau^{\rm L}_s=b_s+c_s\,I^{\rm R}_s\,,\qquad b_s=\frac{4N}{s}\cos\frac{\pi s}{6}\,,\qquad c_s=\frac{\sqrt3}{\pi}\,\chi(s)\,.
\label{eq:tau-exact}
\ee
The sums over $s$ are understood with Abel's prescription\footnote{That is, the sum is defined by evaluating at $\theta<0$, where it converges, and taking the limit $\theta\to0^-$; equivalently, $\sum_sa_s=\lim_{\eta\to0^+}\sum_sa_se^{-\eta s}$.} at $\theta\to0^-$. We remark that~\eqref{eq:folded} is exact, as it keeps all the exponentials of the tail of the kernel and the full dependence on the other kink. 
The resulting equation is defined on the left half-line,
with a driving term of GGE form and effective times
$\tau_s$. Its relation to the chiral TBA on the whole
line is discussed below.
The right kink enters only through its moments $I^{\rm R}_s$, each of which shifts the time of spin $s$ with a strength $c_s$ fixed by the tail of the Lee--Yang kernel. The same holds for the right kink, with $\mathrm{L}\leftrightarrow\mathrm{R}$.

At leading order in $1/N$ this splitting at the centre of the band is the one used in~\cite{Bajnok:2014fca} to extract the energies of the two chiral halves.
For the ground state, the occupation function $L(\theta)=\log(1+e^{-\epsilon(\theta)})$ is an even function of $\theta$, because the driving term and the kernel are even and thus $\epsilon$ is also even. Hence $I^{\rm R}_s=I^{\rm L}_s\equiv I_s$ and $\tau^{\rm L}_s=\tau^{\rm R}_s\equiv\tau_s$, and~\eqref{eq:folded} can equivalently be written on the half line with the folded kernel: 
\be
\epsilon(\theta)=\nu(\theta)-\int_{-\infty}^{0}d\theta'\,\big[\Phi(\theta-\theta')+\Phi(\theta+\theta')\big]L(\theta')\,.
\label{eq:folded-exact}
\ee
Expanding the image term $\Phi(\theta+\theta')$ in exponentials, which converges for $\theta+\theta'<0$, gives back~\eqref{eq:folded} with $\tau_s=b_s+c_sI_s$: the dependence on the opposite kink is entirely a shift of the coefficients of the driving term.

\subsubsection{Kink coordinates and direct feedback}

Consider first the ground state of the chain. By~\eqref{eq:folded},
the equation on the left half-line has the effective
driving term $\sum_s\tau_s e^{s\theta}$, with
$\tau_s=b_s+c_sI_s$.

As in Section~\ref{sec:chiral}, we use a kink coordinate, now centred with the effective scale, $\theta'=\theta+\log\tau_1$ (for the bare scale $\tau_1=2\sqrt3N$ this is $\theta+a$); it normalises the coefficient of $e^{\theta'}$ to unity. The rescaled
times are
\be
t_s=\frac{\tau_s}{\tau_1^{s}}\,,\qquad t_1=1\,.
\label{eq:times-tau}
\ee 
We call $\tau_1$ the scale of the kink: through $\theta'=\theta+\log\tau_1$ it fixes the position of the kink.
The quantities $t_s$ are the effective GGE times seen by the left kink. They differ from the bare lattice times because, by~\eqref{eq:tau-exact}, both $\tau_1$ and the higher $\tau_s$ contain the moments of the opposite kink.
In this coordinate, the driving term of~\eqref{eq:folded} becomes $\sum_st_s\,e^{s\theta'}$, which is the driving term of the chiral GGE equation~\eqref{eq:chiralTBA} with times $t_s$. The two equations differ only in the range of the convolution: the left Fermi region, $\theta'<\log\tau_1$, in~\eqref{eq:folded}, and the whole line in~\eqref{eq:chiralTBA}. The moments differ in the same way: $I_s$ is an integral over the left Fermi region, $M_s[t]$ over the whole line. Both differences involve $L$ only beyond the centre of the band, where it is exponentially small in $N$, so at fixed spin $s$ they are exponentially small as well. The moments of the left kink are therefore those of the chiral GGE, whose vacuum values and responses to the times were computed from the cubic in Section~\ref{sec:chiral-charges}:
\be
I_s=\tau_1^{-s}\,M_s[t]\,,
\label{eq:closure}
\ee
order by order in $1/N$. Inserting~\eqref{eq:closure} in $\tau_s=b_s+c_sI_s$ and dividing by $\tau_1^s$, we obtain one equation for the times,
\be
t_s=b_s\,\tau_1^{-s}+c_s\,\tau_1^{-2s}\,M_s[t]\,.
\label{eq:coupled}
\ee
The first term is the bare lattice contribution to the effective time. The second is the direct feedback from the opposite kink: its spin-$s$ moment shifts the spin-$s$ time of the left kink, with the factor $\tau_1^{-2s}$, where $2\log\tau_1$ is the separation of the two Fermi points in the symmetric ground state.

\subsubsection{Couplings for asymmetric states} 

The chiral TBAs near each Fermi point need not be in the same state, and in this case they
are no longer
related by parity. Thus the local construction must be
carried out separately in the left and right regions. 
The two kinks have scales $\tau^{\rm L}_1$ and
$\tau^{\rm R}_1$, with local coordinates
$\theta'_{\rm L}=\theta+\log\tau^{\rm L}_1$ and
$\theta'_{\rm R}=-\theta+\log\tau^{\rm R}_1$, and
effective times $t^{\rm L}_s$ and $t^{\rm R}_s$.
The relation~\eqref{eq:closure} applies separately to
the two local chiral descriptions:
\be 
I^a_s=(\tau^a_1)^{-s}M_s^a[t^a]\,,\quad \text{ with } a={\rm L},{\rm R}\,.\ee 
Here $M^{\rm L}_s[t]$ and $M^{\rm R}_s[t]$ are the one-kink functions of Section~\ref{sec:chiral-charges}, for the chiral state of the corresponding kink. Their expansions in the times are fixed by their values and responses at zero higher times.
The two local chiral descriptions are therefore coupled by
\be
\begin{gathered}
t^{\rm L}_s=\frac{b_s}{(\tau^{\rm L}_1)^{s}}+c_s\,e^{-s\Delta}\,M_s^{\rm R}\big[t^{\rm R}\big]\,,\qquad
t^{\rm R}_s=\frac{b_s}{(\tau^{\rm R}_1)^{s}}+c_s\,e^{-s\Delta}\,M_s^{\rm L}\big[t^{\rm L}\big]\,,\\[4pt]
\Delta=\log\tau^{\rm L}_1+\log\tau^{\rm R}_1\,,
\end{gathered}
\label{eq:coupled-LR}
\ee
with $\Delta$ the separation of the two kink centres in the $\theta$ coordinate. Since the functions $M^{\rm L,R}_s[t]$ are known, these form a closed set of equations for the scales and the times.
For a state symmetric under the
interchange of the two regions, in particular for the
ground state, we have
$t^{\rm L}_s=t^{\rm R}_s\equiv t_s$ and
$\tau^{\rm L}_1=\tau^{\rm R}_1\equiv\tau_1$, and
\eqref{eq:coupled-LR} reduces to~\eqref{eq:coupled}.

\subsection{The effective length and the times}
\label{sec:gluing-Neff}
\label{sec:gluing-spin5}

The coupling~\eqref{eq:coupled} determines
the effective scale $\tau_1$ and the higher GGE times
entering the local description.
Its spin-$1$ component is (setting $t_1=1$):
\be 
\tau_1^2=2\sqrt3\,N\,\tau_1+\frac{\sqrt3}{\pi}\,M_1[t]\,.
\label{eq:tau1-eq}
\ee
This equation fixes the effective scale
$\tau_1$ in terms of $N$ and the spin-$1$ moment
entering the feedback from the opposite kink.
If $M_1[t]=0$
the scale reduces to its bare value
$\tau_1=2\sqrt3\,N$.

Defining the effective length  $\Neff$ via $\tau_1\equiv2\sqrt3\,\Neff$,~\eqref{eq:tau1-eq} becomes
\be
\Neff\,(\Neff-N)=\frac{M_1[t]}{4\sqrt3\,\pi}\,.
\label{eq:Neff-closure}
\ee
This means that the feedback between the two Fermi regions renormalises the scale from $N$ to $\Neff$. In the symmetric ground state, the corresponding kink positions are therefore shifted to
$\theta\simeq\mp\log(2\sqrt3\,\Neff)$. Equivalently we may write
\be
\Neff=N+\frac{I_1}{2\pi}\,,\qquad I_1=\frac{M_1[t]}{2\sqrt3\,\Neff}\,,
\label{eq:Ttbarlike}
\ee
with $I_1$ the spin-$1$ moment associated with one kink. At leading order we can replace $M_1[t]$ by its conformal value~\eqref{eq:cftCasimir},
\be
M_1[t]\simeq M^{(0)}_1=\frac{\pi^2}{15}=2\pi\cdot\frac{\pi c_{\rm eff}}{12}\,,
\ee
up to a relative correction of order $N^{-4}$ due to the GGE times. Equation~\eqref{eq:Neff-closure} then becomes a quadratic equation for $\Neff$,
\be
\begin{gathered}
\Neff=N+\frac{\alpha}{\Neff}\,,\quad\text{with}\quad  \alpha=\frac{\pi}{60\sqrt3}\,,\\
\text{or}\quad \Neff=\frac{N+\sqrt{N^2+4\alpha}}{2}\,.
\end{gathered}
\label{eq:Neff}
\ee
Its expansion
\be
\log\Neff=\log N+\frac{\alpha}{N^2}-\frac32\frac{\alpha^2}{N^4}+\dots
\label{eq:Neff-exp}
\ee
shows that the two Fermi points, located at
$\theta=\pm\log(2\sqrt3\,\Neff)$ in the symmetric
ground state, are shifted with respect to
$\pm\log(2\sqrt3\,N)$ by $\alpha/N^2$ at first order
and by $-\frac32\alpha^2/N^4$ at second order.
The coefficient $\alpha$ is fixed by the conformal
spin-$1$ moment and by the tail of the kernel, with no
free parameter.

\subsubsection{The \texorpdfstring{$\TTbar$}{TTbar}-like structure}

Let us now look at the expression for the effective length of the spin chain and give a physical interpretation to each of the quantities: 
\be
\Neff=N+\mathsf t\,E_{\rm Cas}\,,\qquad
E_{\rm Cas}=-\frac{\pi c_{\rm eff}}{6\,\Neff}\,,\qquad
\mathsf t=-\frac1{4\sqrt3}\,,
\label{eq:Neff-TT}
\ee
We have recognized the conformal Casimir energy, evaluated at the effective length $\Neff$. So each region  sees the bare length shifted by the Casimir energy of the state, with the fixed coefficient $-1/(4\sqrt3)$. This has the same self-consistent effective-length structure as the finite-volume flow generated by a $\TTbar$ deformation~\cite{Smirnov:2016lqw,Cavaglia:2016oda}, for which the energy of a zero-momentum state on a circle of circumference $R$ can be expressed in terms of an effective length depending on the energy itself. In the present case, the analogous relation involves $R=N$ and the fixed coefficient $\mathsf t=-1/(4\sqrt3)$.

Here the change of scale is not put in by hand: it is generated by the tail of the Lee--Yang kernel, and its strength is fixed with no free parameter. $\TTbar$-like structures for spin chains have been discussed in~\cite{Pozsgay:2019ekd,Marchetto:2019yyt,Jiang:2023}, where the deformation is introduced explicitly. Here, instead, a similar effective-length relation emerges from the finite-size coupling between the two Fermi regions of the undeformed chain.

\subsubsection{The exact equation for the effective length}

Equation~\eqref{eq:Neff} uses the conformal value of the
spin-$1$ moment entering the feedback from the opposite
kink. To go beyond it, we use a sum rule for the TBA equation folded at the centre of the band, derived in Appendix~\ref{app:sumrule} with the dilogarithm trick \cite{Zamolodchikov:1989cf}. 
It relates the effective times of the folded equation
to the moments associated with the same kink:
\be
\sum_{s\equiv\pm1\,(\text{mod }6)}s\,\tau_s\,I_s=\frac{\pi^2}{15}+{\rm Li}_2\!\left(-e^{-\epsilon(0)}\right)\,.
\label{eq:sumrule}
\ee
The constant on the right-hand side comes from the plateau of $L$ outside the band. The dilogarithm comes from the centre of the band, where the folded equation ends, and is exponentially small in $N$. Dropping this term, if only the spin-one time is kept, the sum rule reduces to the conformal Casimir identity
\be
\tau_1 I_1=\frac{\pi^2}{15}=2\pi\frac{\pi c_{\rm eff}}{12}\,.
\ee
The spin-one term of~\eqref{eq:sumrule} fixes the effective length, and the higher spins correct it. Setting $\tau_1=2\sqrt3\,\Neff$ and $I_1=2\pi(\Neff-N)$, the spin-one term equals $4\sqrt3\,\pi\,\Neff(\Neff-N)$, and~\eqref{eq:sumrule} becomes the exact equation
\be
\Neff\,(\Neff-N)=\alpha\Big[1+\frac{15}{\pi^2}\,{\rm Li}_2\!\big(-e^{-\epsilon(0)}\big)-\frac{15}{\pi^2}\,G\Big]\,,\qquad G=\sum_{s\ge5}s\,\tau_s\,I_s\,,
\label{eq:Neff-exact}
\ee
where $\alpha$ was defined in \eqref{eq:Neff}. Here $G$ collects the higher-spin contributions. Setting $G=0$ and dropping the dilogarithm gives back~\eqref{eq:Neff}.
For the symmetric ground state the two kinks have
the same moments, and inserting
$\tau_s=b_s+c_sI_s$ in $G$ gives
\be
G=\sum_{s\ge5}s\,b_sI_s+\sum_{s\ge5}s\,c_sI_s^2\,.
\ee
The first sum comes from the driving term alone. 
The second comes from the coupling between the two Fermi
regions: the spin-$s$ moment associated with one kink shifts
the effective spin-$s$ time of the opposite region, with
strength $c_s$. This has the structure of a generalised
$T\bar T$-type coupling.
\subsubsection{The self-consistency equations for the times}

In the symmetric ground state the solutions of the two chiral TBAs are
related by parity and therefore have the same effective
times and moments. A single set of variables is sufficient
to describe the two local chiral expansions. The relations of the previous subsections then close: the scale $\tau_1=2\sqrt3\,\Neff$ of~\eqref{eq:Neff-closure} and the times of~\eqref{eq:coupled} are determined by the unknowns $\Neff$ and $t_s$, $s=5,7,11,\dots$, which obey the self-consistency equations
\be
\begin{gathered}
\tau_1=2\sqrt3\,\Neff\,,\qquad
\Neff\,(\Neff-N)=\frac{M_1[t]}{4\sqrt3\,\pi}\,,\qquad
M_1[t]=\frac{\pi^2}{15}-\sum_{s\ge5}s\,t_s\,M_s[t]\,,\\[4pt]
\text{with}\quad t_s=b_s\,\tau_1^{-s}+c_s\,\tau_1^{-2s}\,M_s[t]\,,\qquad s=5,7,11,\dots\,.
\end{gathered}
\label{eq:closed-system}
\ee
The first equation defines $\Neff$ through $\tau_1$, and the second is~\eqref{eq:Neff-closure} with $I_1=\tau_1^{-1}M_1[t]$. The third is the sum rule~\eqref{eq:sumrule} in the kink variables, where $\tau_sI_s=t_sM_s[t]$, without the dilogarithm. The last is the expression for the higher times in~\eqref{eq:coupled}. Given the functions $M_s[t]$, these equations determine $\Neff$ and all the times. We solve them by iteration; we start from the bare scale $\tau_1=2\sqrt3N$ and the bare times $b_s\tau_1^{-s}$, which decay as $N^{1-s}$, so that each step gains powers of $1/N$. Writing $\bar t_s=b_s\tau_1^{-s}$ for the bare times at the effective scale, we expand $t_s=\bar t_s+c_s\tau_1^{-2s}M_s[t]$ and $M_s[t]$ in the times, using the vacuum charges and responses of Section~\ref{sec:chiral-charges}. Each term then has a definite order in $1/N$:
\begin{align}
M_1[t]=\frac{\pi^2}{15}&-5\,\bar t_5\,M^{(0)}_5 &&(N^{-4})\notag\\
&-7\,\bar t_7\,M^{(0)}_7 &&(N^{-6})\notag\\
&-5\,\bar t_5^{\,2}\,\frac{\partial M_5}{\partial t_5} &&(N^{-8})\notag\\
&-11\,\bar t_{11}M^{(0)}_{11}-12\,\bar t_5\bar t_7\,\frac{\partial M_5}{\partial t_7}
-5\,c_5\,\tau_1^{-10}\big(M^{(0)}_5\big)^2 &&(N^{-10})\notag\\
&+O(N^{-12})\,,
\label{eq:M1-orders}
\end{align}
The last term is the direct feedback: it changes $t_5$ at relative order $N^{-6}$, and thus first contributes at order $N^{-10}$; its effect on $t_7$ is of relative order $N^{-8}$ and enters only at $N^{-14}$. 

Since $\Neff=N+O(1/N)$, the spin-one equation gives
$\Neff-N\simeq M_1[t]/(4\sqrt3\,\pi N)$ at leading order, so a term of
order $N^{-k}$ in $M_1[t]$ shifts $\Neff$ at order $N^{-(k+1)}$. Solving
the spin-one equation  
for $\Neff$, and inserting the first three lines of~\eqref{eq:M1-orders} 
therefore gives $\Neff$ through order $N^{-9}$,
\be
\begin{aligned}
\Neff&=\frac12\left[N+\sqrt{N^2+\frac{M_1[t]}{\sqrt3\,\pi}}\right]\\
&=N+\frac{\alpha}{N}-\frac{\alpha^2}{N^3}
+\frac1{N^5}\Big(2\alpha^3+\frac{M^{(0)}_5}{2\pi\,(2\sqrt3)^5}\Big)\\
&\quad+\frac1{N^7}\Big(-5\alpha^4-\frac{7\,M^{(0)}_5}{103680}
+\frac{M^{(0)}_7}{2\pi\,(2\sqrt3)^7}\Big)\\
&\quad+\frac1{N^9}\Big(14\alpha^5+\frac{\alpha\,M^{(0)}_5}{2880}
-\frac{M^{(0)}_7}{138240}
-\frac{1}{10\pi\,(2\sqrt3)^9}\,\frac{\partial M_5}{\partial t_5}\Big)
+O(N^{-11})\,,\quad \alpha=\frac{\pi}{60\sqrt3}\,.
\end{aligned}
\label{eq:Neff-series}
\ee
The terms in $\alpha$ alone are the expansion of~\eqref{eq:Neff}, the
terms in $M^{(0)}_5$ and $M^{(0)}_7$ come from the spin-5 and spin-7
times, and the last one comes from the response of the spin-5 charge,
that is, from the spin-9 charge of Section~\ref{sec:chiral-charges}.

The effective length also corrects the times. With
$\tau_1=2\sqrt3\,\Neff$ the bare times are $\bar t_5=-N/(720\,\Neff^5)$ and $\bar t_7=-N/(12096\,\Neff^7)$, and expanding
$\Neff$ gives
\be
\bar t_5=-\frac{1}{720\,N^4}\Big(1-\frac{5\alpha}{N^2}
+\frac{20\alpha^2}{N^4}\Big)+O(N^{-10})\,,\qquad
\bar t_7=-\frac{1}{12096\,N^6}\Big(1-\frac{7\alpha}{N^2}\Big)+O(N^{-10})\,.
\label{eq:times-N5}
\ee
The dictionary of Section~\ref{sec:chiral-dict} holds through relative order
$N^{-8}$, and to this order the influence of the opposite Fermi region
on the local cubic reconstruction is entirely encoded in $\Neff$.

The series~\eqref{eq:Neff-series} neglects the terms of order $N^{-10}$ and beyond in $M_1[t]$, which shift $\Neff$ at order $N^{-11}$; the last line of~\eqref{eq:M1-orders} gives the first of them. Contributions from the centre of the band, such as the dilogarithm in~\eqref{eq:Neff-exact}, are exponentially small in $N$ and do not affect the algebraic $1/N$ expansion.

\subsection{The lattice cubic near the Fermi points}
\label{sec:gluing-cubics}
\label{sec:gluing-tests}

The objective of this section is to construct, near each Fermi point, the cubic oscillator whose Stokes data reproduce the local large-$N$ expansion of the lattice $Y$-function order by order in $1/N$. The corresponding differential equation is
\be
\Big[-\frac{d^2}{dx^2}+x^3+Q(E)\,x-P(E)\Big]\psi(x)=0\,. 
\ee
We determine $Q(E)$ and $P(E)$ as series in $1/N$. Near a Fermi point, the coefficients of this cubic are
given by the local GGE reconstruction of
Section~\ref{sec:chiral-dict}, expressed in terms of
the effective times. For the spin chain these times
are those of~\eqref{eq:times-N5}, up to terms of order
$N^{-10}$. Inserting these gives the $1/N$ expansion of the coefficients:
\begin{align}
Q(E)&=-E^4\bigg[\frac{\cA^5}{360\,\cB\,N^4}
+\Big(\frac{\cA^6}{145152}-\frac{\alpha\,\cA^5}{72\,\cB}\Big)\frac1{N^6}\notag\\
&\hspace{3em}+\Big(\frac{637\,\cA^{10}}{2239488000\,\cB^2}-\frac{\alpha\,\cA^6}{20736}+\frac{\alpha^2\,\cA^5}{18\,\cB}\Big)\frac1{N^8}\bigg]+O(N^{-10})\,, \notag\\[6pt]
P(E)&=E\bigg[1+\frac{637\,\cA^5}{9331200\,\cB\,N^4}
+\Big(\frac{55913\,\cA^6}{5643509760}-\frac{637\,\alpha\,\cA^5}{1866240\,\cB}\Big)\frac1{N^6}\notag\\
&\hspace{3em}+\Big(\frac{5168743489\,\cA^{10}}{1044855521280000\,\cB^2}-\frac{55913\,\alpha\,\cA^6}{806215680}+\frac{637\,\alpha^2\,\cA^5}{466560\,\cB}\Big)\frac1{N^8}\bigg]\notag\\
&\quad-E^6\bigg[\frac{\cA^6}{10080\,N^6}-\Big(\frac{7\,\cA^{10}}{55987200\,\cB^2}+\frac{\alpha\,\cA^6}{1440}\Big)\frac1{N^8}\bigg]+O(N^{-10})\,.
\end{align}
We can now take stock. The strategy outlined in
Section~\ref{sec:strategy} was to solve the inverse
problem locally near a Fermi point, order by order in
$1/N$, while including the coupling to the opposite region
through the tail of the kernel. The resulting local
expansion of the lattice cubic is now explicit: its
coefficients $Q(E)$ and $P(E)$ are series in $1/N$,
with effective times and scale fixed by $N$ through the
self-consistency equations above. The terms containing
$\alpha$ encode the leading feedback from the opposite
kink, while the remaining terms are those of the chiral GGE at the bare times $b_s\,(2\sqrt3N)^{-s}$.
The coefficient $\alpha$ is fixed by the conformal
spin-$1$ moment and by the tail of the kernel, so no
new parameter is introduced.

The pointwise finite-$N$ reconstruction of Section~\ref{sec:inv-numerics} provides a complementary numerical check of the local expansion above. The solutions selected independently from the left and right Fermi-point limits remain real and smooth throughout the interval sampled numerically and are mapped into one another by parity, as in~\eqref{eq:QP-parity}. Near each Fermi point, the corresponding numerical reconstruction agrees with the local large-$N$ expansion obtained above. 

\section{The finite-size energy and the local charges}
\label{sec:energy}

With the local cubic reconstructed near each Fermi point, we can compute the corresponding chiral conserved charges. In the ODE/IM correspondence, these are obtained from the decaying part of the WKB periods, while their dependence on $N$ follows from the effective GGE times determined by the lattice problem. The local charges of the chain are instead logarithmic derivatives of the transfer matrix at a point outside the physical strip. The $Y$-system relates this point to points inside the strip, where, up to the driving term, $\log Y$ is fixed by the moments of the two kinks. The energy and the higher charges of the chain are therefore linear combinations of these moments, with coefficients set by the chain. Below we compute them as expansions in $1/N$ and compare with numerical Bethe ansatz results and, for the leading terms, with the effective Hamiltonian of~\cite{Lukyanov:1997bp}.

\subsection{Observables of the chain from Stokes data}
\label{sec:gluing-energy}

As we have seen in Section \ref{sec:chain-ba}, the energy of a Bethe state can be expressed as a sum over the contribution of Bethe roots $\cE=\sum_je(\lambda_j)$, where the energy of a single root is
\be
e(\lambda)=-\sin\gamma\,p'(\lambda)=-\frac{2\sin^2\gamma}{\cosh2\lambda-\cos\gamma}\,.
\ee
The higher local conserved charges are the sums over the roots of the derivatives of $e$ with respect to the rapidity,
\be
\cQ_n=\sum_je^{(n-2)}(\lambda_j)\,,\qquad n=2,3,4,\dots\,,\qquad \cQ_2=\cE\,.
\label{eq:Qn-def}
\ee

\subsubsection{The Hamiltonian} 

In our conventions, the transfer matrix of the XXZ spin chain $\Lambda$ evaluated at $v=i\gamma/2$ is proportional to the lattice translation operator, and the Hamiltonian is thus
\be
\cE=i\sin\gamma\,\big(\log\Lambda\big)'\big|_{v=i\gamma/2}-N\cos\gamma\,.
\label{eq:E-from-Lambda}
\ee
This expression coincides with the charge $\cQ_2$ \eqref{eq:Qn-def}. The charge $\cQ_3$ is proportional to the energy current, and the $\cQ_n$ are the higher logarithmic derivatives of $\Lambda$ at the same point. Recall that in the variable $\theta=\pi v/\gamma$ the point $v=i\gamma/2$ is mapped to $\theta=i\pi/2$, and that the transfer matrix is related to the $Y$-function  by~\eqref{eq:Y-Lambda}:
\be
\log\Lambda=\log Y-N\log g+{\rm const}\,.
\label{eq:logLambda}
\ee
The term $-N\log g$ is the same for every state and linear in $N$, so its derivatives contribute only to the bulk terms, which are proportional to $N$; we collect these below. The finite-size corrections therefore come from $\log Y$ alone. 
The charges require derivatives evaluated at $\theta=i\pi/2$, outside the physical
strip. We can use the $Y$-system \eqref{eq:Ysystem} to express them in terms of $Y$ functions evaluated at $\theta=i\pi/6$, writing
\be
\log Y\big(\delta+\tfrac{i\pi}{2}\big)=\log Y\big(\delta+\tfrac{i\pi}{6}\big)-\log Y\big(\delta-\tfrac{i\pi}{6}\big) + \log (1+ \tfrac{1}{Y(\delta+\frac{i\pi}{6})})\,.
\ee
Here $\delta$ is a small parameter, to be set to zero after taking the derivatives. Since $Y$ has a pole of order $N$ at the wall $\theta=i\pi/6$, the function $1/Y$ vanishes there as $\delta^N$, so the last term and its first $N-1$ derivatives vanish as $\delta\to0$. Hence, for any fixed charge $\mathcal Q_n$, this term does not contribute once $N\ge n$. The remaining terms lie within the physical strip, where
$\log Y$ is the sum of the lattice driving term and the
two chiral contributions to the asymptotic expansion of
the lattice Stokes data:
\be
\log Y(\theta)=N\ell(\theta)+\sum_sc_s\Big[I^{\rm L}_s\,e^{-s\theta}+I^{\rm R}_s\,e^{s\theta}\Big]\,,\qquad c_s=\frac{\sqrt3}{\pi}\,\chi(s)\,,
\ee
where $I^{\rm L}_s=I^{\rm R}_s=I_s=\tau_1^{-s}M_s[t]$ for the ground state. 
Near the centre of the band, i.e. around $\theta=0$, this sum is understood in Abel's sense; in this way, the resulting expressions for the energy and the charges are exact.
At $\theta=\delta\pm i\pi/6$, we therefore have
\be
    \log Y\big(\delta+\tfrac{i\pi}{6}\big)-\log Y\big(\delta-\tfrac{i\pi}{6}\big)= N\Big[\ell\big(\delta+\tfrac{i\pi}{6}\big)
    -\ell\big(\delta-\tfrac{i\pi}{6}\big)\Big] +4i\sum_sc_sI_s\sinh(s\delta)\,\sin\frac{\pi s}{6}\,.
\ee
Its derivative in $\delta$, evaluated at $\delta=0$, inserted in~\eqref{eq:E-from-Lambda}, with $v=\gamma\theta/\pi$, gives
\be
\cE=N\,e_\infty-4v_F\sum_s s\,\sin\frac{\pi s}{6}\;c_s\,I_s\,,\qquad v_F=\frac{\pi\sin\gamma}{\gamma}\,.
\label{eq:E-sin}
\ee
Here $v_F$ is the Fermi velocity of the chain, the velocity of its low-energy excitations.
In~\eqref{eq:E-sin}, the terms proportional to $N$ come from the driving term $N\ell$, from the term $-N\log g$ in $\log\Lambda$ and from $-N\cos\gamma$ in~\eqref{eq:E-from-Lambda}. Together they give $N e_\infty$, where $e_\infty$ is the energy per site of the infinite chain:
\be
\begin{aligned}
e_\infty&=\int_{-\infty}^{\infty}e(\lambda)\,\rho_\infty(\lambda)\,d\lambda\\
&=-\cos\gamma-\sin\gamma\Big[\frac{4\pi}{\sqrt3\,\gamma}-2\cot\frac{\gamma}{3}-\cot\frac{2\gamma}{3}\Big]\,,
\end{aligned}
\label{eq:einf}
\ee
with $\rho_\infty$ being the density of roots of the infinite chain\footnote{It follows from the Bethe equations in the limit of an infinite chain: differentiating the counting function $Z(\lambda)$ in \eqref{eq:Zlog} gives a linear integral equation for $\rho_\infty$, with $p'(\lambda)$ as source and the kernel $\Theta'(\lambda)$, which is solved by a Fourier transform.} \cite{Takahashi:1999}:
\be
\rho_\infty(\lambda)=\frac{1}{2\gamma\cosh(\pi\lambda/\gamma)}\,.
\ee
Since $\cos\frac{\pi s}{6}=\sqrt3\,\chi(s)\sin\frac{\pi s}{6}$ for $s\equiv\pm1\pmod6$, using \eqref{eq:tau-exact} we may rewrite the spin chain energy as
\begin{align}
\cE&=N\,e_\infty-\frac{4v_F}{\pi}\sum_s s\cos\frac{\pi s}{6}\;I_s\nonumber\\
&=N\,e_\infty-\frac{v_F}{\pi}\,\frac{M_1[t]}{\Neff}-\frac{4v_F}{\pi}\sum_{s\ge5}s\cos\frac{\pi s}{6}\;I_s\,,
\label{eq:energy-moments}
\end{align}
where in the second line we used $I_1=M_1[t]/(2\sqrt3\,\Neff)$.
At $t=0$ the two kinks carry the two chiralities: their spin-one moments are $M^{\rm L}_1=-4\pi^2\big(h-\frac{c}{24}\big)$ and $M^{\rm R}_1=-4\pi^2\big(\bar h-\frac{c}{24}\big)$, equal for the ground state. Keeping only the $s=1$ term we find \be \cE=N\,e_\infty +\frac{2\pi v_F}{\Neff}\big(h+\bar h-\frac{c}{12}\big)+\dots \,, \ee which reproduces the conformal finite-size spectrum of the chain~\cite{Blote:1986,Affleck:1986,Cardy:1986}, 
with $h,\bar h$ being the conformal weights and $c$ the central charge, and corrects it by replacing $N$ with the effective length $\Neff$. The remaining terms, with $s\ge5$, are the contributions of the higher local charges of the two kinks. Since $I_s=\tau_1^{-s}M_s[t]$ with $\tau_1\propto N$, the spin-$s$ term is of order $N^{-s}$, so the first of them enters at $N^{-5}$, together with the corrections to $M_1[t]$ from the higher times.

\subsubsection{The higher charges} 

For the higher charges \eqref{eq:Qn-def} we obtain a similar formula:
\be
\cQ_n=N\,\sq_n-\frac{2v_F}{\pi}\Big(\frac{\pi}{\gamma}\Big)^{n-2}\sum_s s^{n-1}\cos\frac{\pi s}{6}\Big[I^{\rm L}_s+(-1)^n\,I^{\rm R}_s\Big]\,,
\label{eq:Qn-moments}
\ee
where we defined the bulk density of the charge
\be
\sq_n=\int_{-\infty}^{\infty}e^{(n-2)}(\lambda)\,\rho_\infty(\lambda)\,d\lambda\,,\qquad \sq_2=e_\infty\,,
\ee
and $I^{\rm L}_s$ and $I^{\rm R}_s$ are the moments
associated with the left and right kinks. For the symmetric ground state they are equal, and hence the charges with odd $n$ vanish for $N\ge n$. We note three properties of~\eqref{eq:Qn-moments}. First, the $s$-th term is of order $N^{-s}$; once the moments are expanded in $1/N$, the sum becomes an asymptotic series. Second, $\cos\frac{\pi s}{6}$ vanishes at $s=3$, so no spin-3 moment enters. Third, the spin-$s$ moment enters $\mathcal Q_n$ with a factor $(\pi/\gamma)^{n-2}s^{n-1}$: the leading term, from $s=1$, is the same for all even $n$ up to $(\pi/\gamma)^{n-2}$, and the higher the charge, the more strongly it weights the higher spins.

For $N\ge n$, \eqref{eq:energy-moments} and~\eqref{eq:Qn-moments} are exact, with the sums understood in Abel's sense. The spin-one moment can then be eliminated with the exact sum rule~\eqref{eq:sumrule}, which brings in its non-perturbative term. For the ground state, the charges with odd $n$ vanish, and for even $n$
\begin{align}
\cQ_n=&N\,\sq_n-\frac{\pi v_F}{6\,\Neff}\Big(\frac{\pi}{\gamma}\Big)^{n-2}\Big[c_{\rm eff}+\frac{6}{\pi^2}\,{\rm Li}_2\big(-e^{-\epsilon(0)}\big)-\frac{6}{\pi^2}\sum_{s\ge5}s\,\tau_sI_s\Big]\notag\\
&-\frac{4v_F}{\pi}\Big(\frac{\pi}{\gamma}\Big)^{n-2}\sum_{s\ge5}s^{n-1}\cos\frac{\pi s}{6}\;I_s\,,
\label{eq:Qn-exact}
\end{align}
which for $n=2$ is the energy. The dilogarithm is of order $e^{-\epsilon(0)}\sim e^{-2.6N}$, smaller than any power of $1/N$, as are the corrections to the closure relation~\eqref{eq:closure}. In the large-$N$ expansions that follow we omit both, and work with~\eqref{eq:Qn-moments} and the moments $I_s=\tau_1^{-s}M_s[t]$.

\subsection{Asymptotic \texorpdfstring{$1/N$}{1byN} expansions}
\label{sec:energy-chain}
\label{sec:energy-charges}

Equation~\eqref{eq:Qn-moments} expresses the charges of
the chain as weighted sums, over spin $s$, of the
moments $I^{\rm L,R}_s$ associated with the two kinks. In the local coordinates, the moments are
given by the chiral GGE quantities $M_s[t]$ obtained from
the local cubic reconstruction. In particular, by~\eqref{eq:closure} we have
$I^{\rm L,R}_s=(\tau_1^{\rm L,R})^{-s}
M_s[t^{\rm L,R}]$, and the charges
of the chain become
\be
\cQ_n=N\,\sq_n-\frac{2v_F}{\pi}\Big(\frac{\pi}{\gamma}\Big)^{n-2}\sum_s s^{n-1}\cos\frac{\pi s}{6}\Big[(\tau_1^{\rm L})^{-s}M^{\rm L}_s+(-1)^n\,(\tau_1^{\rm R})^{-s}M^{\rm R}_s\Big]\,,
\label{eq:Qn-closed}
\ee
where $\tau_1^{\rm L,R}=2\sqrt3\,N^{\rm L,R}_{\rm eff}$
are the scales associated with the two kinks, and
we use the shorthand
$M^{\rm L,R}_s\equiv M_s[t^{\rm L,R}]$ for the
corresponding chiral moments. The scales and the times are
fixed by the self-consistency equations~\eqref{eq:coupled-LR},
so $\cQ_n$ depends only on $N$ and on the state.
Equation~\eqref{eq:Qn-closed} contains no explicit cross
terms between the left and right contributions; their coupling
is encoded in the dependence of $\tau_1^{\rm L,R}$ and
$t^{\rm L,R}_s$ on the moments of the other kink. We denote by
\be
\hL=h-\frac{c}{24}\,,\qquad \hR=\bar h-\frac{c}{24}
\ee
the eigenvalues of $L_0-c/24$ and $\bar L_0-c/24$ associated with the left and right chiral sectors, respectively. At zero times the spin-one moments are $M^{\rm L,R}_1=-4\pi^2\mathsf h_{\rm L,R}$. 

We expand~\eqref{eq:Qn-closed} in $1/N$. 
In the explicit series below, the moments $M^{\rm L,R}_s$ and the responses $\partial_{t_5}M^{\rm L,R}_5$ are evaluated at zero higher times, in the chiral state of the corresponding kink, since the effect of the times is already contained in the explicit powers of $1/N$. For the ground state these data are given by~\eqref{vacuummoments} and~\eqref{eq:M5-response}. The series then depends on $n$ only through the factors $s^{n-1}$ with which the spin-5 and spin-7 moments enter, $w_5=5^{n-1}$ and $w_7=7^{n-1}$. 
For $n$ odd the charge is given by
\be
\begin{aligned}
\frac{\cQ_n}{v_F\,(\pi/\gamma)^{n-2}}=&\ \frac{2\pi}{N}\,\big(\hL-\hR\big)+\frac{(w_5-1)\,\big(M^{\rm L}_5-M^{\rm R}_5\big)}{288\pi}\,\frac1{N^5}\\
&+(w_5-1)\,\frac{5\sqrt3}{864}\,\big(M^{\rm L}_5\,\hR-M^{\rm R}_5\,\hL\big)\frac1{N^7}+(w_7-1)\,\frac{M^{\rm L}_7-M^{\rm R}_7}{3456\pi}\,\frac1{N^7}\\
&+\Big[5\pi\,(w_5-1)\Big(\frac{\big(M^{\rm L}_5-M^{\rm R}_5\big)\,\hL\hR}{864}+\frac{M^{\rm L}_5\,\hR^2-M^{\rm R}_5\,\hL^2}{288}\Big)\\
&\qquad+(w_7-1)\,\frac{7\sqrt3}{10368}\,\big(M^{\rm L}_7\,\hR-M^{\rm R}_7\,\hL\big)\\
&\qquad-\frac{(w_5-1)\,\big(\partial_{t_5}M^{\rm L}_5-\partial_{t_5}M^{\rm R}_5\big)}{207360\,\pi}\Big]\frac1{N^9}+O(N^{-11})\,.
\end{aligned}
\label{eq:Qn-odd}
\ee
The leading term is proportional to
$\hL-\hR=h-\bar h$, which is the conformal spin, and the odd
charges vanish when the two chiral sectors carry the same
moments, as in the symmetric ground state. 
Every other term carries a factor $w_s-1$, because the spin-$s$ moment enters $\cQ_n$ twice: directly, with weight $w_s$, and through the effective length with weight $-1$. The odd charges therefore measure the asymmetry between
the two chiral sectors in the spin-$5$ and spin-$7$
moments, as well as in the spin-$9$ response
$\partial_{t_5}M_5$.

For $n$ even the charge is given by
\begin{align}
\frac{\cQ_n-N\,\sq_n}{v_F\,(\pi/\gamma)^{n-2}}=&\ \frac{2\pi}{N}\,\big(\hL+\hR\big)+\frac{4\pi^2}{\sqrt3}\,\frac{\hL\,\hR}{N^3}\notag\\
&+\Big[\frac{(w_5-1)\,\big(M^{\rm L}_5+M^{\rm R}_5\big)}{288\pi}+\frac{4\pi^3}{3}\,\hL\hR\,\big(\hL+\hR\big)\Big]\frac1{N^5}\notag\\
&+\Big[\frac{\sqrt3\,(5w_5-7)}{864}\,\big(M^{\rm L}_5\,\hR+M^{\rm R}_5\,\hL\big)+\frac{(w_7-1)\,\big(M^{\rm L}_7+M^{\rm R}_7\big)}{3456\pi}\notag\\
&\qquad+\frac{4\sqrt3\,\pi^4}{9}\,\hL\hR\,\big(\hL^2+3\hL\hR+\hR^2\big)\Big]\frac1{N^7}\notag\\
&+\Big[\pi\,(5w_5-9)\Big(\frac{\big(M^{\rm L}_5+M^{\rm R}_5\big)\,\hL\hR}{864}+\frac{M^{\rm L}_5\,\hR^2+M^{\rm R}_5\,\hL^2}{288}\Big)\notag\\
&\qquad+\frac{\sqrt3\,(7w_7-9)}{10368}\,\big(M^{\rm L}_7\,\hR+M^{\rm R}_7\,\hL\big)-\frac{(w_5-1)\,\big(\partial_{t_5}M^{\rm L}_5+\partial_{t_5}M^{\rm R}_5\big)}{207360\,\pi}\notag\\
&\qquad+\pi^5\Big(\frac49\,\big(\hL^4\hR+\hL\hR^4\big)+\frac83\,\big(\hL^3\hR^2+\hL^2\hR^3\big)\Big)\Big]\frac1{N^9}\notag\\
&+O(N^{-11})\,.
\label{eq:Qn-even}
\end{align}
The terms built from $\hL$ and $\hR$ alone come from the dressing $N\to\Neff$ and are the same for all even $n$. Through order $N^{-9}$ the series is built from the eigenvalues $\hL$, $\hR$, $M^{\rm L,R}_5$ and $M^{\rm L,R}_7$ of KdV
charges, except for the terms
$\partial_{t_5}M^{\rm L,R}_5$ at order $N^{-9}$. In the ground
state these responses are proportional to the vacuum eigenvalues of the spin-9
charge $J_{5,9}$ of the deformed KdV hierarchy. For an excited
chiral sector, the corresponding eigenvalues and responses of
that state must be used instead. Since $J_{5,9}$ is not a KdV
charge, its contribution cannot be written in terms of KdV
charges and their bilinears at the conformal point. 

The energy is the case $n=2$, with $w_5=5$ and $w_7=7$. Its first terms are
\be
\begin{aligned}
\frac{\cE-N\,e_\infty}{v_F}=&\ \frac{2\pi}{N}\,\big(\hL+\hR\big)+\frac{4\pi^2}{\sqrt3}\,\frac{\hL\,\hR}{N^3}\\
&+\Big[\frac{M^{\rm L}_5+M^{\rm R}_5}{72\pi}+\frac{4\pi^3}{3}\,\hL\hR\,\big(\hL+\hR\big)\Big]\frac1{N^5}\\
&+\Big[\frac{\sqrt3}{48}\,\big(M^{\rm L}_5\,\hR+M^{\rm R}_5\,\hL\big)+\frac{M^{\rm L}_7+M^{\rm R}_7}{576\pi}\\
&\qquad+\frac{4\sqrt3\,\pi^4}{9}\,\hL\hR\,\big(\hL^2+3\hL\hR+\hR^2\big)\Big]\frac1{N^7}+O(N^{-9})\,.
\end{aligned}
\label{eq:E-expansion}
\ee
Since the moments $M_s$ grow with $s$, the sum in~\eqref{eq:Qn-moments} is only asymptotic in $1/N$, but for the sizes and orders used here we are well below the optimal truncation order.
The first two terms are the conformal energy and the $\TTbar$ term. They agree with the effective Hamiltonian of~\cite{Lukyanov:1997bp}, where at the Lee--Yang point, the correction of first order in its $T\bar T$ coupling $\lambda_+$ reads, in our notation\footnote{In the conventions of~\cite{Lukyanov:1997bp}, our chain has $\beta^2=2/5$ and twist $\theta=1/5$. The coupling $\lambda_-$ multiplies $I_3(p)+I_3(\bar p)$, which vanishes because the spin-3 vacuum eigenvalue of the Lee--Yang model is zero, and the $\cos(\tilde\varphi/\beta)$ term has a vanishing leading finite-size contribution.},
\be
-\frac{8\pi^3v_F}{N^3}\,\lambda_+\,\hL\,\hR\,,\qquad \lambda_+=-\frac{1}{2\pi\sqrt3}\,,
\ee
and this is exactly the term of order $N^{-3}$ in \eqref{eq:E-expansion}. 
The terms beyond order $N^{-3}$ and the self-consistency equation for $\Neff$ are obtained here directly from the ODE/IM reconstruction of the finite-size expansion.

\subsection{Numerical verification and coupling effects}
\label{sec:coupling-role}

Each local description involves an effective
length shifted by the spin-$1$ moment of the opposite kink,
$N\to N^{\rm L,R}_{\rm eff}$, with strengths
\be
\alpha_{\rm L}=\frac{M^{\rm L}_1}{4\sqrt3\,\pi}=-\frac{\pi}{\sqrt3}\,\hL\,,\qquad
\alpha_{\rm R}=\frac{M^{\rm R}_1}{4\sqrt3\,\pi}=-\frac{\pi}{\sqrt3}\,\hR\,,
\label{eq:alphaLR}
\ee
so that 
\be 
N^{\rm L}_{\rm eff}=N+\alpha_{\rm R}/N^{\rm R}_{\rm eff}~,\quad \text{and}\quad N^{\rm R}_{\rm eff}=N+\alpha_{\rm L}/N^{\rm L}_{\rm eff}~. 
\ee 
The moments in \eqref{eq:alphaLR} are taken at zero times.
For the symmetric ground state both reduce to the constant
$\alpha$ of Section~\ref{sec:gluing-Neff}. The terms
of~\eqref{eq:Qn-even} built from $\hL$ and $\hR$ alone
are generated by this effective-length shift, while the
remaining terms involve the higher-spin chiral moments
obtained from the local cubic reconstruction.

\subsubsection{Comparison with exact values}

To test the series we compare it with exact values, obtained by solving the Bethe equations numerically for the ground state, for $N$ up to $96$. The energy and the charges are the sums over the roots, $\sum_je^{(n-2)}(\lambda_j)$, minus the bulk term $N\sq_n$. Table~\ref{tab:coupling} gives the difference between the exact energy and the series. For each row we retain the indicated chiral moments and responses, solve the resulting self-consistency equations, and evaluate the energy.
\begin{table}[!tbp]
\centering
\begin{tabular}{lcccc}
\toprule
 & $N=12$ & $N=24$ & $N=48$ & $N=96$\\
\midrule
coupling off, any charges & $5.8\times10^{-6}$ & $7.3\times10^{-7}$ & $9.1\times10^{-8}$ & $1.1\times10^{-8}$ \\
coupling on, $M_1$ only & $1.3\times10^{-6}$ & $3.9\times10^{-8}$ & $1.2\times10^{-9}$ & $3.8\times10^{-11}$ \\
\quad $+\,M_5$ & $3.4\times10^{-8}$ & $2.6\times10^{-10}$ & $2.0\times10^{-12}$ & $1.6\times10^{-14}$ \\
\quad $+\,M_7$ & $1.4\times10^{-9}$ & $2.6\times10^{-12}$ & $5.0\times10^{-15}$ & $9.8\times10^{-18}$ \\
\quad $+\,\partial_{t_5}M_5$ & $1.0\times10^{-10}$ & $4.6\times10^{-14}$ & $2.2\times10^{-17}$ & $1.1\times10^{-20}$ \\
\quad all charges through spin 11 & $1.2\times10^{-11}$ & $1.3\times10^{-15}$ & $1.5\times10^{-19}$ & $1.8\times10^{-23}$ \\
\bottomrule
\end{tabular}
\caption{Error of the energy $\cE-N\,e_\infty$ of the ground state, exact minus approximate. With the coupling switched off, the dominant error comes from the missing order-$N^{-3}$ term, while higher-spin charges modify only subleading contributions. With the coupling included, the charges are added cumulatively, one at a time, up to the spin-$9$ response $\partial_{t_5}M_5$ and all charges through spin $11$, which include $M_{11}$ and $\partial_{t_7}M_5=\partial_{t_5}M_7$.}
\label{tab:coupling}
\end{table}

To isolate the effect of the coupling, we also evaluate the
series with it switched off, setting $\Neff=N$ and keeping
the bare times. Without the coupling, higher-spin terms change
the subleading corrections but cannot remove the leading
$N^{-3}$ discrepancy associated with the missing $\TTbar$
feedback. With the coupling included, the successive higher-spin contributions retained in the expansion improve the agreement at the expected successive orders in $1/N$.
The direct shift of the spin-$s$ time associated with the left 
kink by the spin-$s$ moment of the right one 
first enters the energy at order $N^{-11}$, through the
product $M^{\rm L}_5M^{\rm R}_5$. This has the same structure as the spin-$5$ coupling appearing in generalised $\TTbar$ flows~\cite{Smirnov:2016lqw,Conti:2019dxg,Hernandez-Chifflet:2019sua,LeFloch:2019wlf}.

\section{An excited state with asymmetric kinks}
\label{sec:gluing-excited}

The expansions of Section~\ref{sec:energy-chain} depend on
the state through the moments $M^{\rm L}_s$ and
$M^{\rm R}_s$ associated with the left and right kinks, which
need not be equal. For the symmetric ground state the two
local chiral sectors are related by parity, so the odd
charges vanish and the cross terms are symmetric. For an
asymmetric state, the odd charges no longer vanish and
cross terms such as $M^{\rm L}_5\,\hR$ and
$M^{\rm R}_5\,\hL$ are no longer equal. Such states
therefore provide a direct test of the left--right
structure of the construction. The simplest example has the left chiral sector in the Lee--Yang ground state associated with the primary $\phi$ and the right chiral sector in its first descendant $L_{-1}\phi$, so that
\be
\hL=-\frac1{60}\,,\qquad \hR=\frac{59}{60}\,.
\ee
\subsection{The excited state and its TBA}
\label{sec:exc-tba}
This excited state is described by $M-1$ real Bethe roots, plus one Bethe root on the line ${\rm Im}\,\lambda=\pi/2$. The TBA equation has, besides the lattice driving term, a pair of complex conjugate sources $\eta,\bar\eta$:
\be
\epsilon(\theta)=N\ell(\theta)+\log\frac{S(\theta-\eta)}{S(\theta-\bar\eta)}-\int_{-\infty}^{\infty} d\theta'\,\Phi(\theta-\theta')\,L(\theta')\,,\qquad
S(u)=\frac{\sinh u+i\sin\frac\pi3}{\sinh u-i\sin\frac\pi3}\,,
\label{eq:exc-TBA}
\ee
with $S$ the Lee--Yang $S$-matrix~\cite{Zamolodchikov:1989cf}, and the position of the source is fixed by the exact quantisation condition $1+Y(\eta)=0$, which can be rewritten as
\be
\exp\Big[N\ell(\eta)-\int_{-\infty}^{\infty} d\theta'\,\Phi(\eta-\theta')\,L(\theta')\Big]=S(\eta-\bar\eta)\,.
\label{eq:exc-quant}
\ee
This is the lattice form of the excited-state TBA of ~\cite{Bazhanov:1996aq,Dorey:1996re}; the equivalent equation for the RSOS model was derived from the lattice in~\cite{Bajnok:2014fca}, where the pair $\eta,\bar\eta$ corresponds to a short two-string of zeros of the transfer-matrix eigenvalue.\footnote{The equation follows from the $Y$-system by the same argument as Proposition~3.2.8 of~\cite{Hilfiker:2018}, adapted to the meromorphic lattice $Y$, and $1+Y(\eta)=0$ is the corresponding local quantisation condition, cf.\ Lemma~3.2.10.}

\begin{figure}[t]
\centering
\includegraphics[width=.80\textwidth]{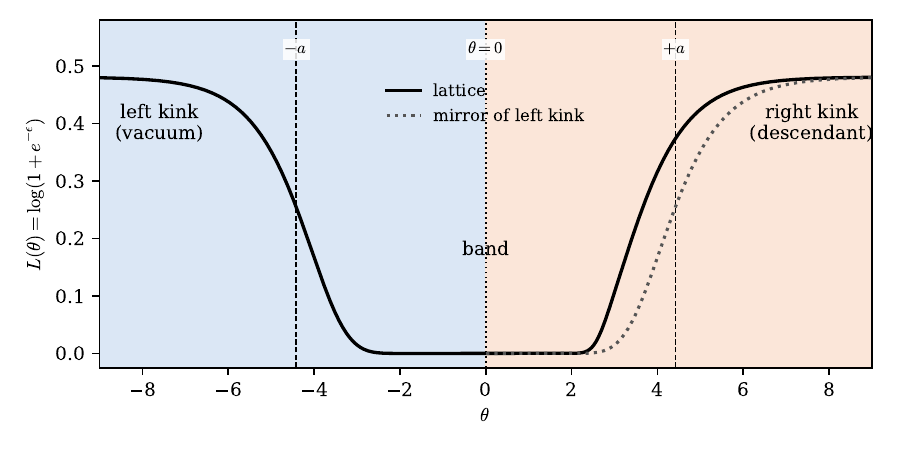}
\caption{The function $L=\log(1+e^{-\epsilon})$ for the state of the chain in which the right kink is in the first descendant, at $N=24$ (full line), together with the mirror image of the left profile (dotted). The right kink is displaced towards the centre of the band relative to this mirror image, so the two kinks are no longer related by parity. The dashed lines mark the Fermi points of the ground state at the same $N$, $\theta=\pm\log(2\sqrt3\,\Neff)$, and the dotted line at $\theta=0$ splits the line into the left and right regions of the text.}
\label{fig:kinks-asym}
\end{figure}

This state is not parity invariant (see Figure~\ref{fig:kinks-asym}), so after splitting the convolution at $\theta=0$ as in Section~\ref{sec:gluing-fold}, we cannot use folding. Since we
choose the excitation to lie in the right chiral sector,
the source has $\text{Re}(\eta)>0$ and therefore
contributes to the moments associated with the right
kink:
\be
I^{\rm L}_s=\int_{-\infty}^{0}e^{s\theta}L(\theta)\,d\theta\,,\qquad
I^{\rm R}_s=\int_{0}^{\infty}e^{-s\theta}L(\theta)\,d\theta-\frac{4\pi}{s}\,e^{-s\,{\rm Re}\,\eta}\sin\big(s\,{\rm Im}\,\eta\big)\,.
\label{eq:exc-moments}
\ee
Each Fermi region is then described locally by a chiral GGE, with effective times fixed by
\be
(\tau^{\rm L}_1)^s\,t^{\rm L}_s=b_s+c_sI^{\rm R}_s\,,\qquad (\tau^{\rm R}_1)^s\,t^{\rm R}_s=b_s+c_sI^{\rm L}_s\,,
\ee
which is~\eqref{eq:tau-exact} with the moments~\eqref{eq:exc-moments} of the excited state.
The residues in the last term of~\eqref{eq:exc-moments} depend on the position of the source, fixed by~\eqref{eq:exc-quant}, which moves when an effective time of the corresponding chiral sector is varied. The source contribution is included explicitly in the modified moments~\eqref{eq:exc-moments}. With these moments, the derivation of Section~\ref{sec:gluing-energy} carries over, and the energy and the local charges are again given by~\eqref{eq:Qn-moments}.

\subsection{The descendant ODE and apparent singularities}
\label{sec:exc-ode}

In the excited-state ODE/IM construction~\cite{Bazhanov:2003ni} (see also~\cite{Fioravanti_2005}, and~\cite{Kotousov:2025qfun} for the XXZ chain in the scaling limit), excited states are described by supplementing the conformal potential with apparent singularities, namely double poles with trivial monodromy. For the level-one state $L_{-1}\phi$, the potential becomes
\be
-\psi''+\Big[x^3-2\frac{d^2}{dx^2}\log\big(x^5-z\big)-E\Big]\psi=0\,,\qquad z=-\frac43\,,
\label{eq:ode-level1}
\ee
with five poles at $x_k^5=z$. These five poles are the orbit of a single parameter $z$ under the $\mathbb Z_5$ symmetry $x\to e^{2\pi i/5}x$ of the cubic, and $z=-\frac43$ is fixed by the requirement of trivial monodromy around them. 

When the times are switched on, the potential acquires the coefficients $Q(E)$ and $P(E)$,
and the poles move to new positions $x_k(Q,P)$ determined by requiring that the monodromy
around each singularity remains trivial. For a potential of the form
\be
V(x)=x^3+Qx-P+2\sum_j\frac{1}{(x-x_j)^2},
\ee
with distinct poles and no simple-pole terms, requiring the Frobenius solutions at $x_k$ to be free
of logarithmic terms gives
\be
r_k'(x_k)
=3x_k^2+Q
-4\sum_{j\ne k}\frac{1}{(x_k-x_j)^3}
=0,
\ee
where $r_k(x)$ denotes the regular part of the potential at
$x_k$. This condition follows from the resonance of the two local Frobenius exponents and determines the pole positions implicitly as functions of $Q$ and $P$, on the branch connected to the conformal solution.

At the conformal point ($t_s=0,\,\,s>1$), the relevant coefficients of the WKB periods of the apparent-singularity ODE~\eqref{eq:ode-level1} can be evaluated exactly, giving the conformal moments:
\be
\begin{aligned}
M^{\rm R}_1&=-\frac{59}{24}\,\frac{\pi}{\sqrt3}\,\cA\,\cB\,,\quad
M^{\rm R}_5=-\frac{14605357}{1119744}\,\frac{\pi}{\sqrt3}\,\cA^6\,,\quad 
M^{\rm R}_7=\frac{15263212459}{376233984}\,\frac{\pi}{\sqrt3}\,\cA^7\cB\,.
\end{aligned}
\label{eq:exc-charges}
\ee
To determine the finite-size corrections through order $N^{-9}$, one also needs the response 
of the spin-5 charge to the deformation time, $\partial_{t_5}M_5|_{t=0}$. In principle, this 
can be obtained by perturbing the positions of the apparent singularities in~\eqref{eq:ode-level1} 
under the GGE deformation. However, it is algebraically much more direct to extract this 
linear response from the eigenvalues of the deformed charge hierarchy computed in 
Appendix~A of~\cite{Karimi:2026kdv}. The response is proportional to the charge itself:
\be
\partial_{t_5}M_5=\xi\,\frac{\cA^5}{\cB}\,M_5\,,\qquad
\xi^{\rm L}=-\frac{104630435}{461376}\,,\qquad
\xi^{\rm R}=-\frac{31284500065}{306535104}\,.
\label{eq:exc-response}
\ee
As in~\eqref{eq:M5-response}, $\xi\,\cA^5/\cB$ is the ratio $(\varsigma/t_5)\,J_{5,9}/I_5$ of the eigenvalues of the non-KdV charge $J_{5,9}$ and the KdV charge $I_5$ in the corresponding chiral state. 

\subsection{The energy and the charges of the descendant}
\label{sec:exc-energy}

Using the charges of the previous subsection,~\eqref{eq:Qn-odd} and~\eqref{eq:Qn-even} give the energy and the local charges of the descendant, which we compare with the Bethe ansatz. In the conformal limit, this state is the descendant $L_{-1}\phi$ of the Lee--Yang ground-state primary $\phi$. In particular, the left kink solves the ground-state TBA and the right one solves the excited-state TBA, with $\hL=-\frac1{60}$ and $\hR=\frac{59}{60}$. The relevant moments of the excited state have been listed above. 
The leading terms are what the Bethe ansatz leads us to expect: the energy exceeds that of the ground state by $2\pi v_F/N$, and the leading term of the first odd charge is proportional to the conformal spin of the state, carried entirely by the excited kink.

The corrections show how the chain distinguishes the two kinks. In the $\TTbar$ term the sign of $\hL\hR$ is reversed compared to the ground state:
the excited right kink has $\hR>0$, so $\alpha_{\rm R}<0$ and the effective length $N^{\rm L}_{\rm eff}$ is shortened, while the ground-state left kink, with $\hL<0$, increases the effective length $N^{\rm R}_{\rm eff}$. The first correction to the odd charge starts at order $N^{-5}$, where it measures the difference between the spin-$5$ moments of the two kinks. At order $N^{-9}$ the response of the spin-$5$ charge enters, much larger for the kink associated to the excited state than for the vacuum one, making this state a sensitive test of the non-KdV charge.

As for the ground state, without the coupling the leading error is controlled by the missing $\TTbar$-type feedback, while including the successive higher-spin contributions improves the agreement at the expected successive orders in $1/N$, both for the energy and for $\cQ_3$ (see Table~\ref{tab:coupling-exc}).
\begin{table}[!tbp]
\centering
\begin{tabular}{lcccc}
\toprule
 & $N=12$ & $N=24$ & $N=48$ & $N=96$\\
\midrule
coupling off, all charges & $-3.5\times10^{-4}$ & $-4.3\times10^{-5}$ & $-5.4\times10^{-6}$ & $-6.7\times10^{-7}$ \\
coupling on, $M_1$ only & $-4.0\times10^{-4}$ & $-1.3\times10^{-5}$ & $-4.0\times10^{-7}$ & $-1.2\times10^{-8}$ \\
\quad $+\,M_5$ & $9.4\times10^{-6}$ & $7.5\times10^{-8}$ & $5.9\times10^{-10}$ & $4.6\times10^{-12}$ \\
\quad $+\,M_7$ & $-1.9\times10^{-7}$ & $-3.8\times10^{-10}$ & $-7.5\times10^{-13}$ & $-1.5\times10^{-15}$ \\
\quad $+\,\partial_{t_5}M_5$ & $5.2\times10^{-9}$ & $2.6\times10^{-12}$ & $1.3\times10^{-15}$ & $6.2\times10^{-19}$ \\
\midrule
$\cQ_3$, all of the above & $-7.9\times10^{-9}$ & $-4.0\times10^{-12}$ & $-2.0\times10^{-15}$ & $-9.5\times10^{-19}$ \\
\bottomrule
\end{tabular}
\caption{Error of the energy $\cE-N\,e_\infty$ of the descendant state, exact minus approximate. Without the coupling, adding higher-spin charges changes subleading corrections but does not remove the leading $N^{-3}$ discrepancy. With the coupling, the charges are added cumulatively, one at a time, up to the spin-$9$ response $\partial_{t_5}M_5$. The last row is the error of the odd charge $\cQ_3$ with the coupling and all of these charges.}
\label{tab:coupling-exc}
\end{table}

\section{The Hamilton--Jacobi structure of the coupling}
\label{sec:hj}

In two dimensions the $T\bar T$ deformation moves the energy levels of a theory on a cylinder of circumference $R$ according to the inviscid Burgers equation~\cite{Smirnov:2016lqw,Cavaglia:2016oda}, with a source term containing the momentum $P$,
\be
\partial_tE=E\,\partial_RE+\frac{P^2}{R}\,,
\label{eq:Burgers}
\ee
where $t$ is the deformation parameter. For $P=0$ this is the equation for the gradient $E=\partial_RS$ of the free Hamilton--Jacobi equation 
\be \partial_tS=\frac12(\partial_RS)^2\,,
\label{def:HamiltonJacobi}
\ee with $t$ as the time and $R$ as the coordinate. In holography, the flow of the effective action of the deformed theory is the Hamilton--Jacobi equation of the radial evolution of the gravitational action~\cite{McGough:2016lol}, and the flow of the energies is a component of the Einstein equations at the cutoff~\cite{Kraus:2018xrn}. Generalised deformations involving higher-spin conserved charges lead to analogous coupled flow equations for the finite-volume spectrum~\cite{Hernandez-Chifflet:2019sua,LeFloch:2019wlf}.
In this section we show that the self-consistency equations of the previous sections can be recast as a Hamilton--Jacobi equation, which makes the analogy with the $T\bar T$ flows precise, and which gives a systematic way to extend our results to higher orders in $1/N$.

\subsection{The generating function of a single kink}
\label{sec:hj-potential}

Take one kink with independent times $\tau_s$, $s\equiv\pm1\pmod6$, and write $t_s=\tau_s/\tau_1^s$. Its moments are $I_s=\tau_1^{-s}M_s[t]$, and its driving term $\nu=\sum_s\tau_se^{s\theta}$. Then the TBA equation is the stationarity condition for the  functional of the pseudoenergy $\epsilon(\theta)$
\be
\mathcal J[\epsilon]=\int_{-\infty}^{\infty} d\theta\Big[\nu\,L-\frac12\,L\,(\Phi*L)-\epsilon\,L+{\rm Li}_2\big(-e^{-\epsilon}\big)\Big]\,,
\label{eq:hj-J}
\ee
whose variation is $\delta\mathcal J=\int d\theta\,L'(\epsilon)\,\big[\nu-\Phi*L-\epsilon\big]\,\delta\epsilon$, with $L=\log(1+e^{-\epsilon})$ and $L'$ its derivative with respect to $\epsilon$. 
The on-shell functional requires a fixed subtraction, since its integrand tends to
$-\pi^2/15$ as $\theta\to-\infty$. We therefore define the regularised on-shell
functional
\be
\mathcal F(\tau)=
\int_{-\infty}^{\infty}d\theta\,
\left[j(\theta;\tau)+\frac{\pi^2}{15}\Theta(-\theta)\right],
\label{def:regularised_func}
\ee
where $j(\theta;\tau)$ denotes the integrand of~\eqref{eq:hj-J} evaluated on the
TBA solution and $\Theta$ is the step function. The subtraction is independent of the times. Since the times enter
$\mathcal J$ only through the driving term $\nu$, varying $\mathcal F$ with respect
to a time $\tau_s$ gives
\be
\frac{\partial\mathcal F}{\partial\tau_s}=\int_{-\infty}^{+\infty} d\theta \left[e^{s\theta}L(\theta)+\frac{\delta \mathcal J}{\delta \epsilon}\frac{\delta \epsilon}{\delta \tau_s}\right]=I_s\,,
\label{eq:hj-potential}
\ee
where the second term in the square bracket vanishes due to the fact that $\delta \mathcal J/\delta \epsilon= 0$ on the TBA solution.
Thus $\mathcal F$ is the generating function of the charges of the chiral GGE. It is the analogue, for the TBA equation of the kink, of the critical value of the Yang--Yang functional of the Bethe roots, whose derivatives with respect to the parameters give the energy and the other observables~\cite{Yang:1966ty, Nekrasov:2011bc, Lukyanov:2011wd}. 
When only $\tau_1$ is non-zero, the driving term is $\tau_1e^{\theta}=e^{\theta+\log\tau_1}$, so changing $\tau_1$ translates the kink. Since the subtraction in~\eqref{def:regularised_func} is not translated, this shifts $\mathcal F$ by
\be
\mathcal F(\tau_1)-\mathcal F(1)=\frac{\pi^2}{15}\,\log\tau_1\,.
\ee
More generally, translating the rapidity by $\log\lambda$ rescales all the times, $\tau_s\to\lambda^s\tau_s$, so that
\be
\mathcal F(\lambda^s\tau_s)=\mathcal F(\tau_s)+M^{(0)}_1\log\lambda\,,\qquad M^{(0)}_1=\frac{\pi^2}{15}\,,
\label{eq:hj-scaling}
\ee
and the times enter $\mathcal F$ only through $\tau_1$ and the rescaled times $t_s=\tau_s/\tau_1^s$:
\be
\mathcal F=M^{(0)}_1\log\tau_1+\mathcal H(t)\,.
\ee
Differentiating~\eqref{eq:hj-scaling} at $\lambda=1$ and using~\eqref{eq:hj-potential} gives
\be
\sum_s s\,\tau_s\,I_s=\frac{\pi^2}{15}\,,
\ee
the large-$N$ form of the sum rule~\eqref{eq:sumrule}, without the exponentially small term from the centre of the band\footnote{The same argument, applied to the functional whose stationary point is the folded equation~\eqref{eq:folded} on the half line $\theta<0$, gives the exact sum rule~\eqref{eq:sumrule}. Rescaling the times is again a translation of the rapidity, but the endpoint of the half line, at the centre of the band, is not translated, and the variation of the on-shell functional with respect to the endpoint contributes the dilogarithm. This is an alternative to the derivation of Appendix~\ref{app:sumrule}.}. For $s\ge5$, \eqref{eq:hj-potential} also gives
\be
\frac{\partial\mathcal F}{\partial\tau_s}=\tau_1^{-s}\,\frac{\partial\mathcal H}{\partial t_s}=I_s=\tau_1^{-s}M_s[t]\,,
\ee
so $\mathcal H$ generates the chiral moments computed by the cubic,
\be
M_s[t]=\frac{\partial\mathcal H}{\partial t_s}\,,\qquad s\ge5\,.
\ee
Expanding $\mathcal H$ in the times,
\be
\mathcal H(t)=\sum_{s\ge5}M^{(0)}_s\,t_s+\frac12\sum_{p,q\ge5}\partial_{t_q}M_p\;t_pt_q+\dots\,,
\label{eq:hj-F}
\ee
its coefficients are the data of Section~\ref{sec:chiral-charges}: the conformal charges $M^{(0)}_s$ and their responses to the times. Since they are derivatives of a single function, the responses are symmetric,
\be
\partial_{t_q}M_p=\partial_{t_p}M_q\,.
\ee
For the chain the higher times are small and negative, where the chiral GGE on the whole line is not defined; $\mathcal H$ is then used only through this expansion about the conformal point, order by order in $1/N$.

For an excited chiral sector, the moments receive an additional contribution from the source, whose position is fixed by the quantisation condition~\eqref{eq:exc-quant} and therefore shifts when a time is varied. In this case we assume that the generating relation~\eqref{eq:hj-potential} continues to hold with the source following its quantisation condition. A derivation including the source position as an additional variational variable would be required to establish this relation directly for excited states.

\subsection{The coupled kinks and the Hamilton--Jacobi equation}

For a single kink, the potential $\mathcal F$ generates its moments through \eqref{eq:hj-potential}. We seek a potential $\mathcal W$ with the same property for the coupled system, and an equation describing how $\mathcal W$ changes when the couplings vary.
To follow how $\mathcal W$ depends on the coupling, we replace the fixed couplings $c_s$ by free parameters $\hat c_s$. We also introduce separate bare times for the two kinks:
\be
\tau^{\rm L}_s=b^{\rm L}_s+\hat c_sI^{\rm R}_s\,,\qquad
\tau^{\rm R}_s=b^{\rm R}_s+\hat c_sI^{\rm L}_s\,.
\label{eq:hj-feedback}
\ee
Together with $I_s^a=\partial_{\tau_s^a}\mathcal F_a$, where $a={\rm L},{\rm R}$ labels the two kinks, these equations determine the times and moments for each choice of $b^{\rm L}$, $b^{\rm R}$ and $\hat c$. The physical chain is the point
\be
b^{\rm L}_s=b^{\rm R}_s=b_s=\frac{4N}{s}\cos\frac{\pi s}{6}\,,\qquad
\hat c_s=c_s=\frac{\sqrt3}{\pi}\,\chi(s)\,.
\label{physicallatticetimes}
\ee
We then define
\be
\mathcal W=\mathcal F_{\rm L}(\tau^{\rm L})+\mathcal F_{\rm R}(\tau^{\rm R})-\sum_s\hat c_s\,I^{\rm L}_sI^{\rm R}_s\,.
\label{eq:hj-W}
\ee
To see the role of the last term, differentiate the times
$
d\tau_s^{\rm L}=db_s^{\rm L}+I_s^{\rm R}\,d\hat c_s+\hat c_s\,dI_s^{\rm R}\,,$
and similarly with ${\rm L}\leftrightarrow{\rm R}$. Since
$d\mathcal F_a=\sum_s I_s^a\,d\tau_s^a$, we obtain
\be
\begin{aligned}
d(\mathcal F_{\rm L}+\mathcal F_{\rm R})
=\sum_s\Big[
&I_s^{\rm L}\,db_s^{\rm L}
+I_s^{\rm R}\,db_s^{\rm R}
+2I_s^{\rm L}I_s^{\rm R}\,d\hat c_s
+\hat c_s\big(I_s^{\rm L}\,dI_s^{\rm R}
+I_s^{\rm R}\,dI_s^{\rm L}\big)
\Big].
\end{aligned}
\ee
Subtracting $d\big(\sum_s\hat c_s I_s^{\rm L}I_s^{\rm R}\big)$ cancels the terms containing $dI_s^{\rm L,R}$ and leaves
\be
d\mathcal W=\sum_s\Big[
I_s^{\rm L}\,db_s^{\rm L}
+I_s^{\rm R}\,db_s^{\rm R}
+I_s^{\rm L}I_s^{\rm R}\,d\hat c_s
\Big].
\ee
Hence we see that
\be
\partial_{b_s^{\rm L}}\mathcal W=I_s^{\rm L}\,,\qquad
\partial_{b_s^{\rm R}}\mathcal W=I_s^{\rm R}\,,\qquad
\partial_{\hat c_s}\mathcal W=I_s^{\rm L}I_s^{\rm R}\,.
\ee
Substituting the first two identities into the third gives
\be
\partial_{\hat c_s}\mathcal W
=\big(\partial_{b_s^{\rm L}}\mathcal W\big)
 \big(\partial_{b_s^{\rm R}}\mathcal W\big)\,.
\label{eq:hj}
\ee
Equation~\eqref{eq:hj} has the form of a Hamilton--Jacobi equation with several evolution parameters,
\be
\frac{\partial S}{\partial\sigma_s}+H_s\Big(q_i,\,p_j=\frac{\partial S}{\partial q_j}\Big)=0\,,
\ee
where the index $i$ runs over the pairs $(a,s)$. The action is the potential $\mathcal W$, the evolution parameters are $\sigma_s=-\hat c_s$, the coordinates are the bare times $b^a_s$, and the conjugate momenta are the charges $p^a_s=\partial_{b^a_s}\mathcal W=I^a_s$. The Hamiltonian depends only on the momenta, and is the product of the left and right charges of spin $s$, $H_s=p^{\rm L}_sp^{\rm R}_s$. This has the same Hamilton--Jacobi structure as the generalised $T\bar T$ flows discussed at the beginning of this section. A related Hamilton--Jacobi description arises for the Yang--Yang function of a single system, with the times of its commuting Hamiltonians as evolution parameters (see for instance~\cite{Nekrasov:2011bc}).

\subsection{The Hamilton--Jacobi recursion}
\label{sec:hj-orders}
We solve~\eqref{eq:hj} as a power series in the coupling. By setting $\hat c_s=\lambda c_s$, we get:
\be
\frac{d \mathcal W}{d\lambda}
=\sum_s c_s\,\partial_{\hat c_s}\mathcal W
=\sum_s c_s \big(\partial_{b_s^{\rm L}}\mathcal W\big)
\big(\partial_{b_s^{\rm R}}\mathcal W\big)\,,
\ee
where for the second equality we used~\eqref{eq:hj}.
Now writing $\mathcal W=\sum_{n\ge0}\lambda^n\mathcal W_n$, keeping the bare times fixed,  we arrive at
\be
\mathcal W_{n+1}=\frac1{n+1}\sum_sc_s\sum_{j=0}^{n}\big(\partial_{b^{\rm L}_s}\mathcal W_j\big)\big(\partial_{b^{\rm R}_s}\mathcal W_{n-j}\big)\,,\qquad
\mathcal W_0=\mathcal F_{\rm L}(b^{\rm L})+\mathcal F_{\rm R}(b^{\rm R})\,.
\label{eq:hj-recursion}
\ee
So each order of the expansion in $\lambda$ of $\mathcal W$ is determined by the lower ones and by the one-kink functions $\mathcal F_a$. At $\lambda=0$ the kinks are uncoupled, and their moments are those of the chiral GGEs at the bare times. The first correction is
\be
\mathcal W_1=\sum_sc_s\,e^{-s\Delta_0}\,M^{\rm L}_sM^{\rm R}_s\,,\qquad \Delta_0=\log b^{\rm L}_1+\log b^{\rm R}_1\,,
\ee
with the moments $M^a_s$ evaluated at the bare times; its $s=1$ term is the $T\bar T$ term of Section~\ref{sec:gluing}. At the physical point~\eqref{physicallatticetimes}, $b^{\rm L}_1=b^{\rm R}_1=2\sqrt3N$, so that
\be
e^{-s\Delta_0}=\big(2\sqrt3\,N\big)^{-2s}\,.
\ee
Each further order brings in one more derivative of the one-kink functions, so that $\mathcal W_n$ is a polynomial in the moments and in their responses to the times, up to order $n-1$. Since every power of the coupling comes with such a factor, at least of order $N^{-2}$, a given order in $1/N$ requires only finitely many terms, all fixed by the data of Section~\ref{sec:chiral-charges}: the conformal moments and their responses. The recursion thus replaces the iterative solution of the self-consistency equations by algebra on these data.

This is the main use of $\mathcal W$: a single function encodes all the moments of the coupled system. Its only input is the pair of one-kink functions $\mathcal F_{\rm L}$ and $\mathcal F_{\rm R}$, which make up $\mathcal W_0$ and encode the moments of each kink on its own, through their expansions in the times~\eqref{eq:hj-F}; the recursion then adds the interaction. Once $\mathcal W$ has been computed to a given order, every moment, including the effect of the interaction between the two kinks, follows by differentiation:
\be
I^a_s=\frac{\partial\mathcal W}{\partial b^a_s}\,,
\ee
evaluated at the physical point~\eqref{physicallatticetimes}. The energy and the local charges of the chain follow from~\eqref{eq:energy-moments} and~\eqref{eq:Qn-moments}: apart from the bulk terms, they are linear combinations of these moments.

\subsubsection{Examples}
We illustrate the recursion in three cases.

\paragraph{Energy at higher orders.} For the ground state, the recursion needs as input the conformal moments $M^{(0)}_s$ and their responses to the times, which we obtained by computing the WKB periods of the local cubic to high order. With these data it reproduces the energy obtained directly in Section~\ref{sec:energy-chain} through $N^{-9}$, and extends it through $N^{-15}$. The coefficients through $N^{-11}$ are compared with the Bethe-ansatz energies in Table~\ref{tab:coupling}; those of $N^{-13}$ and $N^{-15}$ have been checked in the same way, using the exact energies at higher numerical precision.

\paragraph{Spin one, ground state.} Keeping only $s=1$ and selecting the ground state, $\mathcal F_a=M\log\tau_a$ with $M\equiv M^{(0)}_1=\frac{\pi^2}{15}$. With $c=c_1=\sqrt3/\pi$, the recursion~\eqref{eq:hj-recursion} gives
\be
\mathcal W_n=(-1)^{n+1}\,\frac{C_n}{n}\,\frac{c^{\,n}M^{n+1}}{(b_{\rm L}b_{\rm R})^n}\,,\qquad n\ge1\,,
\ee
where $C_n$ are the Catalan numbers $1,1,2,5,14,\dots$. On the symmetric line $b_{\rm L}=b_{\rm R}=b$, the moment of either kink is
\be
I=\left.\frac{\partial\mathcal W}{\partial b_{\rm L}}\right|_{b_{\rm L}=b_{\rm R}=b}
=\frac Mb\sum_{n\ge0}(-1)^nC_n\,x^n\,,\qquad x=\frac{cM}{b^2}\,,
\label{eq:I_recursion_spinone}
\ee
the expansion of the solution of $I\,(b+cI)=M$, which is the one-kink relation $I=M/\tau$ with the feedback~\eqref{eq:hj-feedback} for $s=1$. At the physical point $b=2\sqrt3\,N$, so that $x=\alpha/N^2$, and the energy of the two kinks is
\be
\frac{\cE-N\,e_\infty}{v_F}=-\frac{2\sqrt3}{\pi}\,I=-\frac{\pi}{15N}\sum_{n\ge0}(-1)^nC_n\Big(\frac{\alpha}{N^2}\Big)^n=-\frac{\pi}{15\,\Neff}\,,
\label{eq:spin_one_energy}
\ee
where the series is summed with $\sum_{n\ge0}(-1)^nC_n\,y^n=\big(\sqrt{1+4y}-1\big)/(2y)$, which gives $\Neff(\Neff-N)=\alpha$, that is~\eqref{eq:Neff}. The recursion thus generates the change of scale of Section~\ref{sec:gluing} order by order.

\paragraph{Spin one, unequal states.} At spin one, the same algebraic relations extend to unequal chiral states upon replacing $M$ by $M^a_1=-4\pi^2\mathsf h_a$ in each sector. 
The feedback~\eqref{eq:hj-feedback} for $s=1$, with
$b_1=2\sqrt3\,N$ and $c_1=\sqrt3/\pi$, gives two
different effective lengths for the two 
descriptions
\be 
N^{\rm L,R}_{\rm eff}=\tau^{\rm L,R}_1/(2\sqrt3)\,.
\ee  
Since $I^a_1=M^a_1/\tau^a_1=2\pi\alpha_a/N^a_{\rm eff}$, with $\alpha_a$ defined in \eqref{eq:alphaLR}, we obtain
\be
N^{\rm L}_{\rm eff}=N+\frac{\alpha_{\rm R}}{N^{\rm R}_{\rm eff}}\,,\qquad N^{\rm R}_{\rm eff}=N+\frac{\alpha_{\rm L}}{N^{\rm L}_{\rm eff}}\,.
\ee
Let $e_a=2\pi\mathsf h_a/N^a_{\rm eff}$ be the energy of kink $a$, and $\mathsf t=-1/(4\sqrt3)$ the value of the coupling at the physical point. Then the two equations above can be written as
\be
N^{\rm L}_{\rm eff}=N+2\mathsf t\,e_{\rm R}\,,\qquad N^{\rm R}_{\rm eff}=N+2\mathsf t\,e_{\rm L}\,.
\ee
Defining $\epsilon_\pm=e_{\rm L}\pm e_{\rm R}$ we get\footnote{The relative momentum $\epsilon_-$ also determines the phase of $Y$ at $\theta=\pm i\pi/2$, through $Y(\pm i\pi/2)=e^{\pm i\epsilon_-}$; see Appendix~\ref{app:momentum}.}
\be
N\,\epsilon_-=2\pi(\hL-\hR)\,,\qquad N\,\epsilon_++\mathsf t\big(\epsilon_+^2-\epsilon_-^2\big)=2\pi(\hL+\hR)\,,
\ee
and, differentiating with respect to  $\mathsf t$ at fixed $\hL$ and $\hR$,
\be
\partial_{\mathsf t}\,\epsilon_+=\epsilon_+\,\partial_N\epsilon_++\frac{\epsilon_-^2}{N}\,.
\label{eq:hj-burgers}
\ee
Equation~\eqref{eq:hj-burgers} has the Burgers form characteristic of a $\TTbar$ flow with non-zero momentum~\cite{Smirnov:2016lqw,Cavaglia:2016oda}. In the present setting it is the
spin-$1$ sector of the Hamilton--Jacobi equation
~\eqref{eq:hj}. Including the higher-spin terms gives the full coupled system governing the finite-size expansion considered here, encoded by the Hamilton--Jacobi equation~\eqref{eq:hj}, with the same multi-time structure as the generalised $T\bar T$ flows.

\section{Conclusions}
\label{sec:conclusions}

A spin chain of finite length is expected to be described, at low energies, by the conformal field theory of its scaling limit~\cite{Cardy:1984, Cardy:1986, Blote:1986,Affleck:1986}, deformed by irrelevant operators~\cite{Cardy:1987,Reinicke:1987,Alcaraz:1987uk, Lukyanov:1997bp}. The lattice TBA equation encodes the finite-size corrections exactly, although in a form that is generally solved numerically, while the effective field-theory description is usually available only to the first few orders, since new irrelevant couplings enter at successive orders and must be matched separately. In this work we have made the deformation explicit for the twisted XXZ chain at the Lee--Yang point, using two ingredients. Near each Fermi point, the deformation of each chirality is a generalised Gibbs ensemble of the Lee--Yang theory, with times fixed by the length of the chain. The ODE/IM correspondence for such ensembles gives its charges analytically, through a cubic oscillator. The two chiral sectors are coupled through the tail of the TBA kernel, and the resulting coupling has the structure of a generalised $T\bar T$-type interaction, with coefficients fixed by the kernel. Together these give the energy and the local charges of the chain as asymptotic series in $1/N$, with no free parameters, and organise the finite-size corrections in terms of deformations of the continuum theory.

For a given state, the self-consistency equations determine the effective lengths and the GGE times in terms of the length $N$ of the chain and the corresponding chiral data. At spin one, the coupling has the same self-consistent effective-length structure as a $T\bar T$ finite-size flow, with the bare length $N$ replaced by the effective length $\Neff$, while at higher spins it takes the form of analogous higher-spin couplings~\cite{Smirnov:2016lqw,Cavaglia:2016oda,Conti:2019dxg,Hernandez-Chifflet:2019sua,LeFloch:2019wlf}. The chiral moments generated by the deformed cubic are not expressible solely in terms of the undeformed Lee--Yang KdV charges. They are naturally organised by the deformed KdV hierarchy~\cite{Karimi:2026kdv}, and at order $N^{-9}$ the energy already receives a contribution from a conserved quantity that is not part of the undeformed KdV family.

We tested the series against the Bethe roots, for the ground state and for the first descendant. The coupling between the two chiralities is essential: without it the expansion misses the change of the effective length, while including the successive higher-spin contributions extends the agreement to progressively higher orders in $1/N$.
The coupling can also be written as a Hamilton--Jacobi equation, whose Hamiltonian is the product of the left and right charges of the same spin, as in the generalised $T\bar T$ flows. Its recursion determines the coefficients of the coupled expansion from the two one-kink generating functions. For the symmetric ground state these are related by parity, so that the data of a single Fermi point are sufficient; we used the recursion to extend the ground-state energy from order $N^{-9}$ to order $N^{-15}$. When the left and right chiral sectors are in different states, the spin-one sector has the Burgers form characteristic of a $T\bar T$ flow with non-zero momentum.

The numerical reconstruction of Section~\ref{sec:inv-numerics} provides a complementary check of the relation between the two local cubic descriptions. The reconstructions selected from the left and right Fermi-point limits remain real and smooth over the interval sampled numerically and are mapped into one another by parity. This supports the consistency of the two local descriptions across the band, while an analytic construction of the inverse problem over the full interval remains open.

It is interesting to compare these results with the ground-state energy of a confining string in QED$_3$, computed in~\cite{Aharony:2024ctf} and tested on the lattice in~\cite{Caselle:2025vhx}. There the massive bulk photon corrects the Casimir energy of a string of length $L$ by a term $\frac{1}{\pi L}{\rm Li}_2\big(e^{-mL}\big)$, which comes from massive modes winding around the circle. In our case, the exact sum rule~\eqref{eq:sumrule} and the energy~\eqref{eq:energy-moments} combine into (see equation \eqref{eq:Qn-exact}):
\be
\cE=N\,e_\infty-\frac{\pi v_F}{6\,\Neff}\Big[c_{\rm eff}+\frac{6}{\pi^2}\,{\rm Li}_2\big(-e^{-\epsilon(0)}\big)\Big]+\dots\,,
\ee
where the dots denote the higher-spin contributions. The dilogarithm thus corrects the Casimir term in the same way, but it originates at the centre of the band, at the scale of the lattice, and has the opposite statistics compared to that of \cite{Aharony:2024ctf}. Whether it admits a similar interpretation, for instance in terms of excitations wrapping the chain, remains to be understood.

We have worked at a special anisotropy and twist, where the fusion hierarchy truncates~\cite{Bazhanov:1988bb,Belavin:2001}. There the finite-size corrections that we computed are expressed through the eigenvalues of the deformed local charges alone, which makes the analysis particularly simple. 
A possible route beyond this special point is to start from a non-linear integral equation for the Bethe roots~\cite{Klumper:1991,Destri:1992qk,Destri:1994bv,Klumper:1992vt}, which does not rely on a truncation of the fusion hierarchy and also gives access to the Baxter $Q$-function. It would be interesting to investigate whether the ODE reconstruction developed here can be extended within that framework. We have treated one excited state, the first descendant, whose ODE description is obtained by supplementing the cubic oscillator with apparent singularities; states at higher levels are a natural next step.

The finite-size local charges that we computed are the natural input for the study of generalised Gibbs ensembles of the chain itself, as in the chiral current-carrying eigenstates of~\cite{Sedrakyan:2026chiral}. More generally, an ODE description at finite size could offer a new tool for other finite-size problems in integrable spin chains, such as the Kondo effect in the Heisenberg and XX chains studied with exact methods, for example in~\cite{Kattel:2023kondo,Kattel:2024kondoXX}.

\section*{Acknowledgements}

We thank Max Downing, Faisal Karimi and Adarsh Sudhakar for collaboration on closely related projects and for many discussions, and Davide Fioravanti and Marco Rossi for discussions. We are grateful to Natan Andrei, Ara Sedrakyan and Tigran Sedrakyan for useful discussions, and to the organisers of the workshop EMERGO-2026 at the Max Planck Institute for the Physics of Complex Systems, Dresden, in June 2026, where these discussions took place, as well as to the Institute for hospitality. RT is supported by the INFN project SFT. SKA thanks the Dipartimento di Fisica, Universit\`a di Torino, for hospitality. NP acknowledges support from the Italian Ministry of University and Research (MUR)
under the FIS grant BootBeyond (CUP: D53C24005470001), and from the INFN projects SFT and ST\&FI. NP and RT participate in the project HORIZON-MSCA-2023-SE-01-101182937-HeI. We used the AI assistants Claude (Anthropic) and ChatGPT (OpenAI) to cross-check our results and numerical calculations, as well as to assist in editing the text. 

\appendix

\section{Analytic properties of the transfer-matrix eigenvalue}
\label{app:Lambda}

Section~\ref{sec:chain} builds the lattice TBA equation on
three properties of the eigenvalue $\Lambda(v)$ of the transfer
matrix:
\begin{enumerate}
    \item $\Lambda$ has no poles. \item After removing the known factor $f^N$, the function
$\widetilde Y=Y/f^N$ has no zeros or poles in the physical
strip. \item At $\gamma=3\pi/5$, $\Lambda$
satisfies a functional relation akin to the
Lee--Yang $Y$-system.
\end{enumerate}
These properties are standard in the analysis of the XXZ transfer matrix at roots of unity, and they underlie the derivation of the TBA equations in~\cite{Klumper:1991jsp,Klumper:1992pearce,Bajnok:2014fca}. They can be derived solely from the TQ equation~\cite{Baxter:1982zz} together with the Bethe equations, which will appear as the condition that $\Lambda$ be free of poles. In order to fix our conventions and to make our work self-contained, in this appendix we derive these properties for the twisted chain. We work at $\gamma=3\pi/5$ and $\varphi=2\pi/5$, with $M=N/2$ Bethe roots, and we assume that the roots of the ground state are real.

The spectral parameter $v$ is related to the rapidity variable by $\theta=\pi v/\gamma$. The Bethe roots $\lambda_j$ are the zeros of
\be
Q(v)=\prod_{j=1}^{M}\sinh(v-\lambda_j)\,,
\ee
and the eigenvalue $\Lambda(v)$ of the transfer matrix satisfies Baxter's TQ equation
\be
\Lambda(v)\,Q(v)=e^{i\varphi/2}\sinh^N\!\big(v+\tfrac{i\gamma}2\big)\,Q(v-i\gamma)+e^{-i\varphi/2}\sinh^N\!\big(v-\tfrac{i\gamma}2\big)\,Q(v+i\gamma)\,.
\ee
Dividing by $Q(v)$ we write $\Lambda$ as the sum of two terms, $\Lambda(v)=A(v)+D(v)$, with
\be
\begin{aligned}
A(v)&=e^{i\varphi/2}\,\sinh^N\!\big(v+\tfrac{i\gamma}2\big)\,\frac{Q(v-i\gamma)}{Q(v)}\,,\\
D(v)&=e^{-i\varphi/2}\,\sinh^N\!\big(v-\tfrac{i\gamma}2\big)\,\frac{Q(v+i\gamma)}{Q(v)}\,.
\end{aligned}
\label{eq:app-TQ}
\ee
This formula can also be obtained from the algebraic Bethe ansatz, see for example \cite{Slavnov:2018kfx} for a review.
\subsection{Regularity, and the structure of $\Lambda$}
\label{app:regularityLambda}
The transfer matrix for the XXZ model of spin-$\tfrac12$ is the trace of a product of six-vertex trigonometric $R$-matrices. Since the matrix elements of the R-matrix are functions of the spectral parameter that are free of poles, transfer matrix eigenvalues must also be free of poles. However,
both terms in~\eqref{eq:app-TQ} have a pole appearing at the Bethe roots $v=\lambda_j$. Therefore, we need to impose that such singularities cancel in $\Lambda(v)$. This requires that
\be
e^{i\varphi/2}\sinh^N\!\big(\lambda_j+\tfrac{i\gamma}2\big)\prod_{k=1}^{M}\sinh(\lambda_j-\lambda_k-i\gamma)
+e^{-i\varphi/2}\sinh^N\!\big(\lambda_j-\tfrac{i\gamma}2\big)\prod_{k=1}^{M}\sinh(\lambda_j-\lambda_k+i\gamma)=0\,.
\ee
In each product the factor with $k=j$ is $\sinh(\mp i\gamma)$ and does not depend on the roots, and the two such factors differ by a sign. Moving the second term to the other side and dividing, we obtain
\be
\left[\frac{\sinh(\lambda_j+i\gamma/2)}{\sinh(\lambda_j-i\gamma/2)}\right]^N=e^{-i\varphi}\prod_{k\neq j}\frac{\sinh(\lambda_j-\lambda_k+i\gamma)}{\sinh(\lambda_j-\lambda_k-i\gamma)}\,,
\ee
which is the Bethe equation~\eqref{eq:BAE}. Thus $\Lambda(v)$ is an entire function of $v$ only if the Bethe equation holds, and this procedure to obtain them is known as the analytic Bethe ansatz.

Next we look at the dependence on $v$. For even $N$, $\sinh^N$ does not change under $v\to v+i\pi$, and $Q(v+i\pi)=(-1)^MQ(v)$ drops out of the ratios, so we get the relation
\be
\label{eq:periodicity_Lambda}
\Lambda(v+i\pi)=\Lambda(v)\,.
\ee
We can automatically satisfy this periodicity by assuming that $\Lambda$ is a function of $z=e^{2v}$. As ${\rm Re}\,v\to+\infty$ the ratios $Q(v\mp i\gamma)/Q(v)$ tend to $e^{\mp iM\gamma}$ and thus the two terms of~\eqref{eq:app-TQ} tend to $e^{\pm i\varphi/2}\,2^{-N}e^{Nv}\,e^{\pm i(N/2-M)\gamma}$. For the ground state $M=N/2$, the phases proportional to $\gamma$ cancel, and we remain with
\be
\Lambda(v)\to2\cos\frac{\varphi}{2}\;2^{-N}\,e^{Nv}\qquad({\rm Re}\,v\to+\infty)\,.
\label{eq:app-Lambda-right}
\ee
The computation at $\text{Re}(v)\rightarrow-\infty$ is identical except that we have $\sinh x\to-\frac12e^{-x}$, and thus gives the same coefficient with $e^{-Nv}$ in place of $e^{Nv}$.

These limits say that $e^{Nv}\Lambda(v)$ is a function of $z$ with at most a pole of order $N$ at infinity and no pole at $z=0$, that is, a polynomial of degree at most $N$ in $z$. By the two limits, its coefficient of $z^N$ and its constant term are both
\be
2^{1-N}\cos\frac{\varphi}{2}\,,
\ee
which is nonzero for $\varphi=2\pi/5$. So $\Lambda$ has exactly $N$ zeros in each period strip ${\rm Re}\,v\in\mathbb R$, $0\le{\rm Im}\,v<\pi$. 

\subsection{The zeros}
\label{app:zeros}

The analyticity hypothesis of Section~\ref{sec:chain-tba} needs $\Lambda$ to have no zeros in the physical strip $|{\rm Im}\,\theta|<\pi/3$, which is $|{\rm Im}\,v|<\gamma/3=\pi/5$.

For real roots, it is easy to see from the definition \eqref{eq:app-TQ} that complex conjugation exchanges ${A(\bar v)}=\overline D(v)$, so $\Lambda$ is real on the real axis and together with the result of the last section, it is a polynomial with real coefficients. Its zeros are therefore real or come in complex-conjugate pairs. The real axis of $z=e^{2v}$ consists of two lines in the $v$ plane, ${\rm Im}\,v=0$ for $z>0$ and ${\rm Im}\,v=\pi/2$ for $z<0$, and we examine the latter. There, we have that $v=u+i\pi/2$ with $u$ real, and using simple trigonometric identities we can see that 
the ratio $A/D$ is a pure phase:
\be
\frac{A}{D}=e^{i\Psi(u)}\,,\qquad
\Psi(u)=\varphi+2N\arg\cosh\big(u+\tfrac{i\gamma}2\big)-2\sum_{j=1}^M\arg\cosh\big(u-\lambda_j+i\gamma\big)\,.
\label{eq:app-Psi}
\ee
Furthermore, on this line $D$ is finite and does not vanish, so
\be
\Lambda=D\,\big(1+e^{i\Psi(u)}\big)
\ee
vanishes exactly where $\Psi(u)\in\pi+2\pi\mathbb Z$.
We now show that $\Psi$ is a strictly increasing function. In fact its derivative is given by
\be
\Psi'(u)=\frac{2N\sin\frac{2\pi}{5}}{\cosh2u-\cos\frac{2\pi}{5}}+\sum_{j=1}^M\frac{2\sin\frac{\pi}{5}}{\cosh2(u-\lambda_j)-\cos\frac{\pi}{5}}\,.
\ee
Every term is positive, because $\cosh2x\ge1$, and therefore $\Psi'>0$. To find the total change of $\Psi$ we integrate, using the identity \be 
\int_{-\infty}^{\infty}dx\,\frac{\sin\alpha}{\cosh2x+\cos\alpha}=\alpha~,\quad \text{for}\quad  |\alpha|<\pi~.
\ee 
Adding the contributions of the integrals, with $\alpha=3\pi/5$ for the first term and $\alpha=4\pi/5$ for each term of the sum, we get
\be
\Psi(+\infty)-\Psi(-\infty)=\frac{6\pi N}{5}+\frac{8\pi M}{5}=2\pi N\,,
\ee
where we used $M=N/2$. Since $\Psi'(u)>0$, each solution of
$\Psi(u)=\pi+2\pi\mathbb Z$ is simple. The total change
$2\pi N$ therefore gives exactly $N$ zeros of $\Lambda$ on
the line ${\rm Im}\,v=\pi/2$. Since $\Lambda$ is a polynomial
of degree $N$ in $e^{2v}$, these zeros exhaust all its zeros
in a period strip.

\subsection{A functional relation for \texorpdfstring{$\Lambda$}{Lambda} at \texorpdfstring{$\gamma=3\pi/5$}{gamma=3pi/5}}
\label{app:closure}

The truncation of the fusion hierarchy at roots of unity was found for RSOS models in~\cite{Bazhanov:1988bb}. For vertex models it was shown for the quantum-group invariant open XXZ chain in~\cite{Belavin:1999}, for the XXZ chain with twisted boundary conditions in~\cite{Belavin:2001}, and for periodic and open boundary conditions in~\cite{Nepomechie:2002}. The analogous closure for the quantum transfer matrix of the untwisted chain at rational anisotropy is proved in~\cite{Kuniba:1998} by direct computation. Here we derive it for the twisted chain, directly from the TQ equation, following the argument of~\cite{Bazhanov:1988bb} for RSOS models.

The reason to expect a relation among the eigenvalues at shifted arguments is the following. For $\gamma=3\pi/5$ five shifts of $v$ by $i\gamma$ add up to $3i\pi$, and also
\be
Q(v+5i\gamma)=(-1)^MQ(v)\,.
\ee
So the shifts of $Q$ close on a cycle of length five, and a product of eigenvalues at shifted arguments contains only a finite number of different terms. To follow them, we abbreviate
\be
s_l=\sinh\big(v+(l-\tfrac12)i\gamma\big)\,,\qquad Q_m=Q(v+im\gamma)\,,\qquad \omega=e^{i\varphi}\,.
\ee
The cycle of length five is expressed by
\be
s_{l+5}^N=s_l^N\,,\qquad Q_{m+5}=(-1)^MQ_m\,,\qquad \omega^5=1\,,\qquad \omega^{5/2}=-1\,.
\label{eq:app-cycle}
\ee
A shift of $v$ by $ik\gamma$ shifts all the indices by $k$
\be
\Lambda(v+ik\gamma)=\omega^{1/2}\,\frac{s_{k+1}^N\,Q_{k-1}}{Q_k}+\omega^{-1/2}\,\frac{s_k^N\,Q_{k+1}}{Q_k}\,,
\label{eq:app-Lambda-shift}
\ee
and the two terms at $k=0$ are $A(v)$ and $D(v)$.
We now multiply the two eigenvalues $\Lambda(v)\Lambda(v+i\gamma)=\big[A(v)+D(v)\big]\big[A(v+i\gamma)+D(v+i\gamma)\big]$. We get a total of four terms:
\be
\begin{aligned}
D(v)A(v+i\gamma)&=s_0^N\,s_2^N\,,&\qquad A(v)A(v+i\gamma)&=\omega\,s_1^Ns_2^N\,\frac{Q_{-1}}{Q_1}\,,\\
A(v)D(v+i\gamma)&=s_1^{2N}\,\frac{Q_{-1}Q_2}{Q_0Q_1}\,,&\qquad D(v)D(v+i\gamma)&=\omega^{-1}\,s_0^Ns_1^N\,\frac{Q_2}{Q_0}\,.
\end{aligned}
\ee
The first is free of $Q$'s. The other three can be written in terms of the functions
\be
f_m=\frac{\omega^m\,s_m^N}{Q_{m-1}\,Q_m}
\ee
as
\be
\Lambda(v)\Lambda(v+i\gamma)=s_0^Ns_2^N+\omega^{-1}\,s_1^N\,Q_{-1}Q_2\,\big(f_0+f_1+f_2\big)\,.
\label{eq:app-product}
\ee
By~\eqref{eq:app-cycle}, $f_{m+5}=f_m$, so $f_0,\dots,f_4$ is a complete cycle, and the last term contains three of its five members. The other two, $f_3$ and $f_4$, are contained in $\Lambda(v+3i\gamma)$, since by~\eqref{eq:app-Lambda-shift} at $k=3$ its two terms contain $s_4^NQ_2/Q_3$ and $s_3^NQ_4/Q_3$, which are $f_4$ and $f_3$ up to overall factors. To bring the product to a form with $\Lambda$ alone we therefore need the sum over the whole cycle,
\be
S(v)=\sum_{m=0}^{4}f_m(v)\,,
\ee
to vanish. This can be proven by following the argument of~\cite{Bazhanov:1988bb} for RSOS models, with trigonometric functions in place of elliptic ones.

First, $S$ has no poles. The denominators of the terms are the products $Q_{m-1}Q_m$, so a pole can only sit at a zero of one of the functions $Q_r$. Each $Q_r$ appears in exactly two terms, $f_r$ and $f_{r+1}$, and the cycle closes because $Q_{-1}=(-1)^MQ_4$, so $f_4$ and $f_0$ share one. Let $v_0$ be a zero of $Q_r$, that is $v_0+ir\gamma=\lambda_j$ for some root, up to a multiple of $i\pi$. The roots are regular, so $Q_{r\pm1}$ do not vanish at $v_0$, and the pole is simple, with residue
\be
\frac{1}{Q_r'(v_0)}\left[\frac{\omega^r\,s_r^N}{Q_{r-1}}+\frac{\omega^{r+1}\,s_{r+1}^N}{Q_{r+1}}\right]_{v=v_0}\,.
\ee
At $v_0$ one has $s_r=\sinh(\lambda_j-\tfrac{i\gamma}2)$, $s_{r+1}=\sinh(\lambda_j+\tfrac{i\gamma}2)$, $Q_{r-1}=\prod_k\sinh(\lambda_j-\lambda_k-i\gamma)$ and $Q_{r+1}=\prod_k\sinh(\lambda_j-\lambda_k+i\gamma)$. Multiplying the bracket by $Q_{r-1}Q_{r+1}$, it vanishes when
\be
\omega\,\sinh^N\!\big(\lambda_j+\tfrac{i\gamma}2\big)\prod_k\sinh(\lambda_j-\lambda_k-i\gamma)+\sinh^N\!\big(\lambda_j-\tfrac{i\gamma}2\big)\prod_k\sinh(\lambda_j-\lambda_k+i\gamma)=0\,,
\ee
which is the condition found in Appendix \ref{app:regularityLambda} for the cancellation of the poles of $\Lambda$, that is the Bethe equation for $\lambda_j$. So the residues vanish, and $S$ is an entire function of $v$.

Second, we see how $S$ changes under a shift of $v$. Shifting $v$ by $i\gamma$ raises every index of $f_m$ by one, and from the definition
\be
f_m(v+i\gamma)=\omega^{-1}f_{m+1}(v)\,.
\ee
Summing over $m=0,\dots,4$ and using $f_5=f_0$ from~\eqref{eq:app-cycle}, we find $S(v+i\gamma)=\omega^{-1}S(v)$, and two shifts give a factor $\omega^{-2}$. The function $S$ also has period $i\pi$, since it is a sum of ratios of products of $N$ factors $\sinh$. Because $2\gamma=\pi+\pi/5$, the shift by $2i\gamma$ is a shift by $i\pi/5$, and
\be
S\big(v+\tfrac{i\pi}5\big)=\omega^{-2}\,S(v)\,.
\ee
In particular $|S|$ is periodic in ${\rm Im}\,v$ with period $\pi/5$.

Third, $S$ is bounded. As ${\rm Re}\,v\to\pm\infty$ the numerator $s_m^N$ of each $f_m$ and its denominator $Q_{m-1}Q_m$, a product of $2M=N$ factors $\sinh$, grow in the same way, so each $f_m$ tends to a constant. Together with the periodicity of $|S|$ in ${\rm Im}\,v$ this makes $S$ bounded on the whole plane. A bounded entire function is constant, by Liouville's theorem, and the constant $c$ satisfies $c=\omega^{-2}c$ with $\omega^{-2}\neq1$. So $c=0$ and $S\equiv0$.

With $S=0$ we have $f_0+f_1+f_2=-(f_3+f_4)$, and with $Q_4=(-1)^MQ_{-1}$ equation~\eqref{eq:app-product} becomes
\be
\Lambda(v)\Lambda(v+i\gamma)=s_0^Ns_2^N-s_1^N\Big[\omega^{2}\,s_3^N\frac{Q_{-1}}{Q_3}+(-1)^M\omega^{3}\,s_4^N\frac{Q_2}{Q_3}\Big]\,.
\ee
By~\eqref{eq:app-Lambda-shift} at $k=3$, the bracket is $(-1)^M\omega^{5/2}\Lambda(v+3i\gamma)=-(-1)^M\Lambda(v+3i\gamma)$, and the product of two eigenvalues closes on $\Lambda$,
\be
\Lambda(v)\,\Lambda(v+i\gamma)=\sinh^N\!\big(v-\tfrac{i\gamma}2\big)\sinh^N\!\big(v+\tfrac{3i\gamma}2\big)+(-1)^M\sinh^N\!\big(v+\tfrac{i\gamma}2\big)\,\Lambda(v+3i\gamma)\,.
\label{eq:Lambda-closure}
\ee

\subsection{The \texorpdfstring{$Y$}{Y}-system}
\label{app:Ysystem}

The relation~\eqref{eq:Lambda-closure} has the shifts $i\gamma=3i\pi/5$ and $3i\gamma=9i\pi/5$, and $\Lambda$ has period $i\pi$. We shift $v\rightarrow v-4i\pi/5$ and use the periodicity relation \eqref{eq:periodicity_Lambda} $\Lambda(v-4i\pi/5)=\Lambda(v+i\pi/5)$ and $\Lambda(v+i\pi)=\Lambda(v)$. This gives
\be
\Lambda\big(v-\tfrac{i\pi}5\big)\,\Lambda\big(v+\tfrac{i\pi}5\big)=\sinh^N\!\big(v-\tfrac{11i\pi}{10}\big)\sinh^N\!\big(v+\tfrac{i\pi}{10}\big)+(-1)^M\sinh^N\!\big(v-\tfrac{i\pi}2\big)\,\Lambda(v)\,.
\ee
Since $N$ is even, $\sinh^N(v-\tfrac{11i\pi}{10})=\sinh^N(v-\tfrac{i\pi}{10})$ and $(-1)^M\sinh^N(v-\tfrac{i\pi}2)=\cosh^Nv$. So
\be
\Lambda\big(v-\tfrac{i\pi}5\big)\,\Lambda\big(v+\tfrac{i\pi}5\big)=\Big[\sinh\big(v-\tfrac{i\pi}{10}\big)\sinh\big(v+\tfrac{i\pi}{10}\big)\Big]^N+\cosh^Nv\;\Lambda(v)\,.
\label{eq:app-Lambda-shifted}
\ee
To bring this into a more familiar form, introduce the function
\be
g(v)=\frac{-i\cosh v}{\sinh\big(v-\tfrac{i\pi}{10}\big)\,\sinh\big(v+\tfrac{i\pi}{10}\big)}\,.
\ee
Taking the $N$th power of this function and using trigonometric identities, one can check that the identity involving $\Lambda$ takes the form:
\be
\big[g^N\Lambda\big]\big(v-\tfrac{i\pi}5\big)\,\big[g^N\Lambda\big]\big(v+\tfrac{i\pi}5\big)=1+(-1)^M\,\big[g^N\,\Lambda](v)\,.
\ee
Defining $Y=(-1)^Mg^N\Lambda$, and recalling that $v=3\theta/5$, we obtain the $Y$-system in~\eqref{eq:Ysystem}:
\be
Y\big(\theta+\tfrac{i\pi}{3}\big)\,Y\big(\theta-\tfrac{i\pi}{3}\big)=1+Y(\theta)\,.
\ee

\subsection{The behaviour of \texorpdfstring{$Y$}{Y} at large \texorpdfstring{$\theta$}{theta}}
\label{app:Yasymptotics}

The limit of $\Lambda$ found in Appendix \ref{app:regularityLambda} and the explicit form of $g$ from Appendix \ref{app:Ysystem} give the behaviour of $Y$ at large $|{\rm Re}\,\theta|$, which is that of $v=\gamma\theta/\pi$. As ${\rm Re}\,v\to\pm\infty$ every $\sinh$ and $\cosh$ in $g$ is dominated by its exponential, and
\be
g(v)\to-2i\,e^{\mp v}\,,\qquad g^N(v)\to(-1)^M\,2^N\,e^{\mp Nv}\,,
\ee
where in the second we assumed that $N$ is even. By~\eqref{eq:app-Lambda-right} and the corresponding limit at the other end, $\Lambda(v)\to2\cos\tfrac{\varphi}{2}\;2^{-N}e^{\pm Nv}$. In the product $Y=(-1)^Mg^N\Lambda$ the exponentials cancel, and the factors $(-1)^M$ cancel too, obtaining
\be
Y(\theta)\to2\cos\frac{\varphi}{2}=\frac{1+\sqrt5}{2}\qquad(\theta\to\pm\infty)\,.
\ee

\subsection{The value of \texorpdfstring{$Y$}{Y} at \texorpdfstring{$\theta=\pm i\pi/2$}{theta=±iπ/2}}
\label{app:momentum}

In Section~\ref{sec:chain-walls} the order of the zeros at $\theta=\pm5i\pi/6$ depends on the value of $Y$ at $\theta=\pm i\pi/2$. This value is a root of unity, fixed by the momentum of the state.

At $\theta=i\pi/2$ one has $v=i\gamma/2$, and the second term of the TQ equation vanishes due to the factor $\sinh^N(v-\tfrac{i\gamma}2)$. Setting $\gamma=3\pi/5$, we can check that  $g^N(\tfrac{i\gamma}2)\sinh^N(i\gamma)=(-1)^N=1$ for even $N$. Hence
\be
Y\big(\tfrac{i\pi}{2}\big)=(-1)^M\,e^{i\varphi/2}\prod_{j=1}^M r_j\,,\qquad r_j=\frac{\sinh(\lambda_j+\tfrac{i\gamma}2)}{\sinh(\lambda_j-\tfrac{i\gamma}2)}\,.
\ee
At $v=-i\gamma/2$ the first term of the TQ equation vanishes, and the same steps give $Y(-i\pi/2)=(-1)^Me^{-i\varphi/2}\prod_jr_j^{-1}$.

To see that this is a root of unity we multiply the Bethe equations,
\be
r_j^N=e^{-i\varphi}\prod_{k\neq j}\frac{\sinh(\lambda_j-\lambda_k+i\gamma)}{\sinh(\lambda_j-\lambda_k-i\gamma)}\,,
\ee
over all $j$. The factor for the pair $(j,k)$ is the inverse of the one for $(k,j)$, so the right-hand sides combine to $e^{-iM\varphi}$. 
For $M=N/2$ this says that $\big(e^{i\varphi/2}\prod_jr_j\big)^N=1$, and since $N$ is even, for any solution of the Bethe equations in the sector $M=N/2$,
\be
Y\big(\pm\tfrac{i\pi}{2}\big)^N=1\,,
\ee
so that we may write $Y(\pm i\pi/2)=e^{\mp2\pi i k/N}$ with $k\in\mathbb Z$.
The product $\prod_j r_j$ determines the phase associated
with shifting the spin chain by one lattice site. In the
present twisted normalisation, however, the phase
$e^{i\varphi/2}$ in $Y$ must also be kept explicitly. We
therefore define $k$ from the phase of $Y$, so that the
corresponding lattice momentum is $P=2\pi k/N$.

For real roots $k$ is the sum of the quantum numbers of the Bethe equations. For the ground state $k=0$ and $Y(\pm i\pi/2)=1$. For the state in which the right kink carries a
descendant of level one, $k=\bar h-h=1$. Thus the 
momentum label $k$ is the negative of the conformal spin
$h-\bar h$. In this state
$Y(\pm i\pi/2)=e^{\mp2\pi i/N}$; then $1+Y(\pm i\pi/2)\neq0$ for $N>2$, and the zeros at $\pm5i\pi/6$ keep order $N$. The same momentum label determines the relative momentum $\epsilon_-$ of the two kinks in Section~\ref{sec:hj}. Indeed, from $N\epsilon_-=2\pi(\hL-\hR)$ and $k=\hR-\hL$, we have $\epsilon_-=-2\pi k/N$, so that $Y(\pm i\pi/2)=e^{\pm i\epsilon_-}$.

\section{Classical and quantum periods of the cubic}
\label{app:classes}

This appendix collects the geometry of the cubic curve that is used in Sections~\ref{sec:inverse} and~\ref{sec:chiral}. 

\subsection{The cycles}

Consider the curve
\be
w^2=V(x)=x^3+Qx-P
\ee
with real coefficients, $P>0$, and
$4Q^3+27P^2>0$, so that we are in the chamber with one
real turning point $x_0$ and a complex-conjugate pair
$x_\pm$. We stay away from the discriminant locus
$4Q^3+27P^2=0$. We use three cycles: $\gamma_0$, which encircles the pair $x_+$ and $x_-$, and $\gamma_\pm$, which encircle $x_0$ and $x_\pm$ respectively. With the orientations fixed by the identification with the Stokes multipliers of Section~\ref{sec:inv-periods}, they satisfy the homology relation
\be
\gamma_++\gamma_-=\gamma_0\,.
\ee
The classical period of a cycle is
\be
\Pi^{(0)}_\gamma(Q,P)=\oint_\gamma\sqrt{V(x)}\,dx\,.
\ee
The higher periods are defined by the WKB expansion of the solution of $\psi''=V\psi$. Writing $\psi=\exp\int S\,dx$, the function $S$ obeys the Riccati equation $S'+S^2=V$, whose solution in the WKB parameter is
\be
S=\sum_{n\ge0}S_n\,,\qquad S_0=\sqrt V\,,\qquad
S_n=-\frac{1}{2S_0}\Big(S'_{n-1}+\sum_{i=1}^{n-1}S_iS_{n-i}\Big)\,.
\ee
The first odd term is
$S_1=-\tfrac14 V'/V$, which is a logarithmic derivative
and gives the winding contribution associated with the WKB
prefactor. This contribution is already accounted for in the
normalisation above. The higher odd terms $S_{2k+1}$ with
$k\geq1$ are total derivatives and do not contribute to the
quantum periods below. The non-vanishing higher periods are
\be
\Pi^{(2k)}_\gamma(Q,P)=\oint_\gamma S_{2k}\,dx\,,\qquad k=1,2,\dots\,.
\ee
Each $S_{2k}$ is a finite sum of terms of the form $x^aV^{1/2-m}$ with coefficients polynomial in $Q$, and the integral of each term over a cycle is a derivative of the classical period with respect to $Q$ and $P$, as explained in Section \ref{quantumcorrections} below. The higher periods are therefore differential operators in $(Q,P)$ acting on the classical period, which is the Picard--Fuchs form~\eqref{eq:PF} used in the main text.

\subsection{The classical periods in closed form}

Expanding the square root in powers of $Q$, the classical period can be written as:
\be
\Pi^{(0)}_\gamma(Q,P)=\sum_{n\ge0}\frac{Q^n}{n!}\,\big(\tfrac12\big)\big(\tfrac12-1\big)\cdots\big(\tfrac12-n+1\big)\oint_\gamma x^n\,(x^3-P)^{\frac12-n}\,dx\,.
\ee
With the substitution $x=P^{1/3}t$ each term becomes
\be
\Pi^{(0)}_\gamma=P^{5/6}\sum_{n\ge0}\frac{\kappa^n}{n!}\,\big(\tfrac12\big)\big(\tfrac12-1\big)\cdots\big(\tfrac12-n+1\big)\,J_n^{\gamma}\,,\qquad
\kappa=\frac{Q}{P^{2/3}}\,,
\label{eq:appA-kappa}
\ee
where $J_n^\gamma$ is the integral of $t^n(t^3-1)^{1/2-n}$ over the cycle, which encircles two of the three cube roots of unity.

The cycle integrals are Beta functions. For the cycle around the complex pair $\omega_3=e^{2\pi i/3}$ and $\omega_3^2$,
\be
J_n^{\gamma_0}=\begin{cases}\displaystyle\frac{2}{\sqrt3}\,(-1)^{\lfloor n/3\rfloor}\,B\!\left(\frac{n+1}{3},\frac32-n\right) & n\not\equiv2\pmod3\\[8pt]
0 & n\equiv2\pmod3\end{cases}
\label{eq:appA-Jn}
\ee
for the orientation fixed above. The Beta function with a negative second argument is understood by analytic continuation (the contour integral regularises the endpoint divergences). For the cycles around $(1,\omega_3^{\pm1})$ the same value is multiplied by $-e^{\mp2\pi i(n+1)/3}$.

Inserting~\eqref{eq:appA-Jn} in~\eqref{eq:appA-kappa} and using $\Gamma(n-\frac12)\Gamma(\frac32-n)=-\pi(-1)^n$, the coefficient of $\kappa^n$, including the factor $2/\sqrt3$, is
\be
\Pi^{(0)}_{\gamma_0}=P^{5/6}\sum_{\substack{n\ge0\\ n\not\equiv2\,(3)}}a_n\,\kappa^n\,,\qquad
a_n=\sqrt{\frac{\pi}{3}}\;\frac{(-1)^{\lfloor n/3\rfloor}\,\Gamma\!\big(\frac{n+1}{3}\big)}{n!\;\Gamma\!\big(\frac{11}{6}-\frac{2n}{3}\big)}\,,
\label{eq:appA-Piseries}
\ee
with, in particular,
\be
a_0=\cA=\frac{2}{\sqrt3}\,B\!\left(\tfrac13,\tfrac32\right)\,,\qquad
a_1=\frac{\cB}{2}\,,\qquad \cB=\frac{2}{\sqrt3}\,B\!\left(\tfrac23,\tfrac12\right)\,.
\ee
Writing $n=3m+r$ with $r=0,1$, the ratio of consecutive terms in each residue class is rational in $m$,
\be
\frac{a_{n+3}\,\kappa^{n+3}}{a_n\,\kappa^n}=-\,\frac{4\,\big(m+\frac r3-\frac5{12}\big)\big(m+\frac r3+\frac1{12}\big)}{27\,\big(m+\frac{r+2}{3}\big)\big(m+\frac{r+3}{3}\big)}\;\kappa^3\,,
\ee
which is the ratio of the terms of a Gauss hypergeometric function of
\be
z=-\frac{4\kappa^3}{27}=-\frac{4Q^3}{27P^2}\,.
\ee
For $r=0$ the parameters are $(a,b;c)=(-\frac5{12},\frac1{12};\frac23)$ and for $r=1$ they are $(-\frac1{12},\frac5{12};\frac43)$, so that each class sums to ${}_2F_1(a,b;c;z)$ with the prefactor given by its first term, $\cA$ and $\cB/2$. This is the expression~\eqref{eq:Pi-series} of Section~\ref{sec:inv-periods}.

Finally, the radius of convergence of~\eqref{eq:appA-Piseries}
is $|\kappa|^3=27/4$, corresponding to $|z|=1$. The point
$z=1$ is the discriminant locus
$4Q^3+27P^2=0$. Beyond the radius of convergence, the
hypergeometric representation provides the analytic continuation
of the periods.

\subsection{The quantum corrections}
\label{quantumcorrections}

The higher terms of the WKB series are integrals of expressions in $V$ and its derivatives. For the cubic $V''=6x$ and $V'''=6$, so that every term reduces to integrals of the form
\be
\oint_\gamma x^a\,V^{\frac12-m}\,dx\,.
\ee
These are derivatives of the classical period, because $V$
depends on the coefficients only through
\be
\p_QV=x\,,\qquad \p_PV=-1\,,\qquad
\text{hence}\qquad \p_Q=-x\,\p_P
\quad\text{on functions of }V\,.
\ee
For $m\geq a$ this gives
\be
\oint_\gamma x^a\,V^{\frac12-m}\,dx
\ \propto\
\p_Q^{\,a}\,\p_P^{\,m-a}\,\Pi^{(0)}_\gamma\,.
\ee
When $a>m$, powers of $x$ are first reduced using
$x^3=V-Qx+P$, together with integration by parts, until the
result is expressed in terms with $m\geq a$. The same
Picard--Fuchs reduction then applies.

\section{Derivation of a sum rule}
\label{app:sumrule}
This appendix derives the sum rule~\eqref{eq:sumrule} and the exact equation~\eqref{eq:Neff-exact} for the ground state.
We begin with the folded TBA equation for $\theta<0$:
\be
\epsilon(\theta)=\nu_\tau(\theta)-\int_{-\infty}^{0}d\theta'\,\Phi(\theta-\theta')\,L(\theta')\,,\qquad
\nu_\tau(\theta)=\sum_s\tau_s\,e^{s\theta}\,,
\label{eq:B-folded}
\ee
with
\be
L=\log\big(1+e^{-\epsilon}\big)\,,\qquad I_s=\int_{-\infty}^{0}e^{s\theta}\,L(\theta)\,d\theta\,,
\ee
and the sum over $s$ is understood with Abel's prescription at $\theta\to0^-$. As $\theta\to-\infty$ the driving term vanishes and $\epsilon$ tends to the plateau of the Lee--Yang $Y$-function,
\be
e^{\epsilon_\infty}=\phi\,,\qquad \phi=\frac{1+\sqrt5}{2}\,,\qquad L_\infty=\epsilon_\infty=\log\phi\,.
\ee
We need two identities. The first is local. With
\be
\mathcal G(\epsilon)=2\,{\rm Li}_2\big(-e^{-\epsilon}\big)-\epsilon\,\log\big(1+e^{-\epsilon}\big)
\ee
and $d\,{\rm Li}_2(-e^{-\epsilon})/d\epsilon=\log(1+e^{-\epsilon})$, one finds, for any function $\epsilon(\theta)$,
\be
\epsilon'L-\epsilon L'=\frac{d}{d\theta}\,\mathcal G\big(\epsilon(\theta)\big)\,,
\label{eq:B-local}
\ee
where a prime denotes $d/d\theta$.

The second identity is non-local. Differentiating~\eqref{eq:B-folded} and integrating by parts in $\theta'$, using $\Phi'(\theta-\theta')=-\partial_{\theta'}\Phi(\theta-\theta')$ and the decay of the kernel, which leaves a boundary term only at $\theta'=0$, gives
\be
\nu_\tau'=\epsilon'-\Phi(\theta)\,L(0)+\Phi*L'\,,\qquad
\nu_\tau=\epsilon+\Phi*L\,,
\ee
where $\Phi*f$ denotes the convolution over $\theta'<0$ in~\eqref{eq:B-folded}. Therefore
\be
\nu_\tau'L-\nu_\tau L'=\epsilon'L-\epsilon L'-L(0)\,\Phi(\theta)\,L(\theta)+(\Phi*L')\,L-(\Phi*L)\,L'\,.
\ee
When integrated over $\theta<0$ the last two terms cancel, because $\Phi$ is even and the two double integrals differ only by a relabelling of the variables. With~\eqref{eq:B-local} this leaves
\be
\int_{-\infty}^{0}\big(\nu_\tau'L-\nu_\tau L'\big)\,d\theta=\mathcal G\big(\epsilon(0)\big)-\mathcal G(\epsilon_\infty)-L(0)\int_{-\infty}^{0}\Phi(\theta)\,L(\theta)\,d\theta\,.
\label{eq:B-ident}
\ee
The left side can be computed from the definition of $\nu_\tau$. Since $\int_{-\infty}^0e^{s\theta}\,(sL-L')\,d\theta=2sI_s-L(0)$, it equals $2\sum_ss\,\tau_sI_s-L(0)\sum_s\tau_s$. Evaluating~\eqref{eq:B-folded} at $\theta=0$ and using the evenness of $\Phi$, the terms proportional to $L(0)$ combine into $L(0)\,\epsilon(0)$, which cancels the corresponding term of $\mathcal G(\epsilon(0))$, and
\be
\sum_s s\,\tau_sI_s={\rm Li}_2\big(-e^{-\epsilon(0)}\big)-\frac12\,\mathcal G(\epsilon_\infty)\,.
\ee
At the plateau $\mathcal G(\epsilon_\infty)=2\,{\rm Li}_2(-1/\phi)-\log^2\phi$, and Rogers' value ${\rm Li}_2(-1/\phi)=-\pi^2/15+\frac12\log^2\phi$ gives $\mathcal G(\epsilon_\infty)=-2\pi^2/15$. Hence
\be
\sum_{s\equiv\pm1\,(6)}s\,\tau_sI_s=\frac{\pi^2}{15}+{\rm Li}_2\big(-e^{-\epsilon(0)}\big)\,,
\ee
which is~\eqref{eq:sumrule}. The dilogarithm is the only term that depends on the centre of the band: $\epsilon(0)$ grows linearly with $N$, so it is exponentially small. In the conformal limit this term vanishes, and the sum rule reduces to the dilogarithm computation of the effective central charge in~\cite{Bajnok:2014fca}; at finite $N$ it is what makes the identity exact.

To obtain~\eqref{eq:Neff-exact}, recall that for the ground state $\tau_s=b_s+c_sI_s$, so that $\tau_1=2\sqrt3\,N+(\sqrt3/\pi)\,I_1$. Defining $\Neff$ by $\tau_1=2\sqrt3\,\Neff$ gives $I_1=2\pi(\Neff-N)$, and the spin-one term of the sum rule is $4\sqrt3\,\pi\,\Neff(\Neff-N)$. Moving the terms with $s\ge5$, which we denote by $G$, to the other side and dividing by $4\sqrt3\,\pi$, we find the identity quoted in~\eqref{eq:Neff-exact}:
\be
\Neff\,(\Neff-N)=\frac{\pi}{60\sqrt3}\Big[1+\frac{15}{\pi^2}\,{\rm Li}_2\big(-e^{-\epsilon(0)}\big)-\frac{15}{\pi^2}\,G\Big]\,,\qquad G=\sum_{s\ge5}s\,\tau_sI_s\,.
\ee

\IfFileExists{JHEP.bst}{\bibliographystyle{JHEP}}{\bibliographystyle{unsrt}}
{\small\bibliography{refs}}
\end{document}